# Aspects of the physical principles of the proton therapy with inclusion of nuclear interactions

W. Ulmer

MPI of Physics, Goettingen, Germany

A. Carabe-Fernandez

University of Pennsylvania, Philadelphia, USA

**Abstract**

**The radiotherapy of malignant diseases has reached much progress during the past decade. Thus, intensity modulated radiation therapy (IMRT) and VMAT (Rapidarc) now belong to the standard modalities of tumor treatment with high energy radiation in clinical practice. In recent time, the particle therapy (protons and partially with heavy carbon ions) has reached an important completion of these modalities with regard to some suitable applications. In spite of this enrichment essential features need further research activities and publications in this field: Nuclear reactions and the role of the released neutrons; electron capture of positively charged nuclei at lower projectile energies (e.g. in the environment of the Bragg peak and at the distal end of the particle track); correct dose delivery in scanning methods by accounting for the influence of the lateral scatter of beam-lets. Deconvolution methods can help to overcome these problems, which already occur in radiotherapy of very small photon beams [1 - 8].**

# Introduction

The connection between the initial energy $E_0$ of a projectile particle (e.g., proton, electron, α-particle) and the range $R_{CSDA}$ in a (homogenous) medium (e.g., water) by continuous energy loss, which is referred to as stopping power, represents an important challenge for many questions in radiation physics. Thus, CSDA (<u>c</u>ontinuous <u>s</u>lowing <u>d</u>own <u>a</u>pproximation) only considers the energy transfer from the projectile to the medium by continuous damping of the particle motion; statistical fluctuations (energy straggling) of the residual energy $E(z)$ or its gradient $dE(z)/dz$ at the position z are not taken into account. The phenomenon of the local fluctuation of the energy transfer (range straggling effects) requires the introduction of convolutions based on statistical mechanics and/or quantum statistics (Landau/Vavilov- and Gaussian distributions).

Nevertheless, the damping of particle motion (energy dissipation) by interaction with the environment has become an important subject in many disciplines of physics, as it also pertains to the domain of the thermodynamics of irreversible processes. We have previously shown that the determination of $R_{CSDA}$ results from a generalized (non-relativistic) Langevin equation or by the integration of the Bethe-Bloch equation (BBE). With respect to energy straggling, we shall verify that a quantum statistical base implies a Gaussian convolution and some generalizations for energy straggling (Landau tail, if the relativistic effects are taken into account). The energy-range relation goes back to the Bragg-Kleeman rule (BKR):

$$R_{CSDA} = A \cdot E_0^p \qquad (1)$$

A special case of Eq. (1) is the Geiger rule with p = 1.5, which is assumed to be valid for α-particles with low energy, whereas for therapeutic protons ($E_0$ between 50 and 250 MeV) the power p has been determined to p ≈ 1.7 – 1.8. Since $E_0$ is given in MeV and $R_{CSDA}$ in cm, the dimension of A is cm/MeV$^p$ (p is dimensionless). Based on ICRU49 [9], a least-squares fit [10] of the range $R_{CSDA}$ of protons provided for p = 1.77 and A = 0.0022 cm/MeV$^p$ a standard error of about 2 %, if $E_0 \leq 200$ MeV. A precise knowledge of $p = p(E_0)$ requires a comparison with solutions of the BBE or with Monte-Carlo calculations.

At recent times, the radiotherapy with protons has become a modality with increasing importance, which triggers a lot of work in the field of algorithms for treatment planning. One of the challenges in treatment planning is to find a reasonable compromise between the speed and the accuracy of an algorithm. The fastest dose calculations are based on look-up tables for depth dose and lateral distributions [11 - 16]. Look-up tables often consist of measured data directly or of an analytical model fitted to the measurements in water. A depth scaling mechanism is always applied in order to convert the dose from the water to another medium. A further approach is an iterative numerical calculation of the dose deposited by a proton beam along its path through a medium [17]. Monte-Carlo methods can lead to a high precision – especially in heterogeneous media – but, lacking sufficient speed, they are hardly applied to routine treatment planning [14, 18, 19]. Common to most of the cited papers is that the authors do not provide a rigorous adaptation procedure of the dose model to different properties of the beam-line. Some of the models might not be easily adaptable to another beam-line or an intermediate range/energy at all. An exception to this is the model [10]. However, this model does not correctly describe the build-up, observed in higher energy protons, and the transport of secondary protons. The model [10] results by methods elaborated in section 1.2.

Furthermore, we present a theoretical explanation of the buildup effect, which is measured in higher energy proton beams and is based on secondary nuclear reaction protons as well as skew symmetric energy transfer from protons to electrons (Landau tails). The buildup is not modeled correctly by the Monte-Carlo code PTRAN [20]. Also, none of the other authors, referred to above, thoroughly discuss the origins of buildup effects. In this work, we will deal with the calculation model based on BBE; a simplification is incorporated in the commercial Treatment Planning System Varian-Eclipse$^{TM}$ [3, 21]. The reference system is water. Their outcome is the three-dimensional (3D) dose distribution delivered by the quasi-mono-energetic beam impinging on a water surface with no lateral extension and angular divergence (the term 'quasi-mono-energetic beam' refers to the spectrum as produced by the accelerator and beam-line without any intended range modulation). The beam-let can be separated into a depth-dose component and a lateral distribution. The depth dose, obtained from a quasi-mono-energetic beam, is often called '*pristine Bragg peak*'.

In this work, we also focus our interest to theoretical depth-dose models which contain some fitting parameters referring to the spectral distribution of impinging proton beams (measured pristine Bragg peaks) in order to incorporate beam-line specific characteristics. The measurements have been obtained from different sources and will be described in more detail in the corresponding sections. The lateral component of the beam-let only includes scattering in the patient. It does not require any beam-line specific configuration. The beam-line specific component of the lateral penumbra is included by the lateral distribution of the in-air fluence [22]. The Monte-Carlo code GEANT4 [23] is suitable to study the lateral scattering distributions, the Landau tails, the production of heavy recoil particles, and to obtain some numerical parameters. Many comparisons between Monte-Carlo results and an exact integration of the BBE, and additionally nuclear cross-sections have been previously published [4 - 5].

## 1. Theoretical methods in proton therapy

### 1.1 Abbreviations and definitions

$\beta$: ratio v/c of the particle velocity v and the velocity of light c
m: electron rest mass (electron rest energy $m \cdot c^2 = 0.511$ MeV)
M: proton rest mass ($1836.15655 \cdot m$; proton rest energy $M \cdot c^2 = 938.276$ MeV)
$\mu$: reduced mass of the center-of-mass system 'proton - electron' with $\mu = m + m/M$
$A_N$: mass number of a nucleus
Z: charge number of a nucleus
$\rho$: mass density (g/cm$^3$)
$e_0$: electric charge
q: charge of a projectile particle
In this work, we use the following definition of the standard deviation:

$$\text{dev} = \frac{1}{N} \sum_{k=1}^{N} |x_k - y_k| \quad (2)$$

The variables $x_k$ and $y_k$ may either refer to a set of measured and calculated data or to a comparison of an approximate calculation (e.g., finite order of a power expansion) with an accurate one.

The relativistic terms have to account for the power expansion:

$$\left. \begin{array}{l} (1 \pm x)^r = 1 \pm rx/1! + r(r-1)x^2/2! \\ \pm r(r-1)(r-2)x^3/3! \pm . + (-1)^n r(r-1) \\ ..(r-n+1)x^n/n!, \quad r = \pm \frac{1}{2} \end{array} \right\} \quad (3)$$

The parameter x in Eq. (3) may refer to different substitutions (e.g., $x = p^2/m^2c^2$ resulting from the root $W = mc^2(1+p^2/m^2c^2)^{0.5}$ of the energy-momentum relation $W^2 = p^2c^2 + m^2c^4$).

### 1.2 Phenomenological treatment of energy-range relations of protons (BKR and its generalization - model M1)

In irreversible thermodynamics the Langevin equation (4) and the related solution (5) have been investigated with respect to the energy dissipation to the environment due to the motion of a projectile with velocity v(t):

$$M \cdot d\text{v}/dt = -\gamma \cdot \text{v} \quad (4)$$

The integration of Eq. (4) is yields:

$$v(t) = v_0 \exp(-\gamma t/M) \Rightarrow E(t) = E_0 \exp(-2\gamma t/M)$$

$$z = R_{CSDA} - \tfrac{v_0 M}{\gamma} \exp(-\gamma t/M) \rightarrow R_{CSDA} = \tfrac{v_0 M}{\gamma} = \tfrac{\sqrt{2ME_0}}{\gamma} \quad (5)$$

Only in Eqs. (4 – 5) M may be identified by the mass of a macroscopic 'particle'. Special cases are the Stokes law of a spherical body with $\gamma = 6\pi\eta r$ in a fluid and, in analogous sense, Ohm's law of the electric resistance. A unique feature is the irreversible heat production by damping of the particle motion. The stopping of the particle motion according to Eq. (5) is described by CSDA, which is always assumed for macroscopic processes. From a microscopic view-point, the damping constant $\gamma$ results from numerous collisions with environmental atoms/molecules. The question arises, whether the energy loss of a projectile (proton or an $\alpha$-particle) can be established in a similar way. With respect to Eqs. (4 - 5) we consider the following modification (the power q may be arbitrary, but q = -1 leading to Eq. (4) must be excluded):

$$M \cdot \tfrac{dv}{dt} = -\tfrac{\delta}{v^q} \Rightarrow \int v^q \cdot dv = -\int \tfrac{\delta}{M} dt \quad (6)$$

By taking account of the initial condition t = 0 and v = $v_0$, the integration of Eq. (6) yields:

$$v^{q+1} = -\tfrac{\delta t(q+1)}{M} + v_0^{q+1} \quad (7)$$

The further integration of Eq. (7) via (q+1)$^{th}$ root provides:

$$R_{CSDA} + \int (v_0^{q+1} - \tfrac{\delta t(q+1)}{M})^{\tfrac{1}{q+1}} dt$$

$$= R_{CSDA} - \tfrac{M}{\delta(q+2)} (v_0^{q+1} - \tfrac{\delta t(q+1)}{M})^{\tfrac{q+2}{q+1}} \quad (8)$$

$$\tau = M v_0^{q+1} / \delta(q+1) \quad (8a)$$

The initial condition t = 0 $\Rightarrow$ z = 0 now yields:

$$R_{CSDA} = \tfrac{M}{\delta(q+2)} v_0^{q+2} \quad (9)$$

Replacing the initial velocity $v_0$ by $E_0 = Mv_0^2/2$, Eq. (9) takes the form:

$$R_{CSDA} = A \cdot E_0^p, \quad p = 1 + q/2; \; A = 2^{p-1} \cdot \delta^{-1} \cdot M^{1-p} \quad (10)$$

Eq. (10) agrees with Eq. (1); the Geiger rule results from q = 1 $\Rightarrow$ p = 3/2. The difference between Eq. (5) and Eq. (8a) is the time interval $\tau$ needed to reach z = $R_{CSDA}$. Thus in Eqs. (4, 5) $\tau$ is infinite, whereas for Eq. (8a) $\tau$ is finite. If $t > \tau$ the roots in Eq. (8) become imaginary. The range $R_{CSDA}$ and the connection to p (p $\approx$ 1.7 – 1.8) of therapeutic protons has often been studied [10, 24 - 28]. According to Eq. (10), we may consider p and A as depending on $E_0$. In view of a relativistic treatment, we only consider $p = p(E_0)$ and keep A constant (section 1.4). Eq. (8) provides a tool to calculate E(z) and the stopping power S(z) = dE(z)/dz as a function of the position z:

$$E(z) = A^{1/p} (R_{CSDA} - z)^{1/p} \qquad (11)$$

$$S(z) = dE/dz = -p^{-1} A^{-1/p} (R_{CSDA} - z)^{(1/p-1)} \qquad (12)$$

It should be noted that for protons with $E \ll Mc^2$ a phenomenological description of the motion and energy loss by damping makes physical sense, since quantum mechanical effects like production of '*bremsstrahlung*' are negligible. The release of secondary protons via proton – nucleus interactions is too small to invalidate Eqs. (11 – 12). There are principal differences between non-relativistic and relativistic calculations of S(z). The non-relativistic treatment consists of two steps: Integration of the equation of motion, i.e., dv(t)/dt, providing either dz/dt or E(v(t)) as a function of t. A further integration yields z(t), and, by that, we obtain E(z) and dE(z)/dz. In a relativistic approach, only *ds* (Eq. (13)) is invariant. Since the energy-momentum relation also holds, there is no additional integration. The relativistic extension of the motion with damping and energy dissipation according to Eq. (6) has to take the following principles into account: The relativistic equation of a particle motion under the action of a force $F_\mu$ (though μ refers to an arbitrary component of a four-vector, we can restrict ourselves to the z-component) and the energy-momentum relation are:

$$ds = \sqrt{dz^2 - c^2 dt^2} \qquad (13)$$

$$dp_\mu / ds = F_\mu, \qquad dp_z / ds = F_z \qquad (14)$$

$$W^2 = p_z^2 c^2 + M^2 c^4 \quad and \quad W = Mc^2 + E \qquad (15)$$

The use of the invariant variable *s* accounts for the length contraction besides the relativistic mass dependence. Thus $E_0$ has the same meaning as previously defined: $W_0 = Mc^2 + E_0$. In analogy to Eq. (6) we consider the following damping equation:

$$dp_z / ds = -\eta / p_z^q \qquad (16)$$

Since the dimensions in Eq. (16) are different from those in Eq. (6), we have replaced δ by η. The integration of Eq. (16) goes parallel to previous calculations; additionally, Eqs. (13 - 15) have to be satisfied. Thus we obtain:

$$s = \frac{1}{\eta(q+1)} [p_{z,0}^{q+1} - p_z^{q+1}] \qquad (17)$$

Inserting the solution (17) into Eq. (15) results in:

$$\left. \begin{array}{l} R_{CSDA} = \frac{1}{(q+1)\eta} [(M^2 c^4 + E_0^2 + 2Mc^2 E_0)/c^2 - M^2 c^2]^{(q+1)/2} \\ = A[E_0 + E_0^2 / 2Mc^2]^p, \; A = (2M)^{(q+1)/2} /(\eta + \eta q) \end{array} \right\} (18)$$

From Eqs. (17 - 18), it follows:

$$\left. \begin{array}{l} E + E^2 / 2Mc^2 = [(R_{CSDA} - s)/A]^{1/p} \Rightarrow \\ E(s) = Mc^2 [\sqrt{1 + 2(R_{CSDA} - s)^{1/p} /(Mc^2 A^{1/p})} - 1] \end{array} \right\} (19)$$

$$S(s) = dE/ds = -p^{-1} A^{-1/p} (R-s)^{1/p-1} / \sqrt{1 + 2(R_{CSDA} - s)^{1/p}/(Mc^2 A^{1/p})} \quad (20)$$

The power p according to Eq. (19) is dimensionless: p = (q+1)/2. E(s) and dE(s)/ds according to Eqs. (19 – 20) can be subjected to a series expansion, if $x = 2(R_{CSDA} - s)^{1/p}/(Mc^2 \cdot A^{1/p}) < 1$. This condition is equivalent to $E(s) \leq Mc^2$. The factor A amounts to A = 0.00259 cm/MeV$^p$. If v << c the length-contraction expressed by s can be omitted; therefore, we substitute s by z:

$$E(z) = Mc^2 (\sqrt{1 + 2(R_{CSDA} - z)^{1/p}/(Mc^2 A^{1/p})} - 1) \quad (21)$$

$$S(z) = dE/dz = -p^{-1} A^{-1/p} (R-z)^{1/p-1} / \sqrt{1 + 2(R_{CSDA} - z)^{1/p}/(Mc^2 A^{1/p})} \quad (22)$$

As expected, the expansion of the root of Eq. (22) according to Eq. (3) yields Eq. (12) as low-order approximation. Both Eqs. (12) and (22) show a singularity at $z = R_{CSDA}$, if p > 1. In the relativistic case, this singularity is weakened due to nonsingular relativistic corrections. In section 2 we only deal with results based on section 1.3 (BBE); in [4] results based on model M1 (Eqs. (16 - 22)) have been represented.

### 1.3 Integration of the Bethe-Bloch equation (BBE - model M2)

In the present investigation, we mainly deal with BBE, the model M1 is presented in detail in [4]. Monte-Carlo codes for the computation of electronic stopping power of protons and other charged particles have to be based on the BBE ($E_I$ is the atomic ionization energy, weighted over all possible transition probabilities of atomic/molecular shells, and q denotes the charge number of the projectile (proton). The meaning of the correction terms $a_{shell}$, $a_{Barkas}$, $a_0$ and $a_{Bloch}$ are explained in literature [4, 9, 25, 27 - 30]; therefore we do not present them here. It is also possible to substitute the electron mass m by the reduced mass $m \Rightarrow \mu$. For protons, this leads to a rather small correction (< 0.1 %):

$$-dE(z)/dz = (K/v^2) \cdot [\ln(2mv^2/E_I) - \ln(1-\beta^2) + a_{shell} + a_{Barkas} + a_0 v^2 + a_{Bloch}]$$
$$K = (Z\rho/A_N) \cdot 8\pi q^2 e_0^4 / 2m \quad (23)$$

The complex systems $E_I$ and contributions like $a_{shell}$ and $a_{Barkas}$ can only be approximately calculated by quantum-mechanical models (e.g., oscillator); the latter terms are often omitted and $E_I$ is treated as a fitting parameter, but different values are proposed [9]. The restriction to the logarithmic term leads to severe problems, if either $v \to 0$ or $2m v^2/E_I \to 1$. With regard to the integration procedure, we start with the logarithmic term of Eq. (23) and perform the substitutions:

$$v^2 = 2E/M; \quad \beta_I = 4m/ME_I; \quad E = (1/\beta_I)\exp(-u/2) \quad (24)$$

With the help of substitution (24) - without any correction terms - Eq. (23) leads to the integration:

$$-\int du \exp(-u) \cdot (1/u) = \tfrac{1}{2} K \cdot \beta_I^2 \cdot M \int dz \quad (25)$$

The boundary conditions of the integral (25) are:

$$z = 0 \Rightarrow E = E_0 \text{ (or} : u = -2\ln(E_0 \cdot \beta_I))$$
$$z = R_{CSDA} \Rightarrow E = 0 \text{ (or} : u \Rightarrow \infty) \quad (26)$$

The general solution is given by the Euler exponential integral function Ei(ξ) with P.V. = principal value:

$$\left. \begin{array}{l} \frac{1}{2} K \cdot M \cdot \beta_I^2 \cdot R_{CSDA} = - P.V. \int_{-\xi}^{\infty} u^{-1} \exp(-u) du = Ei(\xi) \\ \xi = 2 \ln(4 m E_0 / M E_I) \quad \text{and} \quad \xi > 0 \end{array} \right\} \quad (27)$$

Some details of Ei(ξ) and its power expansions can be found in [31]. The critical case ξ = 0 results from $E_{critical} = ME_I/4m$ (for water with $E_I$ = 75.1 eV, the critical energy $E_{critical}$ amounts to 34.474 keV; for Pb with $E_I \approx$ 800 eV to ≈ 0.4 MeV). Since the logarithmic term derived by Bethe implies the Born approximation, valid only if the transferred energy $E_{transfer} \gg$ energy of shell transitions, the above corrections, exempting the Bloch correction, play a significant role in the environment of the Bragg peak, and the terms $a_0$ and $a_{shell}$ remove the singularity. With respect to numerical integrations (Monte Carlo), we note that, in the environment of E = $E_{critical}$, the logarithmic term may become crucial (leading to overflows); rigorous cutoffs circumvent the problem. Therefore, the shell correction is an important feature for low proton energies. In similar fashion, we can take account of the *Barka*s correction. Since this correction is also important for low proton energies, it is difficult to make a quantitative distinction to the shell correction, and different models exist in the literature implying overall errors up to 2 % [9]. Using the suggestions of the correction terms according to [9] and the substitutions (24), we obtain:

$$\frac{1}{2} K \cdot M \cdot \beta_I^2 \int dz = \int du \exp(-u) [u + 2\alpha_S + \\ + 2\alpha_{Barkas} (4 \cdot m / E_I)^{p_B} \exp(p_B u / 2) \\ + \alpha_0 (E_I / 2m) \exp(-u/2)]^{-1} \quad (28)$$

A closed integration of Eq. (28) does not exist, but it can be evaluated via a procedure valid for integral operators [32, 33]:

$$[A' + B']^{-1} = A'^{-1} - A'^{-2} B' + A'^{-3} B'^2 - A'^{-4} B'^3 + .. \\ + (-1)^n A'^{-n-1} B'^n \quad (29)$$

A'+B' is equated to the complete denominator on the right-hand side of Eq. (28). The (small) Barkas correction is identified with A' and the other (more important) terms with B':

$$\left. \begin{array}{l} A' = 2\alpha_{Barkas} (4m / E_I)^{p_B} \exp(p_B u / 2) \\ B' = u + 2\alpha_S + \alpha_0 (E_I / 2m) \exp(-u / 2) \end{array} \right\} \quad (30)$$

The integration of Eq. (28) via Eq. (29) leads to standard tasks (a series of exponential functions). In the following, we add $a_{Bloch}$ to the denominator of Eq. (28). In order to use Eq. (29), we define now the non-relativistic energy by: $E_{nr} = 0.5 \cdot Mv^2$ and write the relativistic expression $E_{rel}$ (the rest energy $Mc^2$ is omitted) in terms of an expansion according to Eq. (24):

$$\left. \begin{array}{l} a_{Bloch} = -(\frac{1}{2} q^2 \alpha^2 \beta_I Mc^2 \exp(u/2)) \cdot [1.042 - \frac{1}{2} 0.854 \cdot q^2 \alpha^2 \beta_I Mc^2 \\ \cdot \exp(u/2) + \frac{1}{4} 0.343 q^4 \alpha^4 \beta_I^2 M^2 c^4 \exp(u) \\ + higher - order \ terms] \end{array} \right\} \quad (31)$$

We add the term $a_{Bloch}$ to the term *A'* in Eq. (30); α: Sommerfeld's fine structure constant. All corrections are included by taking account of Eq. (31):

$$A' = 2\alpha_{Barkas} (4m / E_I)^{p_B} \exp(p_B u / 2) + a_{Bloch} \quad (32)$$

$$\frac{1}{2} K \cdot M \cdot \beta_I^2 \int dz = \int du \exp(-u) \cdot [A' + B']^{-1} \quad (33)$$

The integration of Eq. (33) is carried out with the boundary conditions (24). Since these conditions are defined by logarithmic values, which have to be inserted to an exponential function series yielding a

power expansion for $R_{CSDA}$ in terms of $E_0$:

$$R_{CSDA} = \frac{1}{\rho} \cdot \frac{A_N}{Z} \sum_{n=1}^{N} \alpha_n E_I^{pn} E_0^n \quad (N \Rightarrow \infty) \quad (34)$$

The coefficients $\alpha_n$ are determined by the integration procedure and only depend on the parameters of the BBE. For applications to therapeutic protons ($E_0 < 300$ MeV), N = 4 provides excellent results (Figure 1). For water, we have to take $E_I = 75.1$ eV, $Z/A_N = 10/18$, $\rho = 1$ g/cm$^3$; Eq. (35) becomes:

$$R_{CSDA} = \sum_{n=1}^{N} a_n E_0^n \quad (N \Rightarrow \infty) \quad (35)$$

The parameters of Eqs. (34, 34) with N = 4 are displayed in Tables 1 and 2.

| α1 | α2 | α3 | α4 |
|---|---|---|---|
| 6.8469·10-4 | 2.26769·10-4 | -2.4610·10-7 | 1.4275·10-10 |
| P1 | P2 | P3 | P4 |
| 0.4002 | 0.1594 | 0.2326 | 0.3264 |

**Table 1:** Parameters of Eq. (35) if $E_0$ is in MeV, $E_I$ in eV and $R_{CSDA}$ in cm.

| $a_1$ | $a_2$ | $a_3$ | $a_4$ |
|---|---|---|---|
| 6.94656·10$^{-3}$ | 8.13116·10$^{-4}$ | -1.21068·10$^{-6}$ | 1.053·10$^{-9}$ |

**Table 2:** Parameters of Eq. (35), if $E_0$ is in MeV, $E_I$ in eV and $R_{CSDA}$ in cm.

The determination of $A_N$ and Z is not a problem in case of atoms or molecules; weight factors can be used according to the Bragg rule; for tissue heterogeneities, it is already a difficult task. Much more difficult is the accurate determination of $E_I$, which results from transition probabilities of all atomic/molecular states to the continuum (δ-electrons). Thus, according to [9] the stopping powers of protons in different media, there are different values of $E_I$ proposed (for Pb: $E_I = 820$ eV and $E_I = 779$ eV). If we use the average ($E_I = 800.5$ eV), the above formula provides a mean standard deviation of 0.27 % referred to stopping-power data in [9], whereas for $E_I = 820$ eV or $E_I = 779$ eV we obtain 0.35 % - 0.4 %. If we apply the above formula to data of other elements listed in [9], *dev* also amounts to about 0.2 % - 0.4 %. Instead of Eq. (35), we can represent all integrals by Gompertz-functions multiplied with a single exponential function by collection of all exponential functions obtained by the expression of $[A' + B']^{-1}$ and the substitution $\beta_I E = \exp(-u/2)$. A Gompertz-function is defined by:

$$\exp(-\xi \exp(-u/2)) = 1 + \sum_{k=1}^{\infty} \frac{1}{k!} (-1)^k \xi^k \exp(-ku/2) \quad (36)$$

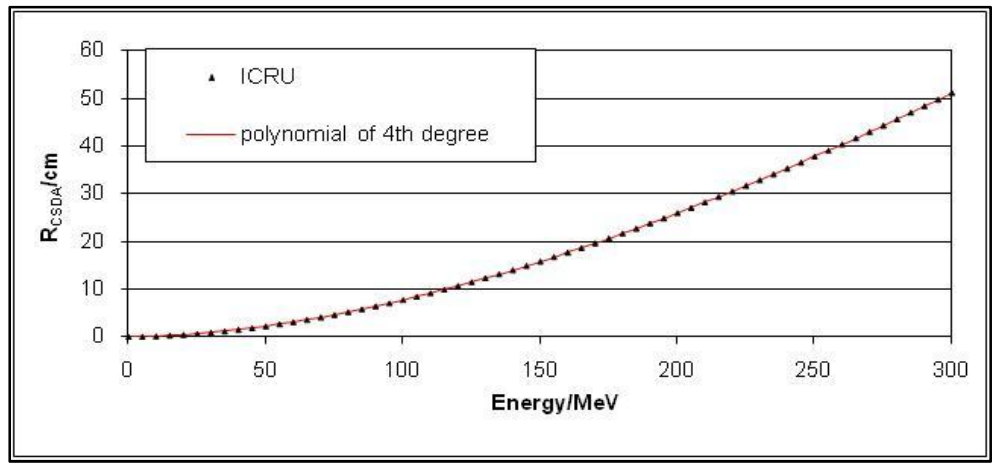

**Figure 1:** Comparison between ICRU49 data of proton $R_{CSDA}$ range (up to 300 MeV) in water and the fourth-degree polynomial (Eq. (35)). The average deviation amounts to 0.0013 MeV.

By inserting the integration boundaries $u = 2 \cdot \ln \cdot 4m \cdot E_0/(M \cdot E_I)$, i.e., $E = E_0$ and $u \to \infty$ ($E = 0$), the integration leads to a sequence of exponential functions; the power expansion (35) is replaced by:

$$R_{CSDA} = a_1 E_0 \cdot [1 + \sum_{k=1}^{N}(b_k - b_k \exp(-g_k \cdot E_0))], (N \to \infty) \quad (37)$$

:If $E_0 < 270$ MeV, the restriction to $N = 2$ provides the same accuracy (Figure 2 as Eq. (35); the parameters are given in Table 3 and $a_1$ in Table 2).

| $b_1$ | $b_2$ | $g_1$ | $g_2$ |
|---|---|---|---|
| 15.14450027 | 29.84400076 | 0.001260021 | 0.003260031 |

**Table 3:** Parameters of Eq. (37); $b_1$ and $b_2$ are dimensionless; $g_1$ and $g_2$ are given in MeV$^{-1}$.

In the following, we shall verify that Eq. (37) provides some advantages with respect to the inversion $E_0 = E_0(R_{CSDA})$. Eqs. (35, 37) can also be used for the calculation of the residual distance $R_{CSDA} - z$, relating to the residual energy $E(z)$; we have only to perform the substitutions $R_{CSDA} \to R_{CSDA} - z$ and $E_0 \to E(z)$ in these formulas. Eq. (35) leads to a series $E_0 = E_0(R_{CSDA})$ in terms of powers. This series expansion is ill-posed and diverging [4], but the inversion procedure of Eq. (37) is successful:

$$\left. \begin{array}{l} E_0 = R_{csda} \sum_{i=1}^{N} c_k \exp(-\lambda_k R_{csda}) \quad (N \to \infty) \\ E(z) = (R_{csda} - z) \sum_{i=1}^{N} c_k \exp(-\lambda_k (R_{csda} - z)) \end{array} \right\} \quad (38)$$

A detailed analysis of the inversion of Eq. (35) leading to a diverging series is given in [4].

A modification of Eq. (38) represents the inverse formula of Eq. (37):

$$\left. \begin{array}{l} c'_k = c_k \cdot (18/10) \cdot Z \cdot \rho \cdot (75.1/E_I)^{q_k} /(A_N \cdot \rho_w) \\ \lambda^{-1}{}'_k = \lambda^{-1}{}_k \cdot (10 \cdot \rho_w /18) \cdot (75.1/E_I)^{p_k} \cdot A_N /(\rho \cdot Z) \\ E(z) = (R_{CSDA} - z) \cdot \sum_{k=1}^{5} c'_k \cdot \exp[-(R_{CSDA} - z) \cdot \lambda'_k] \end{array} \right\} \quad (39)$$

For therapeutic protons, a very high precision is obtained by the restriction to $N = 5$ (Table 4 and Figure 3). One way to obtain the inversion of Eq. (37) is to use a sum of exponential functions with the help of a fitting procedure. Thus it turned out that the restriction to $N = 5$ is absolutely sufficient and yields a very high accuracy. A more rigorous way (mathematically) has been described in the LR of [33].

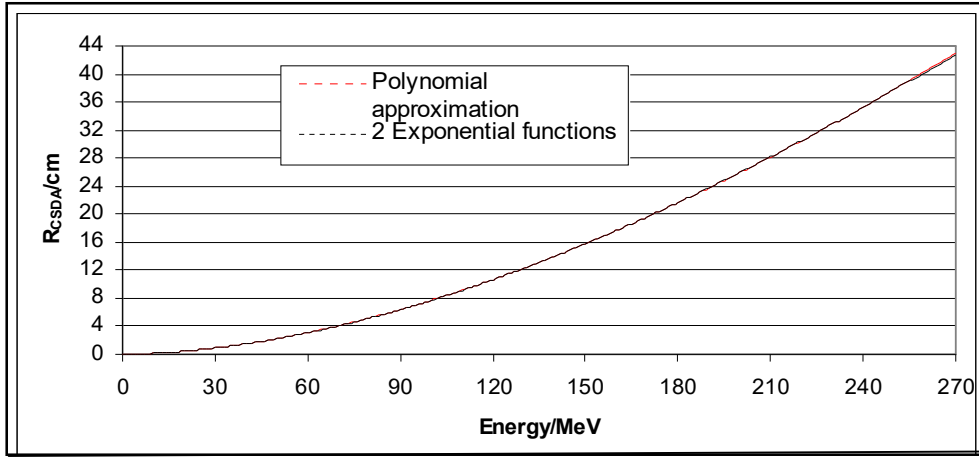

**Figure 2:** $R_{CSDA}$ calculation: comparison between a fourth-degree polynomial (Eq. (35)) and two exponential functions (Eq. (37)).

| $c_1$ | $c_2$ | $c_3$ | $c_4$ | $c_5$ |
|---|---|---|---|---|
| 106.63872 | 25.0472 | 8.80745 | 4.19001 | 9.2732 |
| $\lambda_1^{-1}$ | $\lambda_2^{-1}$ | $\lambda_3^{-1}$ | $\lambda_4^{-1}$ | $\lambda_5^{-1}$ |
| 0.0975 | 1.24999 | 5.7001 | 10.6501 | 126.26 |
| P1 | P2 | P3 | P4 | P5 |
| -0.1619 | -0.0482 | -0.0778 | 0.0847 | -0.0221 |
| q1 | q2 | q3 | q4 | q5 |
| 0.4525 | 0.195 | 0.2125 | 0.06 | 0.0892 |

**Table 4:** Parameters of the inversion Eq. (39) with N = 5 (dimension of $c_k$: cm/MeV, $\lambda_k$: cm$^{-1}$).

The residual energy E(z) appearing in Eq. (39), is the desired analytical base for all calculations of stopping power and comparisons with GEANT4. The stopping power is determined by dE(z)/dz and yields the following expression:

$$\left.\begin{aligned} S(z) &= dE(z)/dz \\ &= -E(z)/(R_{CSDA} - z) + \sum_{k=1}^{N} \lambda_k E_k(z) \; (N \to \infty) \\ E_k(z) &= c_k(R_{CSDA} - z) \cdot \exp[-\lambda_k(R_{CSDA} - z)] \end{aligned}\right\} (40)$$

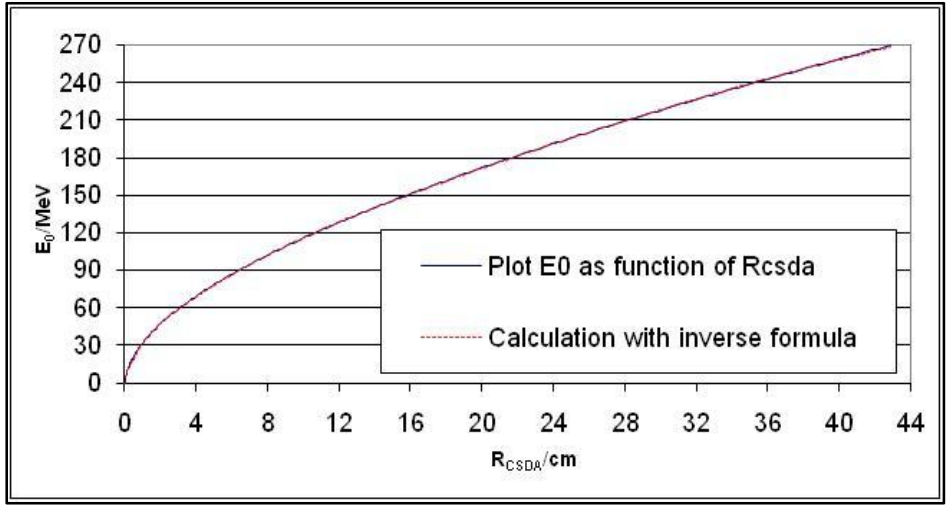

**Figure 3:** Test of Eq. (39) $E_0 = E_0(R_{CSDA})$ by five exponential functions. The mean deviation amounts to 0.11 MeV. The plot results from Figure 1.

The aforementioned restriction to N = 5 is certainly extended to Eq. (40), which can be considered as a representation of the BBE in terms of the residual energy E(z). Due to the low-energy corrections ($a_0$, $a_{shell}$, $a_{Barkas}$), the energy-transfer function dE(z)/dz remains finite for all z (i.e., $0 \leq z \leq R_{CSDA}$). This is, for instance, not true for the corresponding results according to Eqs. (12 – 22) at $z = R_{CSDA}$ due to the singularities resulting from these equations. A change from the interacting reference medium water to any other medium can be carried out by the calculation of $R_{CSDA}$, where the substitutions have to be performed and used in Eqs. (39 - 41):

$$R_{CSDA}(medium) = R_{CSDA}(water) \cdot (Z \cdot \rho / A_N)_{water} \cdot (A_N / Z \cdot \rho)_{medium} \quad (41)$$

It is also possible to apply Eqs. (39 – 41) in a stepwise manner (e.g., voxels of CT). This procedure will not be discussed here, since it requires a correspondence between $(Z \cdot \rho/A_N)_{Medium}$ and information via CT [33 - 36].

### 1.4 Determination of the power $p(E_0)$ of section 1.2 (model M1)

The BKR (Eq. (1)) and its relativistic generalization (18) have been developed on the basis of phenomenological principles and non-relativistic/relativistic equations of motion. The undefined parameters A and p can either be adapted by fits to experimental data or by comparison with $R_{CSDA}$ calculations based on the BBE. Since different numerical values for A and p have been proposed in relation to the considered energy domain $E_0$, we have performed an adaptation of these parameters based on the results (35, 37). Eq. (18) agrees with Eq. (1) in the non-relativistic limit with $E_0 \rightarrow 0$. Therefore, we have fixed A by this request and permitted only a dependence of p on $E_0$. The result of this fit is given in Figure 4; due to relativistic corrections, p turns out to be lower than the appropriate value used in Eq. (1). The dimensionless power p (non-relativistic and relativistic) can be obtained from the expressions:

$$p_{nr} = -5 \cdot 10^{15} \cdot E_0^6 + 5 \cdot 10^{12} \cdot E_0^5 - 2 \cdot 10^2 \cdot E_0^4 +$$
$$+ 4 \cdot 10^7 \cdot E_0^3 - 5 \cdot 10^5 \cdot E_0^2 + 0.0003 \cdot E_0 + 1.6577 \quad (42)$$

$$p_{re} = -4 \cdot 10^{15} \cdot E_0^6 + 4 \cdot 10^{12} \cdot E_0^5 - 2 \cdot 10^9 \cdot E_0^4 +$$
$$+ 4 \cdot 10^7 \cdot E_0^3 - 5 \cdot 10^5 \cdot E_0^2 + 0.0027 \cdot E_0 + 1.6576 \quad (43)$$

Eq. (42) is the non-relativistic case of $p(E_0)$ and Eq. (43) is valid for the relativistic case. It appears to be a simple task to invert Eqs. (6 – 18) to obtain $E_0 = E_0(R_{CSDA})$, yet the energy dependence $p = p(E_0)$ prevents simple calculations; hence, iterative procedures are necessary. If one keeps A and p constant in a certain energy domain [10, 25], then the deviations turn out to be higher for inverse calculations than for the original problem: $R_{CSDA}$ as a function of $E_0$. In Figure 4, we have assumed that A = 0.00259 cm/MeV$^p$. Eqs. (1, 18) are only valid for water. Therefore, the question as to a connection to Eq. (35) has to be considered. If we pass from water to any other (homogeneous) medium, then the Bragg rule suggests the substitutions:

$$A \Rightarrow A \cdot \frac{Z_{water} \cdot \rho_{water} \cdot A_{N,medium}}{A_{N,water} \cdot Z_{medium} \cdot \rho_{medium}} \quad (44)$$

A comparison of Eq. (44) with Eq. (35) showed that this substitution only holds, if the ionization energy $E_{I,water} \approx E_{I,medium}$. If $E_{I,water}$ significantly differed from $E_{I,medium}$, modifications also in the power p in Eqs. (1, 18) are required. For this purpose, we substitute p by $p_{water}$ and write $p_{medium}$ as the modified power:

$$p_{medium} = p_{water} + \frac{1}{2 \cdot p_{water}} \cdot \ln(E_{I,water}/E_{I,medium}) \quad (45)$$

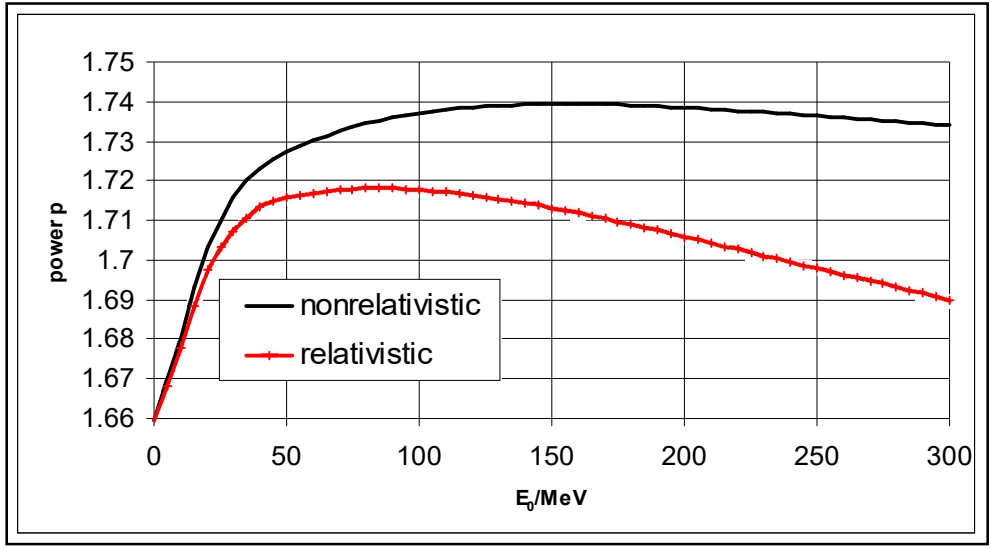

**Figure 4:** The energy dependence of the power p in Eqs. (42 - 43) is determined by Eq. (35).

We have performed various comparisons of Eq. (35) with Eqs. (43 –44) for Pb, Cu, Ca, Al, and Be. The mean standard deviation amounted to 2.16 %. The maximal deviation was obtained for some ranges of protons in Pb with 3.76 %.

### 1.5 Monte-Carlo calculations and résumé of theoretical calculations of nuclear interactions (Appendix)

The Monte-Carlo code GEANT4 is described in detail [23]. The calculation of the stopping power is based on a numerical manipulation of BBE inclusive Bloch corrections. The cutoff amounts to 1 MeV. Since GEANT4 is an open system, all correction terms according to the BBE in accordance with [9] have been implemented. These corrections only imply the removal of singularities, if the total integration procedure is carried out analytically and not by step-by-step calculations of $\Delta E/\Delta z$; these calculations depend on the actual velocity v and have to be performed for each term of the BBE separately. The scatter of protons is treated by the Molière multiple-scatter theory [37, 38] or (optionally) by the Lewis scatter theory [23]. The code also contains a hadronic generator for the simulation of nuclear interaction processes. With respect to proton dose deposition, the basic theory is the BBE and a numerical fit of the Vavilov distribution function. A Vavilov distribution function takes account of the Landau tails; in the limit case of fluctuations of small transfer energies from protons to environmental electrons, a Vavilov distribution function assumes a Gaussian shape. In GEANT4, it is also possible to restrict these fluctuations to a Gaussian shape. This fact is of interest with regard to the role of the Landau tails in the initial plateau (entrance region) of Bragg curves. A theoretical analysis of Gaussian convolutions and their generalizations to account for Landau tails will be given in a later section.

The most important aspect is the hadronic generator and the energy transport of secondary (and higher-order) particles. However, the default nuclear cross-section implemented in GEANT4 is very poor. Instead of using the default routine, which is based on data of Berger et al. [39], we have implemented the cross-section data of $O^{16}$ of Chadwick et al. [40]. Furthermore, we have calculated this nuclear cross-section with the help of the extended nuclear shell theory containing, apart from the strong interaction spin-spin and spin-orbit couplings, the electrostatic interaction and the exchange interactions between the nucleons due to the Pauli principle (see Appendix). Thus, the wave-functions of ground and excited states can be calculated by a perturbed $SU_3$. The calculation of the cross-section due to energy transfer by an external proton is based on well-elaborated principles (determination of the transition probability and density of states). The essential result is given in Figure 5 (oxygen) and Figure 6 (some nuclei of importance).

The decrease of fluence of primary protons can also be calculated from the results of this figure; we will deal with this issue in the next section. Protons with energy lower than the threshold energy $E_{Th}$ cannot surmount the potential wall of $O^{16}$. Figure 5 presenting the total nuclear cross-section, shows that there is a threshold energy $E_{Th} = 7$ MeV (more accurate: 6.997 MeV), which a proton should have to perform nuclear interactions with the $O^{16}$. For proton energies lower than the resonance maximum at $E_{res} = 20.12$ MeV, the primary proton is preferably scattered by the nucleus (the secondary proton is now identical to the deflected primary one); the nucleus is excited, to undergo rotations/oscillations and emission of X-rays of very low energy (around 1 keV), leading to the release of Auger electrons. A complete classification of the total nuclear cross-section of the proton – nucleus interactions including Breit-Wigner formula [4, 28] and its extension by Flügge [41] for therapeutic protons is given in the Appendix.

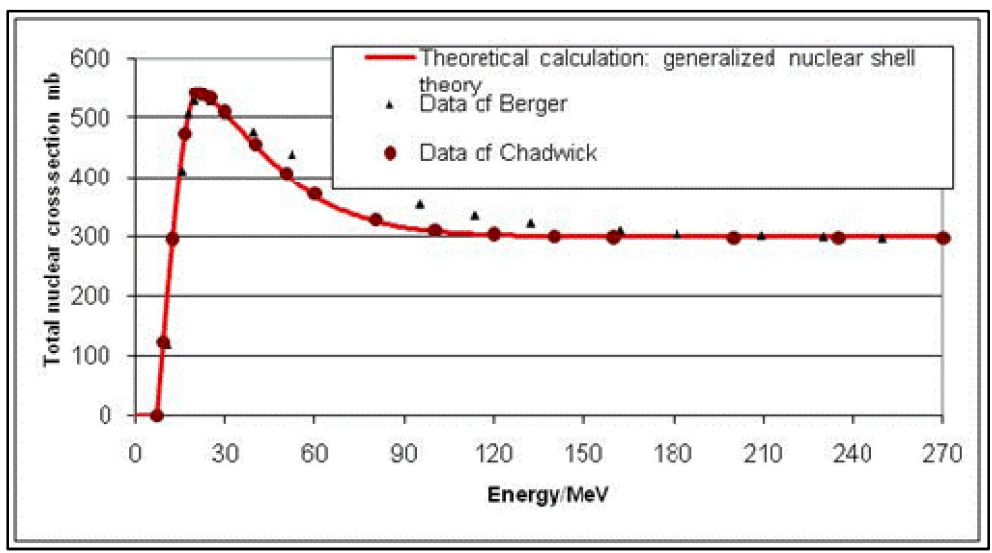

**Figure 5:** The total nuclear cross-section of the proton – nucleus (O) interaction. By 'Data of Berger', we imply 'Data of Berger et al.[39] '. By 'Data of Chadwick', we imply 'Data of Chadwick et al. [40]'.

An analysis of theoretical results and of the nuclear cross-section data of the Los Alamos library resulted in the following connection between $E_{Th}$, $A_N$ and Z:

$$E_{Th} = C \cdot Z^p / A_N^q + D \cdot Z^{p1} / A_N^{q1} \quad (46)$$

The parameters of Eq. (46) are: C = 6.565304, p = -0.10368, q = 0.00481, D = -1,2889, p1 = -0.6597, q1 = -0.6601. With regard to $E_{Th}$, Figure 5 gives an application of Eq. (46). Based on the presented calculations of total nuclear cross-sections $Q^{tot}$ (and of some further cases) and of data available in the Los Alamos library, the following adaptation model is suggested ($E_m$: the characteristic energy of the maximum value $Q^{tot}_{max}$ of $Q^{tot}$, $Q^{tot}_{as}$: the asymptotic value of $Q^{tot}$, approximately equal to the geometric cross-section; $\sigma_{res}$: r.m.s. of the resonance region; $\sigma_{as}$: a characteristic value used in the description before reaching the asymptotic behavior).

$$\left. \begin{array}{l} A_{Th} = \exp(-(E_{Th} - E_{res})^2 / \sigma_{res}^2) \\ E_m = E_{res} - E_{Th}; \quad \sigma_{res} = \sqrt{\pi} \cdot E_m \end{array} \right\} \quad (47)$$

The parameters $E_{res}$, $Q^{tot}_{max}$, $\sigma_{res}$, $I_c$ ($I_c = Q^{tot}_c / Q^{tot}_{max}$), $Q^{tot}_{as}$, and $\sigma_{as}$ are still not defined. A discussion on these parameters (as well as an overview of some theoretical aspects of nuclear physics, e.g., of the nuclear shell theory and extensions will be given in Appendix). The parameters required in Eqs. (47 – 49), referring to Figure 6, are given in Table 5.

$$\left. \begin{array}{l} Q^{tot} = Q^{tot}_{max} \cdot [\exp(-(E - E_{res})^2 / \sigma_{res}^2) - A_{Th}] \cdot (1 - A_{Th})^{-1} \\ \quad (if \ E_{Th} \leq E_{res}) \\ Q^{tot} = Q^{tot}_{max} \cdot \exp(-(E - E_{res})^2 / 2\sigma_{res}^2) \\ \quad (if \ E_{res} < E < E_c); \ E_c = E_{res} + \sqrt{-2 \cdot \ln(I_c)} \end{array} \right\} \quad (48)$$

$$\left. \begin{array}{l} Q^{tot} = Q^{tot}_c - (Q^{tot}_c - Q^{tot}_{as}) \cdot \tanh[(E - E_c)/\sigma_{as}] \ (if \ E > E_c) \\ Q^{tot}_c = I_c \cdot Q^{tot}_{max} \\ \sigma_{as} = \sigma_{res} \cdot (Q^{tot}_c - Q^{tot}_{as}) / (Q^{tot}_{max} \cdot \sqrt{-2 \cdot \ln(I_c)}) \end{array} \right\} (49)$$

| Nucleus | $E_{Th}$/MeV | $E_{res}$/MeV | $\sigma_{res}$/MeV | $\sigma_{as}$/MeV | $Q^{tot}_{max}$/mb | $Q^{tot}_c$/mb | $Q^{tot}_{as}$/mb |
|---|---|---|---|---|---|---|---|
| C | 5.7433 | 17.5033 | 21.1985 | 27.1703 | 447.86 | 426.91 | 247.64 |
| O | 6.9999 | 20.1202 | 23.2546 | 34.1357 | 541.06 | 517.31 | 299.79 |
| Ca | 7.7096 | 25.2128 | 35.6329 | 58.4172 | 984.86 | 954.82 | 552.56 |
| Cu | 8.2911 | 33.4733 | 47.6475 | 93.2700 | 1341.94 | 1308.07 | 752.03 |
| Zn | 8.3213 | 33.9144 | 48.6416 | 96.8560 | 1365.50 | 1332.31 | 766.35 |

**Table 5:** Numerical parameters for the evaluation of Figures 6 - 7 according to Eqs. (47 - 49).

An inspection of Figure 6 and of Eqs. (47 - 49) indicates that two Gaussian distributions are needed between $E_{Th} \leq E \leq E_{res}$ and $E_{res} < E \leq E_c$; this is a result of the different interaction mechanisms of the proton with the nuclei. The parameter $\sigma_{as}$ describes the asymptotic behavior of $Q^{tot}$ for $E > E_c$ and is determined by the condition that, at $E = E_c$, Eqs. (47 - 49) have to be compatible (that is, the functions and their first derivatives must be equal). It is known from nuclear (and even particle) physics that inelastic cross-sections show an exponential decrease represented by the sum of some exponential functions before reaching the asymptotic region. We have verified that a single exponential function is not sufficient and that the tanh- function provided more accurate results. It should be added that an extension expressed by Eqs. (47 - 49), which also includes tunneling effects, is developed in the Appendix.

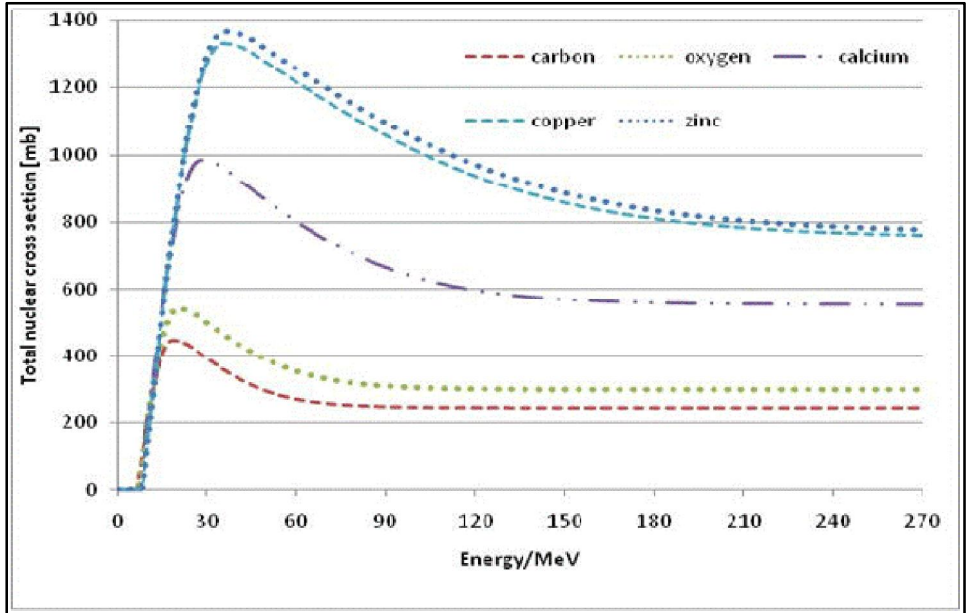

**Figure 6:** The total nuclear cross-section for C, O, Ca, Cu, and Zn [3].

## 1.6 The Fluence decrease of primary protons and nuclear interactions

Using methods described in section 1.5, we can define the fluence decrease of primary protons, the creation of secondary protons, recoil particles, and the contribution of heavy recoil particles to the total stopping power S. The fluence decrease of can be calculated by a method of Segrè [28]:

$$\left.\begin{array}{l} d\Phi/\Phi = Z \cdot \rho \cdot (N_{Avogadro}/A_N) \cdot Q^{tot}(E) \cdot dz(E) \Rightarrow \\ \int_{\Phi_0}^{\Phi} d\Phi/\Phi = \ln(\Phi/\Phi_0) = Z \cdot \rho \cdot (N_{Avogadro}/A_N) \cdot \int_{E_1}^{E_2} Q^{tot}(E) \cdot dE \cdot [dE/dz]^{-1} \end{array}\right\} (50)$$

Note that $dz(E) = [dE/dz]^{-1} \cdot dE$; the boundary condition for the fluence $\Phi$ is chosen by $\Phi_0 = 1$ at the surface. The decrease function of primary protons, obtained via Eq. (50), is given in the subsequent sections.

### 1.6.1 Primary (mono-energetic) protons $\Phi_{pp}$:

It might be surprising that in Eq. (51) the error function $erf(\xi)$ and $\sigma_{pp}$ appear. In principle, the behavior of $\Phi_{pp}$, valid within the CSDA framework, should be a straight line as long as $E = E_{Th}$ is not yet reached. From $E < E_{Th}$ to $E = 0$, $\Phi_{pp}$ should be constant and, at $E = 0$ ($z = R_{CSDA}$), a jump to $\Phi_{pp} = 0$ is expected. However, due to energy/range straggling, proton beams can never remain mono-energetic in the sense of the CSDA. The parameter $\sigma_{pp}$ refers to the half-width of a Gaussian convolution, which introduces 'roundness' in the shape.

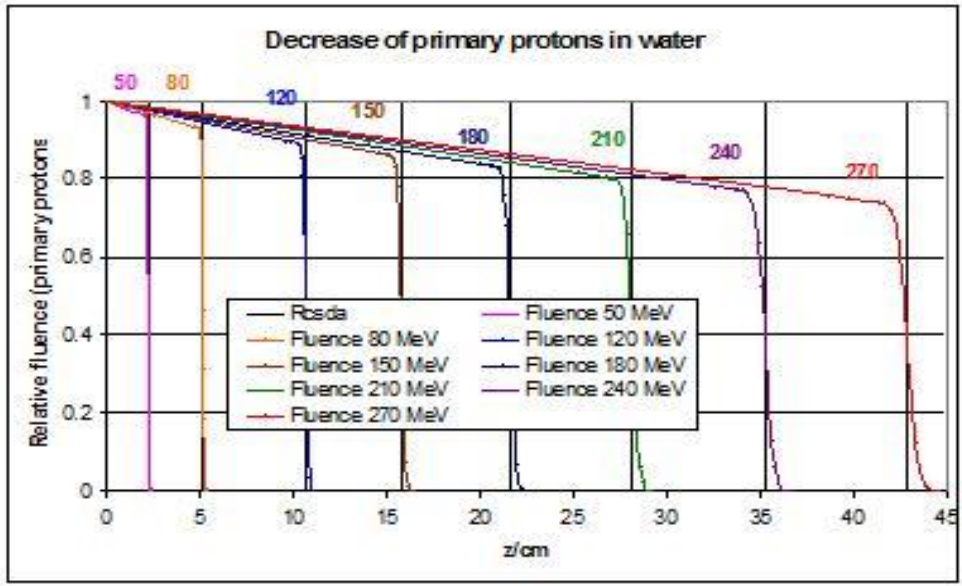

**Figure 7:** The decrease of the fluence of the primary protons due to the nuclear interactions (O); evaluation of Figures 5 - 9 using Gaussian convolutions with nuclear scatter and energy straggling (f = 1.031, $E_{Th}$ = 7 MeV).

$$\left.\begin{array}{l} \Phi_{pp} = \Phi_0 \cdot [1 - (\frac{E_0 - E_{Th}}{Mc^2})^f \cdot \frac{z}{R_{CSDA}}] \\ \cdot [1 + erf((R_{CSDA} - z)/\sigma_{pp})] \tfrac{1}{2} \end{array}\right\} \quad (51)$$

The range of 7 MeV protons is less than 1 mm; therefore, we cannot verify whether $\Phi_{pp}$ is constant in the fluence profile of primary protons. The half-width parameter $\sigma_{PP}$ will be defined in a forthcoming section. An evaluation on the basis of Eq. (51) yields Figure 7. Figure 7 clearly shows that the slope of the straight lines depends on the difference $E_0 - E_{Th}$, if $E_0 \geq E_{Th}$. If $E_0 < E_{Th}$, the expression $(E_0 - E_{Th})^f$ becomes complex and we have to impose $\Phi_{pp} = 1$. There have been attempts [10] to fit the fluence decrease of primary protons by an approximated form, which solely depends on $R_{CSDA}$. Apart from the fact that these fits are complicated, they are rather inaccurate, since they do not involve the threshold energy $E_{Th}$, which decides, whether a nuclear interaction is possible or not.

### 1.6.2 Secondary protons $\Phi_{sp}$:

In this section we separate the whole number of secondary protons by their origins, i.e., we differ between reaction protons $\Phi_{sp,r}$ and non-reaction protons $\Phi_{sp,n}$. Due to the complexity the contributions of reaction protons will be determined in section Appendix. It should be noted that $\sigma_{sp}$ is somewhat different from $\sigma_{pp}$ of Eq. (51). The argument of the error function in Eq. (51) is slightly changed by the additional $z_{shift}$, which results from an average energy loss of the secondary protons. The uncertainty intervals in the value $\square$ = 0.958 are + 0.40 % and – 0.42 %. Thus, $\square$ = 0.958 represents the value with the lowest mean standard deviation.

$$\left.\begin{array}{l} \Phi_{sp,n} = \Phi_0 \cdot [\upsilon' \cdot (\frac{E_0 - E_{Th}}{Mc^2})^f \cdot \frac{z}{R_{CSDA}}] \\ \cdot [1 + erf((R_{CSDA} - z - z_{shift}(E_0))/\sigma_{sp})] \cdot \tfrac{1}{2} \\ \upsilon' = \upsilon - 2 \cdot C_{heavy} \;;\; \upsilon = 0.958 \end{array}\right\} \quad (52)$$

### 1.6.3 Recoil protons/neutrons $\Phi_{rp}$:

$$\Phi_{rp} = \Phi_0 \cdot [\eta \cdot (\tfrac{E_0 - E_{Th}}{Mc^2})^f \cdot \tfrac{z}{R_{CSDA}}] \cdot$$

$$\cdot [1 + erf((R_{CSDA} - z - z_{shift})/\sigma_{rp})] \cdot \tfrac{1}{2} \quad (53)$$

For the recoil factor η, we use η' = 0.042. The remaining parameters are the same as in Eqs. (51 – 52). The fluence decrease of primary protons, resulting from Figures 8 - 9, is analogous to that of water (Figure 7).

A comparison of Figures 8 and 9 shows, that the slope of the fluence decrease is increased in the latter case, in particular for proton energies below 100 MeV. For Cu, the threshold energy $E_{Th}$ now amounts to 8.24 MeV and for Ca to 7.86 MeV. Thus, for proton energies below $E_{Th}$, the fluence remains constant. Due to the rather significant fluctuations, we cannot verify the constancy of the fluence at the end of the track. Eq. (46) can be applied to Figures 5 - 9.

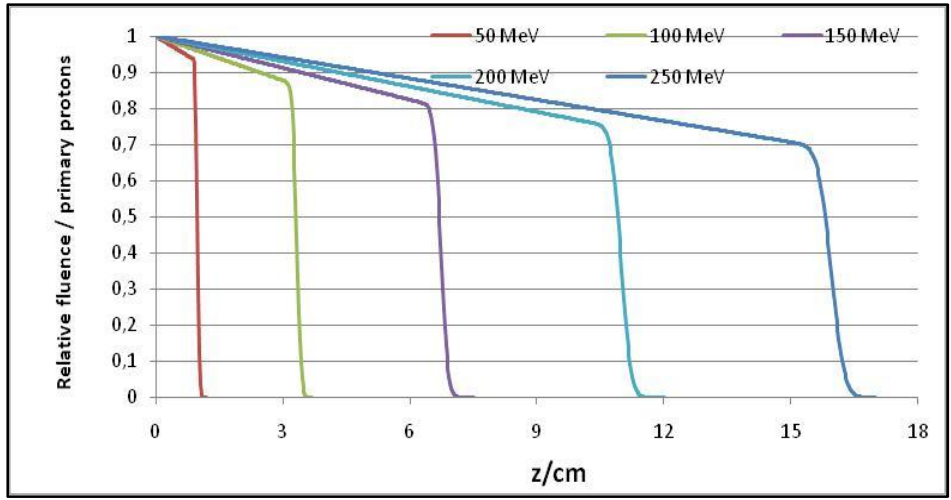

**Figure 8:** Decrease of the fluence of some initial energies of primary protons in calcium. The $R_{CSDA}$ ranges are stated by perpendicular lines.

The power f =1.032, which is valid for water to compute the slope of the straight line in Eqs. (52 – 53), has to be slightly modified (f = 0.755 for Cu and f = 0.86 for Ca), and the general formulas are given by:

$$f(Z, A_N) = a \cdot A_N^{-1} + b \cdot A_N^{-2/3} + c \cdot A_N^{-1/3} + d \cdot Z \cdot A_N^{-1/2} \quad (54)$$

$$\left(\tfrac{E_0 - E_{Th}}{M \cdot c^2}\right)^{1.032} \Rightarrow \left(\tfrac{E_0 - E_{Th}}{M \cdot c^2}\right)^f \quad (55)$$

After these modifications, Eqs. (52 – 53) can be applied. Eq. (54) is closely related to the nuclear collective model (see Appendix ), where the parameters *a, b, c* and *d* are interpreted in terms of different contributions to the total cross section. These parameters have been determined by this model: a = -0.087660001, b = -6.379250217 , c = 5.401490050 and d = - 0.054279999. There are two applications, in which Eqs. (54 - 55) are relevant: 1. Passage of protons through collimators. 2. Passage of protons through bone/metallic implants. In case 2, only a small path length has to be corrected; however, Figures 8 - 9 show that the fluence decrease has also to be corrected to fulfill continuity at the boundaries.

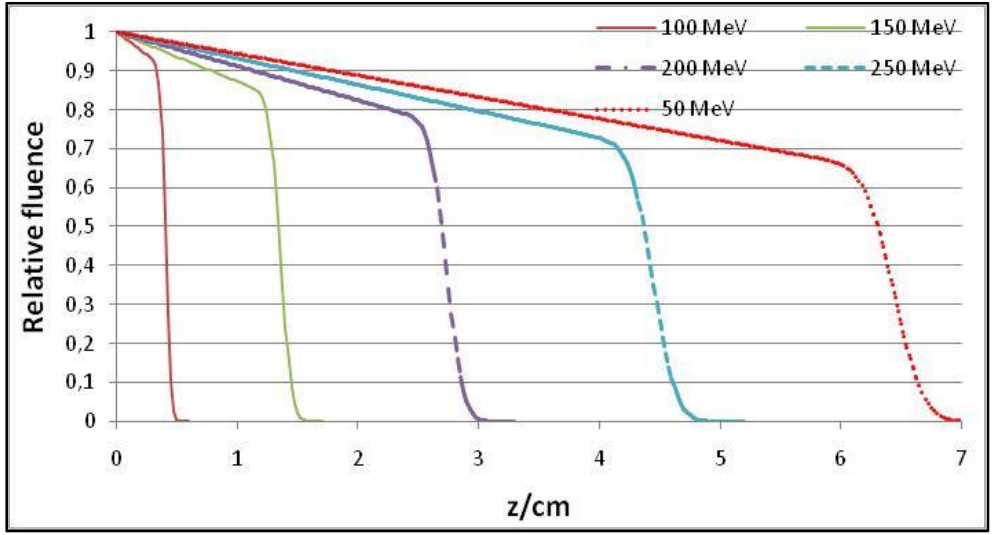

**Figure 9:** Decrease of the fluence of some initial energies of primary protons in copper.

## 1.7 Convolution theory and the energy-range straggling by fluctuations of the energy transfer to environmental electrons

This section is based on some elements of advanced mathematical physics. The result of this section, in a concise form, is that a Gaussian convolution is only valid in the non-relativistic domain of a proton track (environment of the Bragg peak), whereas relativistic corrections have to be included in the initial plateau. It is possible to directly proceed to the section 1.8.

### 1.7.1. Bohr's classical formula of energy straggling

According to Bohr's formalism [27], the formula for energy straggling (or fluctuation) $S_F$ is given by

$$S_F = \frac{1}{\sqrt{\pi}\sigma_E} \exp[-(E - E_{Average})^2 / \sigma_E^2] \quad (56)$$

The fluctuation parameter $\sigma_E$ can be best determined according to [27].

$$\left. \begin{array}{l} \Delta \sigma_E^{\,2} = \Delta z \cdot \tfrac{1}{2} \cdot (Z/A_N) \cdot \rho \cdot f \cdot \frac{2mc^2}{1-\beta^2}(1-\beta^2/2) \\ \text{(for finite intervals } \Delta z), \; f = 0.1535 \; MeV cm^2/g \\ d\sigma_E^{\,2}/dz = \tfrac{1}{2} \cdot (Z/A_N) \cdot \rho \cdot f \cdot \frac{2mc^2}{1-\beta^2}(1-\beta^2/2) \end{array} \right\} \quad (57)$$

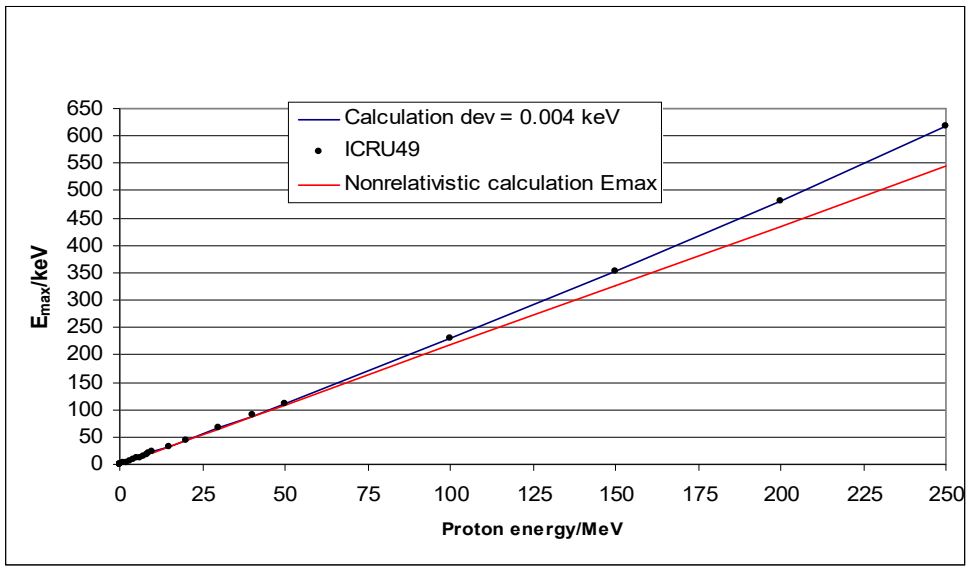

**Figure 10:** Calculation of $E_{max}$ according to Eq. (58). The straight line refers to the non-relativistic limit.

$\Delta\sigma_E^2$ contains as a factor the important magnitude $E_{max}$, that is, the maximum energy transfer from the proton to an environmental electron; it is given by $E_{max} = 2mv^2/(1-\beta^2)$. In a non-relativistic approach, we get $E_{max} = 2mv^2$. $E_{max}$ can be represented in terms of the energy E, and, for the integrations to be performed, we recall the relation $E = E(z)$ according to Eq. (40):

$$E_{max}(in\ keV) = \sum_{k=1}^{4} s_k \cdot E^k \quad (E\ in\ MeV) \quad (58)$$

| $S_1$: 2.176519870758 | $S_2$: 0.001175000049 |
|---|---|
| $S_3$: -0.000000045000 | $S_4$: 0.0000000000348 |

**Table 7:** The parameters $s_k$ for the calculation of $E_{max}$ (Eq. (58)).

This formula is plotted in Figure 10 (the non-relativistic limit is given by the straight line $s_1 \cdot E$); the parameters $s_k$ are displayed in Table 7.

### 1.7.2 Gaussian convolution in configuration space, its operator notation and inclusion of the inverse operator

A Gaussian distribution is often assumed to account for the energy straggling, if the stopping power is given by the CSDA approach. However, this is valid only in the non-relativistic domain, and the Gaussian convolution kernel in the configuration space must not be mixed with Bohr's relation. The convolution kernel reads as:

$$K(s, u-z) = \frac{1}{s\sqrt{\pi}} \exp(-(u-z)^2/s^2) \quad (59)$$

Here the half-width parameter s is arbitrary; this quantity will be identified with the energy/range straggling or the lateral scatter, if Eq. (59) is appropriately modified. In a 3D version, linear combinations of $K(s, u-x)$ and its inversion $K^{-1}$ are also used in scatter problems of photons and electrons [6, 7]. If the dose deposition of protons is calculated by the BBE or by Eqs. (12, 22) based on classical energy dissipation, then the energy fluctuations are usually accounted for by:

$$D(z) = \int D_{R_{CSDA}}(u) K(s, u-z) du \quad (60)$$

This kernel may either be established by non-relativistic transport theory (Boltzmann equation) or, as we prefer here, by a quantum statistical derivation, where a relation to the path-integral formulation of Neumann's density matrix will be obtained [42].

Let $\varphi$ be a distribution function and $\Phi$ a source function, mutually connected by the operator $O^{-1}$ (operator notation of a canonical ensemble):

$$\left.\begin{array}{l} \varphi = \exp(-H/E_{ex})\Phi = O^{-1} \cdot \Phi \\ H = -\frac{\hbar^2}{2m} d^2/dz^2 \\ O^{-1} = \exp(0.25\ s^2\ d^2/dz^2/E_{ex}) \end{array}\right\} \quad (61)$$

The exchange Hamiltonian H in Eq. (61) couples the source field $\Phi$ (proton fluence) with an environmental field $\varphi$ by $O^{-1}$ via exchange interaction with electrons and the parameter $s^2$ can be identified by:

$$s^2 = 2\hbar^2/mE_{ex} \quad (62)$$

It is also possible to use $s^2$ according to Eq. (62) as a free parameter without quantum-statistical background and it must be noted that the operator $O^{-1}$ in Eq. (61) was formally introduced [8] to obtain a Gaussian convolution as Green's function and to derive the inverse convolution. $O^{-1}$ may formally be expanded in the same fashion as the usual exponential function $\exp(\xi)$; $\xi$ may either be a real or complex number. This expansion is referred to as Lie series of an operator function. Only in the thermodynamic equilibrium, can we write $E_{ex} = k_B T$, where $k_B$ is the Boltzmann constant. Eq. (61) can be solved by the spectral theorem (functional analysis). According to this theorem, we have to consider the eigen-value problem:

$$\left.\begin{aligned}
O^{-1} \cdot \Phi &= \gamma \cdot \Phi; \; \Phi_k = \tfrac{1}{\sqrt{2\pi}} \cdot \exp(-ikz) \\
O^{-1} \cdot \Phi_k &= \tfrac{\gamma(k)}{\sqrt{2\pi}} \cdot \exp(ikz) = \tfrac{1}{\sqrt{2\pi}} \cdot \exp(-s^2 k^2 / 4) \cdot \exp(ikz) \\
K(s, u-z) &= \int \Phi_k^*(z) \Phi_k(u) \gamma(k) dk = \\
K(s, u-z) &= \tfrac{1}{s\sqrt{\pi}} \exp(-(u-z)^2 / s^2)
\end{aligned}\right\} \quad (63)$$

The operator $O^{-1} = \exp(0.25 \cdot s^2 \cdot d^2/dz^2)$, which provides the kernel $K(s, u-z)$, has some advantages, e.g. to gain the inverse operator in a very easy way:

$$\left.\begin{aligned}
O^1 &= \exp(-0.25\, s^2 d^2/dz^2 / E_{ex}) \Rightarrow \\
O^1 \cdot O^{-1} &= O^{-1} \cdot O^1 = 1
\end{aligned}\right\} \quad (64)$$

Thus, both operators $O^{-1}$ and $O^1$, require that each function $\Phi$ belongs to the function space $C^\infty$ (or a subspace, e.g. a polynomial of finite order). This can readily be verified by the Lie series expansion [6, 7, 43]:

$$\left.\begin{aligned}
O^{\mp 1} &= \exp(\pm \tfrac{1}{4} \cdot s^2 \cdot d^2/dz^2) \cdot \Phi = \\
&= [1 + \sum_{k=1}^\infty (\pm 1)^k \cdot \tfrac{s^{2k}}{4^k} \cdot \tfrac{1}{k!} \cdot \tfrac{d^{2k}}{dz^{2k}}] \cdot \Phi
\end{aligned}\right\} \quad (65)$$

The function space spanned by the kernel $K(s, u-z)$ is the set of all $L_1$-integrable functions [4, 6 - 8]. The advantage is the applicability in computational problems of only step-wise defined data. The ***inverse kernel $K^{-1}(s, u-z)$*** has been developed and applied in previous studies quoted above; interested readers may consult the papers [6 - 8] and references therein.

The density-matrix formulation of quantum mechanics is based on the definition [42]:

$$\rho(u, z) = \sum_{k=0}^\infty \exp(-E_k / k_B T) \psi^*(k, u) \cdot \psi(k, z) \quad (66)$$

For the statistical motion of free particles ($E_k = \hbar^2 k^2 / 2m$), we obtain $\rho(u - z) = K(\sigma, u - z)$, if $E_{ex} = k_B T$. In the continuous case, the summation over k has to be replaced by an integral. With the help of Eq. (66), various properties (e.g., the partition function) can be calculated. . Thus, the operator-function formalism according to Eqs. (64 – 66) can be regarded as an operator calculus of path-integral kernels. The question now is, which temperature T or exchange energy $E_{ex}$ should be used and whether $E_{ex} = k_B T$ may hold in the energy straggling of protons and, consequently, this parameter may be kept constant along the pathway. For this purpose, we consider some further properties which can easily be derived from the operator notation. It immediately follows from $O^{-1} = \exp(0.25 \cdot s^2 \cdot d^2/dz^2)$ that by n-times repetition of this operator (method of iterated operators), we obtain:

$$(O^{-1})^n = O^{-n} = [\exp(0.25 \cdot s^2 d^2/dz^2)]^n$$
$$= \exp(0.25 \cdot n \cdot s^2 d^2/dz^2) \quad (67)$$

The kernel K resulting from Eq. (67) is:

$$K(s_n, u-z) = \frac{1}{\sqrt{\pi}} \frac{1}{\sigma_n} \exp(-(u-z)^2/s_n^2)$$
$$s_n^2 = n \cdot s^2 \quad (68)$$

A composite application (e.g., the energy-range straggling of polychromatic proton beams) is now given by $O^{-1}(s_1) \cdot O^{-1}(s_2)$ and yields:

$$O^{-1}(s_1) \cdot O^{-1}(s_2) = \exp(0.25 \cdot (s_1^2 + s_2^2) d^2/dz^2)$$
$$\Rightarrow K(s_t, u-z); \quad (s_t^2 = s_1^2 + s_2^2) \quad (69)$$

The advantage of the operator calculus is the straightforward calculation of generalized kernels K, even if these kernels are not Gaussian, e.g., Landau/Vavilov distributions. The question arises as to the order of magnitude following from Eq. (62) for $\sigma^2$, which is expressed there in terms of fundamental constants. For this purpose, we partition the whole proton path from $z = 0$ to $z = R_{CSDA}$ in molecular intervals of water (or other) atoms/molecules. With respect to water molecules ($\rho = 1 g/cm^3$, $N_{Avogadro} = 6.022 \cdot 10^{23}$, $A_N = 18$, $M_{Mol} = 18$ g), the average distance between the centers of mass amounts to $l_A = 8.4446 \cdot 10^{-7}$ cm. The exchange particle is the electron, which undergoes further collisions with environmental electrons, to yield finally a locally stored energy that can be recorded by calorimetric measurements. In the case of electrons $E_{ex} = 1$ eV, the calculated σ amounts to $3.26 \cdot 10^{-7}$ cm, and for $E_{ex} = 0.1$ eV (thermal energy) to $s = 1.03 \cdot 10^{-6}$ cm. It is obvious that this process has to be repeated. This means that $s_n^2 = n \cdot s^2$ enters the convolution kernel as a parameter, which is now $s^2(z)$. It should be noted that the problem of composite convolutions can easily be handled by the operator notation. If the exchange energy $E_{ex}$ would remain always constant, then we should obtain $s_n \sim R_{CSDA}^{1/2}$ at the distal end. However, this is not true, as the locally stored exchange energy is increasing with decreasing proton energy. *In Figures 11 and 12*, we show the 'global average: $E_{av\_global}$', which results from the energy $E_0$ divided by the number of water molecules per unit length, and the 'local average: $E_{av\_local}$', obtained by a subtraction method (starting with the lowest energy $E_0 = 1$ keV). If we compare this result with $E_{max}$ according to Figure 10, we may verify that, in particular for high proton energies, a considerable amount of the proton energy is stored in δ-electrons, and, only by further collisions of these electrons along the track with the electrons of the environmental molecules, is the stored energy locally downgraded to thermal energies. (These processes comprise the starting point of calorimetric measurements). Some specific numerical cases for exchange energies $E_{ex}$ are given in Table 8. With respect to the operator calculus of the Gaussian kernel, we finally add two items. The 3D version of $O^{-1}$ with different σ values in the z direction and x/y plane is given by:

$$O^{-1} = \exp[\tfrac{1}{4}(s^2 \cdot \partial^2/\partial z^2 + \sigma^2(\partial^2/\partial x^2 + \partial^2/\partial y^2))]$$
$$\Rightarrow K(u-z, v-x, w-y) =$$
$$\frac{1}{s\sqrt{\pi}} \frac{1}{\pi \cdot \sigma^2} \exp[-((u-z)^2/\sigma^2 + ((v-x)^2 + (w-y)^2)/\sigma^2)] \quad (70)$$

One should note that, only for proton – electron collisions, is the exchange particle an electron, where a Boltzmann partition function can be applied. For proton – nucleon or electron – electron collisions, due to the Pauli principle, we have to make use of Fermi – Dirac statistics (next section).

| Energy | exchange length σ (in cm) | exchange length σ (in cm) |
|---|---|---|
| Exchange particle | proton | electron |
| 1 MeV | $0.91 \cdot 10^{-12}$ | $3.86 \cdot 10^{-11}$ |
| 1 keV | $0.24 \cdot 10^{-10}$ | $1.03 \cdot 10^{-9}$ |
| 1 eV  | $0.76 \cdot 10^{-8}$  | $3.26 \cdot 10^{-7}$ |
| 0.1 eV| $0.24 \cdot 10^{-7}$  | $1.02 \cdot 10^{-6}$ |

**Table 8:** Exchange length s in dependence energy $E_{ex}$ according to Eq. (61).

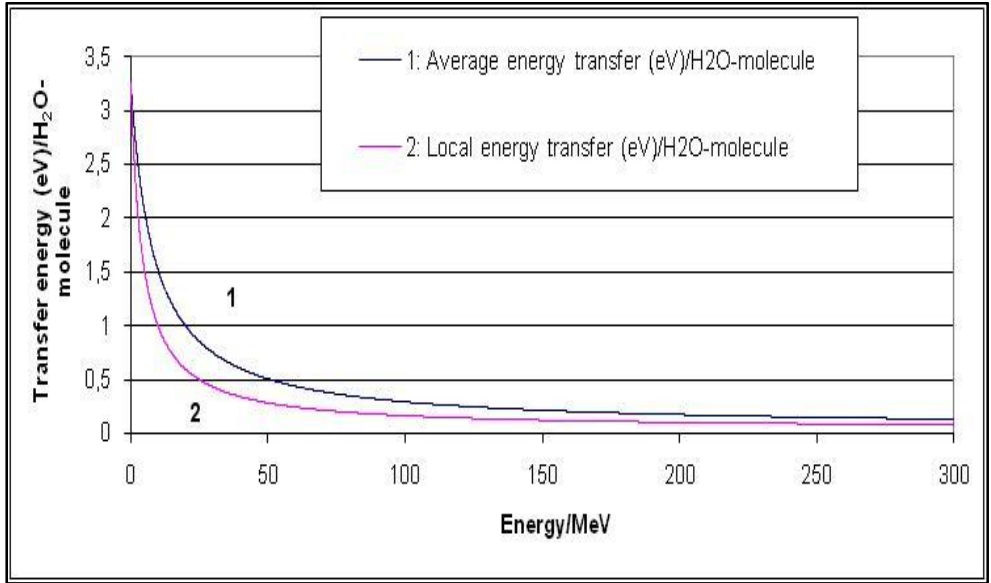

**Figure 11**: The average transfer energy (1) and the local transfer energy (2) versus the total energy E.

Eq. (70) additionally represents the basis for multiple lateral scatter. If it is modified according to principles developed in the next section, i.e., that proton scatter is only *approximately* described by one Gaussian on the x/y plane, we are able to describe the Molière multiple scatter and energy fluctuations in a unique manner. It should be noted that $s^2 + \sigma^2 \sim N_{Avogadro}$ holds now. The general relation $s = s(z)$ does not imply any constraint, since the operator $O^{-1}(s(z))$ yields the kernel $K(s(z), u - z)$. Owing to the linearity of $O^{-1}$, that is, the absence of terms $(O^{-1} \cdot \Phi)^2$, the solution function $\Phi_k \sim \exp(i \cdot k \cdot z)$ is not the only possible one; the multiplication with an arbitrary function g(k) also solves Eq. (63):

$$\Phi_k = g(k)\exp(-i\,k\,z)/\sqrt{2\pi} \quad (71)$$

Based on Eq. (71), some applications have been previously stated [6, 7].

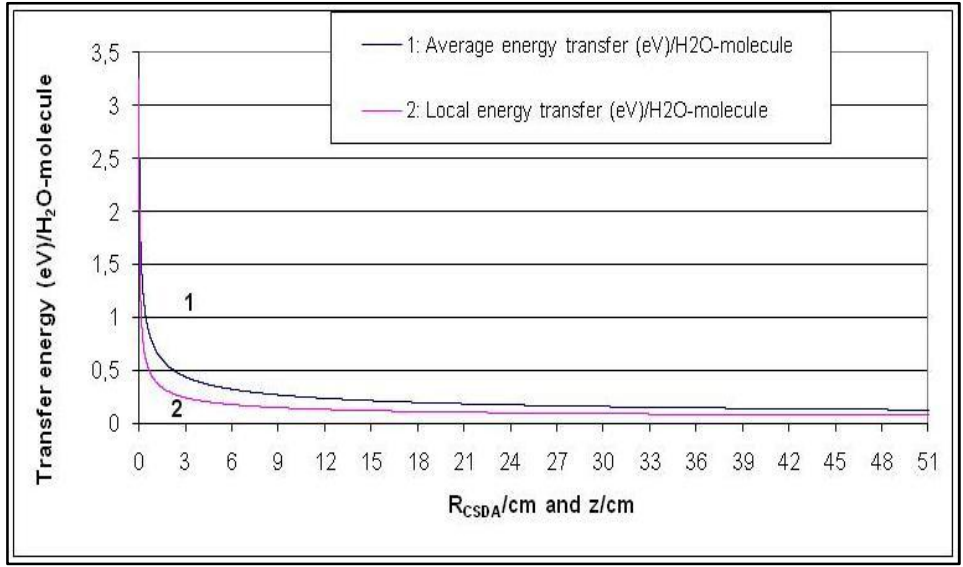

**Figure 12:** The average transfer energy (1) and the local transfer energy (2) from proton to water molecules (CSDA approach).

### 1.7.3 Dirac equation and Fermi-Dirac statistics

In the following, we make use of properties, which one may find in a textbook, e.g., [32]. The Dirac Hamiltonian $H_D$ reads as:

$$\left.\begin{array}{l} H_D = c\vec{\alpha}\,\vec{p} + \beta\,mc^2 \Rightarrow \\ (\beta\,mc^2 + \tfrac{\hbar c}{i}\cdot\vec{\alpha}\cdot\nabla)\psi = E_D\cdot\psi \\ \vec{\alpha} = \begin{pmatrix}0 & \vec{\sigma} \\ \vec{\sigma} & 0\end{pmatrix}\;\beta = \begin{pmatrix}1 & 0 \\ 0 & -1\end{pmatrix};\; H_D^{\,2} = c^2 p^2 + m^2 c^4 \end{array}\right\} \quad (72)$$

With respect to the Fermi-Dirac statistics, we use the notation $\hat{H} = H_D - E_F$ ($E_F$: energy of the Fermi edge). The Fermi distribution function reads:

$$f_F(\hat{H}) = f(\hat{H})\cdot d_s(H_D) \quad (73)$$

The notation $d_s(H_D)$ refers to the density of states corresponding to the energy $E_D$ (or Hamiltonian $H_D$). By use of these definitions, the operator equation of Fermi-Dirac statistics, which will be used, assumes the shape:

$$f_F(\hat{H}) = \frac{1}{1+\exp(H_D/E_{ex})}\cdot d_s(H_D) \quad (74)$$

With regard to Eq. (74) we have to recall that, for the eigen-values E of *continuous* operators H, the following property is valid:

$$\left.\begin{array}{l} H\,\psi = E\,\psi \\ f(H)\,\psi = f(E)\,\psi \end{array}\right\} \quad (75)$$

The operator function f(H) may result from an iteration of H (Lie series [43]); in Section 1.8 (energy straggling), we shall intensively make use of this property. With the restriction to the z-axis and n-

times repetition of the Fermi operator $f_F(\hat{H})$, we obtain:

$$\left.\begin{array}{l}\eta(k)=[mc^2\sqrt{1+\hbar^2 \cdot k^2/m^2c^2}-E_F]/E_{ex}\\ \gamma(\eta(k))=[\tfrac{1}{2}(\eta \cdot E_{ex}+E_F)/mc^2]^n \cdot \exp(-n\eta/2)\cdot \sec h(\eta/2)^n\end{array}\right\} \quad (76)$$

The general convolution kernel $K_F$ is given by:

$$K_F = \tfrac{1}{2\pi}\int_{-\infty}^{\infty}\exp(ik(u-z))\cdot d\gamma(\eta(k)) \quad (77)$$

This integral cannot be evaluated by elementary methods, but it is possible to expand the hyperbolic-secant function $\sec h(\eta/2)$ in terms of a Gaussian multiplied with a specific power expansion. By that, we obtain a rigorously generalized version of the non-relativistic kernels, which are already investigated in section 1.7.2. It is known from textbooks [31] that the polynomial expansion of $\sec h(\xi)$ is given by:

$$\sec h(\xi) = \sum_{l=0}^{\infty} E_{2l} \cdot \xi^{2l}/(2l)! \quad (|\xi|<\pi/2) \quad (78)$$

$E_{2n}$ are the Euler numbers; because of the very restricted convergence conditions, this expansion does not help. In view of the evaluation of the kernel $K_F$, we have derived a more promising expansion without any restriction for convergence [4]:

$$\left.\begin{array}{l}\sec h(\xi) = \exp(-\xi^2)\cdot \sum_{l=0}^{\infty}\alpha_{2l}\cdot \xi^{2l}\\ \alpha_{2l}=E_{2l}/(2l)!+\sum_{l'=1}^{l}(-1)^{l'+1}\cdot \alpha_{2l-2l'}/l'!\end{array}\right\} \quad (79)$$

The coefficients $\alpha_{2l-2l'}$ are recursively defined. Some low-order coefficients are: $\alpha_0 = 1$, $\alpha_2 = 1/2$, $\alpha_4 = 5/4!$, $\alpha_6 = 29/6!$, $\alpha_8 = 489/8!$. It is easy to verify that, due to the Gaussian, the expansion above converges for $|\xi|\leq\infty$. Since $\eta(k)$ is given by a root, some terms are connected with difficulties, but if we expand $\eta(k)$ by the power expansion (3), then – besides the rest energy $mc^2$ – the first expansion term is the non-relativistic contribution; thus, we may consider only Fermi statistics without relativistic terms (Dirac). The higher-order terms are now readily evaluated (up to arbitrary order). The energy distribution function $S_E$ assumes the shape:

$$S_E = N_f \cdot \exp(-(E_n(k)-E_{Average,n})^2/2\sigma_E(n)^2)$$
$$\cdot \sum_{l=0}^{\infty} b_l(n,mc^2)\cdot (E_n(k)/2E_{ex})^l \quad (80)$$

$N_f$ is a normalization factor. Eq. (80) provides a very interesting special case: if we take $l = 0$, we obtain Bohr's classical formula (56) and a connection between $E_{Average}$ and $E_F$. The convolution kernel $K_F$ is determined by the general structure:

$$K_F = N_f \cdot \sum_{l=0}^{\infty} H_l((u-z-z_{shift}(l))/\sigma_n)\cdot B_l(n,mc^2)$$
$$\cdot \exp(-(u-z-z_{shift}(l))^2/2\sigma_n^2) \quad (81)$$

Now $E_{Average}$ and $z_{shift}$ depend in any order on the Fermi edge energy. Detailed descriptions of the above terms are avoided in this paper; Eqs. (80 - 81) represent only a general outline. If we recall that $\sigma_n$ is a function of $z$ (that is, $\sigma(z)$), and take the proportionality $E_F \sim E_{Average}$ into account, we can verify that the Landau – Vavilov theory represents a specific form of a Fermi-Dirac distribution. It should also be emphasized that $E_F$ and consequently $E_{Average}$ depend on the proton track, but in a low order (that is, the proton energy is $E_0 \ll Mc^2$ and the kinetic energy $E_\delta$ of δ-electrons satisfies $E_\delta \ll 2mc^2$), the Bohr approximation formula is a good basis. One might be surprised that, in practical calculations, we can use $\sigma(R_{CSDA})$ instead of the continuously increasing $\sigma(z)$. For homogeneous media, this simplification works, since the residual energy $E(z)$ of a proton is monotonically decreasing. (For heterogeneous media, we have always to take slabs with different densities and material properties like $(Z/A_N)_{medium}$ into account.) The question arises as to when the kernel $K_F$ can be used. The first case pertains to the energy/range straggling of fast electrons as primary projectiles; to avoid unnecessary evaluations of higher order, it is necessary to check the order of $K_F$ by comparison to the measured stopping power. The second case is the calculation of transition probabilities and of the energy transfer of protons passing through nuclei; the Pauli principle for spin and iso-spin gives raise to exchange interactions.

The operators $O^1$ and $O^{-1}$ according to Eqs. (64, 65) can readily be extended to the 3D case (Δ is the 3D Laplace operator):

$$O^1 = \exp(-\tfrac{1}{4} \cdot s^2 \cdot \Delta); \quad O^{-1} = \exp(\tfrac{1}{4} \cdot s^2 \cdot \Delta) \quad (82)$$

Thus $O^{-1}$ and $O^1$ again satisfy: $O^1 \cdot O^{-1} = O^{-1} \cdot O^1 = 1$. These operators are defined by the Lie series:

$$O^{-1} = 1 + \sum_{n=1}^{\infty} \frac{s^{2n}}{n! \cdot 4^n} \Delta^n; \quad O = 1 + \sum_{n=1}^{\infty} \frac{s^{2n}}{n! \cdot 4^n} \cdot (-1)^n \cdot \Delta^n. \quad (83)$$

The integral operator notation of $O^{-1}$ now is the well-known 3D Gaussian kernel. However, the inverse operator $O^1$ is not so simple in the integral operator version $K^{-1}$:

$$K(s, u-x, v-y, w-z) = \tfrac{1}{\sqrt{\pi}^3} \cdot \tfrac{1}{s^3} \cdot \exp\left(-\tfrac{1}{s^2} \cdot ((u-x)^2 + (v-y)^2 + (w-z)^2)\right) \quad (84)$$

$$K^{-1}(s, u-x, v-y, w-z) = O^2 \cdot K(s, u-x, v-y, w-z)$$

$$= \sum_{n=0}^{\infty} (-1)^n \cdot 2^{-n} \cdot \tfrac{1}{n!} \cdot \Delta^n \cdot K(s, u-x, v-y, w-z) \quad (84\,a)$$

The linear combination of Gaussian kernels with different r.m.s parameters s and the related inverse problem has been previously analyzed [6, 7]. A good starting-point is again the operator notation using Lie series (we restrict ourselves to two operators):

$$A^{-1} = c_1 \cdot O_1^{-1}(s_1) + c_2 \cdot O_2^{-1}(s_2); \quad (c_1 + c_2 = 1) \quad (85)$$

Thus the resulting kernel resulting from Eq. (85) simply is $K_A = c_1 \cdot K(s_1) + c_2 \cdot K(s_2)$. The operator A and the kernel $K_A^{-1}$ is obtained by the procedure:

$$A = [c_1 \cdot O_1^{-1}(s_1) + c_2 \cdot O_2^{-1}(s_2)]^{-1}$$

$$= \tfrac{1}{c_1} \cdot O_1^1(s_1) \cdot [1 + \tfrac{c_2}{c_1} \cdot O_2^{-1}(s_2) \cdot O_1^1(s_1)]^{-1} \quad (85\,a)$$

The right hand side of Eq. (85a) now can be expanded by a Lie series, which is formally a geometric series $(1+x)^{-1} = 1 - x + x^2 - x^3 + x^4 ..-+..$ , which can be written as:

$$A = \tfrac{1}{c_1} \cdot O_1^1(s_1) + \tfrac{1}{c_1} \cdot \sum_{n=1}^{\infty} (-1)^n \cdot (c_2/c_1)^n \cdot O_1^{n+1}(s_1) \cdot O_2^{-n}(s_2) \quad (85b)$$

From Eq. (85b) results that $K^{-1}_A$ is derived by the kernel $K^{-1}(s_1)$ - this is a deconvolution kernel - and an infinite sequence of composite kernels $K(s_1, s_2)$, see e.g., [6, 7]. Till the present stage we tacitly have assumed that the configuration space spanned by the kernels $K^{-1}$, $K$, $K_A^{-1}$, $K_A$ and the related operators $O^{-1}$, $O^1$, $A^{-1}$ and $A^1$ is defined by the position space. However, this is an unnecessary restriction, and it is useful to substitute each position coordinate by the coordinate $\xi$, where $\xi$ may also refer to the momentum, to a component of the wave-vector **k** or to the energy E. It is evident that the r.m.s-parameter s has to be appropriately adapted in order to be consistent with the dimension of $\xi$. From the quantum-mechanical view-point, however, it is not always possible to start with Eqs. (61 - 63). This fact is easy to verify with regards to the operators $O^1(E, s_E)$, $O^{-1}(E, s_E)$, $K(s_E, E)$, which formally assume the form:

$$O^{\pm 1}(s_E, E) = \exp[\mp \tfrac{s_E^2}{4} \cdot \partial^2 / \partial E^2] \quad (86)$$

$$K(s_E, E, -E') = \tfrac{1}{\sqrt{\pi}} \cdot s_E^{-1} \cdot \exp(-(E-E')^2 / s_E^2) \quad (86a)$$

$O^{-1}(s_E, E)$ can be associated with a Gauss transformation of the form:

$$\varphi(E) = \int K(s_E, E - E') \cdot \Phi(E') dE' \quad (86b)$$

In order to reach a quantum-mechanical basis of Eqs. (86 - 86b) we wish to consider the conformal Lorentz group; this is a function of the invariant magnitude $x^2 + y^2 + z^2 - c^2 \cdot t^2$, which in our case reads:

$$f_L = \exp(-(x^2 + y^2 + z^2 - c^2 \cdot t^2) / l^2) \quad (87)$$

It is now the goal to transform Eq. (87) in a quantum-mechanical operator version. Thus the replacement of the energy-momentum relation $E^2 = c^2 \cdot \mathbf{p}^2 + m^2 \cdot c^4$ by the well-known quantum-mechanical differential operators directly yields the Klein-Gordon equation:

$$\left. \begin{array}{l} E \to i\hbar \cdot \tfrac{\partial}{\partial t}; \quad \vec{p} \to -i\hbar \cdot \nabla \\ (\Delta - \tfrac{1}{c^2} \partial^2 / \partial t^2) \cdot \psi = \tfrac{m^2 \cdot c^2}{\hbar^2} \cdot \psi \end{array} \right\} \quad (88)$$

The differential operators in Eq. (88) result from the position/time representations of Heisenberg's commutation relations:

$$\left. \begin{array}{l} x_k \cdot p_l - p_l \cdot x_k = i\hbar \cdot \delta_{kl} \quad (k, l = 1, 3) \\ E \cdot t - t \cdot E = i\hbar \end{array} \right\} \quad (88a)$$

An operator representation of the conformal Lorentz group according to Eq. (87) is received by the replacements yielding an operator function of $f_L$:

$$\left.\begin{array}{l}\vec{x}\to i\hbar\cdot\nabla;\ t\to -i\hbar\cdot\partial/\partial E\ \Rightarrow\\ O^1=\exp(-(-\hbar^2\cdot\Delta_p-c^2\cdot(-\hbar^2)\cdot\partial^2/\partial E^2)/l^2)\\ O^1=\exp((\hbar^2\cdot\Delta_p-c^2\cdot\hbar^2\cdot\partial^2/\partial E^2)/l^2);\ s_E^2=4\cdot\hbar^2\cdot c^2/l^2\end{array}\right\} \quad (88b)$$

Thus $O^1$ agrees with Eq. (86), if the operator $\Delta_p$ acting on the momentum space is canceled. However, this is only possible, if the function space $\Phi(E)$ does not depend on the momentum. The implications of this restriction will be accounted for in the Appendix. It is the intension in the Appendix to apply $O^1(s_E, E)$ to the total nuclear cross-section $Q^{tot}(E)$ in order to differ proper nuclear reactions from elastic/inelastic scatter of protons at nuclei.

## 1.8 Calculation methods for pristine Bragg curves by inclusion of energy/range straggling and Landau tails

This section deals with the parameters required for the description of fluctuations, the energy shifts due to the nuclear interactions, and the contributions of heavy recoils to the total stopping power.

### 1.8.1 Energy-range straggling and r.m.s parameters (Eq. (89))

It is obvious that, even for initially mono-energetic proton beam-lets, there are fluctuations in the energy transfer from the protons to the environmental electrons. In first order, we assume that these fluctuations are symmetrical (i.e., Gaussian distributed). In particular, Figure 5 ignores these fluctuations. It is also commonly assumed that the incident-proton spectra are also of Gaussian type. We adopt here these first-order assumptions and add, as a second-order correction, the Landau tails for the description of the complete beam-line. However, these corrections are necessary, if the initial proton energy $E_0$ exceeds 100 MeV; the importance is increasing with $E_0$. We denote r.m.s of mono-energetic proton beam-lets by $\tau_{straggle}$ and r.m.s of the impinging proton beam by $\tau_{in}$. The r.m.s value of the polychromatic energy spectrum $\tau$ is given by:

$$\tau^2 = \tau_{straggle}^2 + \tau_{in}^2 \quad (89)$$

It is a favorite property of Gaussian convolutions that the two successive convolutions lead to the result of one convolution, performed according to Eq. (89). With regard to primary protons, we will give the results of all required convolutions in terms of $\tau$; by setting $\tau_{in} = 0$, we obtain the pristine Bragg curves of initially mono-energetic protons.

It is clear that $\tau_{straggle}(z)$ has to be a monotonically increasing function of z. However, it is usual in various calculation models to restrict $\tau_{straggle}(z)$ to $\tau_{straggle}(R_{CSDA})$. In almost all applications in therapy planning, this restriction is justified, since S(z) in the CSDA framework is a monotonically-increasing function, too, and $\tau_{straggle}(z)$ does not have an influence in the initial plateau of a pristine Bragg curve.

With respect to this calculation, we are closely related to the Bohr approximation, i.e., $E_{Average} = 0$, if E < 10 MeV:

$$\tau_{straggle}^2 = \int_0^{R_{CSDA}} (d\sigma_E^2/dz)[dE/dz]^{-2} dz \quad (90)$$

Using the relativistic formula for E(z) and dE(z)/dz, i.e., Eqs. (21 - 22):

$$\left.\begin{aligned}\tau_{straggle}^2 &= 2 \cdot C_{Bohr} \cdot \int_0^{R_{CSDA}} [p^2 A^{1/p} \cdot (R_{CSDA} - z)^{2-2/p} + (2p^2 A^{1/p} / Mc^2) \cdot (R_{CSDA} - z)^{2-1/p}] dz \\ &= 2 \cdot C_{Bohr} \cdot [p^2 A^{2/p} \cdot (3 - 2/p)^{-1} R_{CSDA}^{3-2/p} + (2p^2 A^{1/p} / Mc^2) \cdot (3 - 1/p)^{-1} \cdot R_{CSDA}^{3-1/p}]\end{aligned}\right\} \quad (91)$$

$C_{Bohr}$ amounts to 0.087 MeV$^2$/cm. If the integrations according to Eq. (91) are only carried out up to a final value E(z) > 0 and with regard to Eq. (12), the two most important contributions are accounted for; we are then able to obtain a fitting formula for $\tau_{straggle}(z)$:

$$\tau_{straggle}(z) = \tau_{straggle}(R_{CSDA}) \cdot \frac{e^{Q_z \cdot z} - 1}{e^{Q_z \cdot R_{CSDA}} - 1} \quad (92)$$

where $Q_z$ = 2.887 cm$^{-1}$. Figure 4 shows that we cannot deal with constant values of p; a good approximation in the relativistic case is:

$$\left.\begin{aligned}p &= 1.71 \quad (if\ E_0 \geq E_{50} = 50\ MeV) \\ p &= 1.66 + 0.05 \cdot E_0 / E_{50} \quad (if\ E_0 \leq E_{50})\end{aligned}\right\} \quad (93)$$

These two restrictions can be used for the evaluation of the result given by Eq. (91). The applicability of the connection (93) is independent of the way in which $\tau_{straggle}(R_{CSDA})$ is calculated. In [33] we have presented an alternative method. We decomposed the total energy straggling into a Gaussian part and a Landau tail. Since the resulting Landau tail represents a comparably small correction, we assume that $\tau_{straggle}$ for the Landau tail is the same as for the Gaussian part. On the other hand, we do not permit any approximation with respect to the evaluation of $[dE(z)/dz - E_{average}]^{-2}$ according to Eqs. (39 – 40). The resulting formula is a power expansion in terms of exponential functions:

$$\left.\begin{aligned}\tau_{straggle}(z) =\ & R_{CSDA} \cdot [c_{m,1} \cdot (1 - \exp(-g_1 \cdot R_{CSDA})) + \\ & + c_{m,2} \cdot (1 - \exp(-g_2 \cdot R_{CSDA}))] \\ & - (R_{CSDA} - z) \cdot [c_{m,1} \cdot (1 - \exp(-g_1 \cdot (R_{CSDA} - z))) + \\ & + c_{m,2} \cdot (1 - \exp(-g_2 \cdot (R_{CSDA} - z)))] + 0(higher\ order)\end{aligned}\right\} \quad (94)$$

The higher-order terms are smaller than 0.001 and hence negligible. In view of a possible step-by-step calculation, we write $\tau_{straggle}(z)$, not simply $\tau_{straggle}(R_{CSDA})$. The dimensionless coefficients $c_{m,1}$ and $c_{m,2}$ are given by: $c_{m,1}$ = 0.4968 and $c_{m,2}$ = 0.1662. For the factors $g_1$ and $g_2$: $g_1$ = 176.752 cm$^{-1}$ and $g_2$ = 112.384 cm$^{-1}$. We emphasize that these values are valid for water. For other media, the coefficients $c_{m,1}$ and $c_{m,2}$ have to be rescaled:

$$\left.\begin{aligned}c_{m,1}(medium) &= c_{m,1} \cdot \sqrt{(Z \cdot \rho / A_N)_{medium} / (Z \cdot \rho / A_N)_{water}} \\ c_{m,2}(medium) &= c_{m,2} \cdot \sqrt{(Z \cdot \rho / A_N)_{medium} / (Z \cdot \rho / A_N)_{water}}\end{aligned}\right\} \quad (95)$$

Based on Eq. (22) with a rigorous account for p = p($E_0$) according to Eq. (43) and without the Bohr approximation, we have also performed a numerical calculation of $\tau_{straggle}(R_{CSDA})$. The results (Figure 13) are rather close to those of Eq. (91). However, it appears that small errors, obtained (for instance) by the Bohr approximation, have no significant importance. It has to be noted that that $\tau_{in}$ represents an additional fitting parameter with importance for all models. In a subsequent section, we will investigate the dependence of this additional parameter on the machine type.

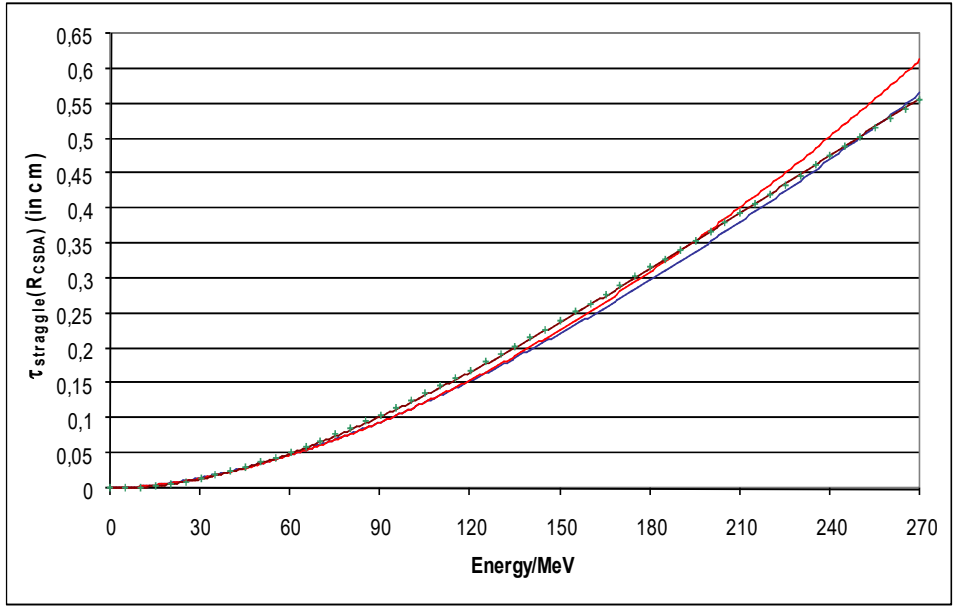

**Figure 13:** The determination of $\tau_{straggle}$ according to [10]; solid blue; according to the relativistic Eq. (22) and restriction (93): solid red; according to the exact integration of the BBE (Eq. (39)): solid brown; calculation with the relativistic Eq.(22) without restriction (93): green crosses.

### 1.8.2 Determination of the parameters for secondary/recoil protons and heavy recoils
In the case of secondary (non-reaction) protons, we have to replace Eq. (89) by:

$$\tau = \sqrt{\tau_{straggle}^2 + \tau_{in}^2 + 0.5 \cdot \tau_{heavy}^2} \qquad (96)$$

The r.m.s value $\tau_{heavy}$ is obtained via $Q^{tot}(E)$ as previously defined:

$$\tau_{heavy} = \begin{cases} 0, & \text{if } E_0 < E_{Th} \\ 0.55411 \cdot \dfrac{E_0 - E_{Th}}{E_{res} - E_{Th}}, & \text{if } E_{Th} < E_0 \leq E_{res} \\ 0.554111 - 0.000585437 \cdot (E_0 - E_{res}), & \text{if } E_{res} < E_0 \end{cases} \qquad (97)$$

The shift of the z-coordinate by $z_{shift}$, due to the reduced energy of secondary protons, is given by:

$$z_{shift} = \begin{cases} 0 & (\text{if } E_0 < E_{res}) \\ \displaystyle\sum_{n=1}^{4} a_n E_{res}^n \cdot \left[1 - \exp\left(-\dfrac{(E_0 - E_{res})^2}{5.2^2 \cdot E_{res}^2}\right)\right] & (\text{if } E_0 \geq E_{res}) \end{cases} \qquad (98)$$

If $E_0 \leq E_{res}$, we have to put $z_{shift} = 0$, but, in practice, this case does not occur. The application of Eq. (98) implies either Eq. (35) or Eq. (37) and the substitution $E_0 \rightarrow E_{res}$. Finally, we have to give the parameters for the determination of the stopping-power contribution of heavy recoils:

$$C_{heavy} = \begin{cases} 0 & (\text{if } E_0 < E_{Th}) \\ 0.0000004264 \cdot (E_0 - E_{Th})^2 & (\text{if } E_0 \geq E_{Th}) \end{cases} \qquad (99)$$

$$S_{heavy}(z, E_0) = \Phi_0 \, C_{heavy} \cdot e^{-\frac{z}{z_{max}}} \cdot \left[1 + \text{erf}\left(\sqrt{2} \cdot \dfrac{R_{csda} - z + \dfrac{E_0}{250} - 1}{\tau_{heavy}}\right)\right] \qquad (100)$$

We emphasize that, with regard to fits necessary to complete Eqs. (96 - 100), we have partially used results obtained by Monte-Carlo calculations with GEANT4.

### 1.8.3 Calculation of the total stopping power
The total stopping power $S_{tot}$ is defined by:

$$\left.\begin{aligned}S_{tot} &= S_{pp} + S_{sp} + S_{rp} + S_{heavy} \\ S_{sp} &= S_{sp,n} + S_{sp,r}\end{aligned}\right\} \quad (101)$$

$S_{pp}$ refers to primary protons, $S_{sp}$ to secondary protons (i.e., $S_{sp,r}$ to reaction and $S_{sp,n}$ to non-reaction protons), $S_{rp}$ to recoil protons, and $S_{heavy}$ to heavy recoil particles. For $S_{pp}$ and $S_{rp}$, $\tau$ is defined by Eq. (96) and, for $S_{sp,n}$, by Eq. (89). The contributions of $S_{sp,r}$ are defined in the Appendix. Due to the presence of the beam-line elements (degrader, modulator wheel, range shifter, etc.), the assumption $\tau_{in} = 0$ is rather meaningless. In this section, we make use of the following abbreviations:

$$\left.\begin{aligned}z_s &= z + z_{shift} \\ F &= R_{CSDA} \cdot [\tfrac{E_0 - E_{Th}}{M \cdot c^2}]^f \, ; \; f = 1.032 \\ F_{\upsilon s} &= F \cdot \upsilon \, ; \; \upsilon = 0.958 \, ; \; \upsilon' = \upsilon - 2 \cdot C_{heavy} \\ F_\upsilon &= F \cdot (1 - \upsilon); \; F_{\upsilon'} = F \cdot \upsilon'\end{aligned}\right\} \quad (102)$$

The convolution problem of every contribution of primary, secondary, and recoil protons has the general structure (S(u) solely refers to results obtained by CSDA approach):

$$S(z, \tau) = \int_{-\infty}^{R_{CSDA}} S(u) \cdot \tfrac{1}{\tau \sqrt{\pi}} \exp(-(u-z)^2 / \tau^2) \, du$$

### 1.8.4 Calculation of $S_{pp}$, $S_{sp,n}$ and $S_{rp}$

**(a) Primary protons $S_{pp}$**
In this subsection, we use $S_{pp} = S_{pp,1} + S_{pp,2}$ and introduce the following abbreviation:

$$Erf \lambda_k = \tfrac{1}{2} \cdot [1 + erf((R_{CSDA} - z - 0.5 \cdot \lambda_k \cdot \tau^2)/\tau)] \\ \cdot \exp(-\lambda_k \cdot (R_{CSDA} - z)) \cdot \exp(0.25 \cdot \tau^2 \cdot \lambda_k^2)$$

$S_{pp,1}$:

$$\left.\begin{aligned}S_{pp,1}(z,\tau) = &-\sum_{k=1}^{5} c_k \cdot [1 + \gamma_k \cdot (R_{CSDA} - z) - 0.5 \cdot \lambda_k^2 \cdot \tau^2] \cdot Erf\lambda_k \\ &-\sum_{k=1}^{5} c_k \cdot \lambda_k \cdot \exp(-\lambda_k \cdot (R_{CSDA} - z)) \cdot \exp(0.25 \cdot \tau^2 \cdot \lambda_k^2) \\ &\cdot \exp[-((R_{CSDA} - z - 0.5 \cdot \lambda_k \cdot \tau^2)/\tau)^2]\end{aligned}\right\} \quad (103)$$

$S_{pp,2}$:

$$S_{pp,2} = F \cdot z \cdot [\sum_{k=1}^{5} c_k \cdot [1 + \lambda_k \cdot (R_{CSDA} - z) - 0.5 \cdot \lambda_k^2 \cdot \tau^2] \cdot Erf\lambda_k]$$

$$+ F \cdot z \cdot \{\sum_{k=1}^{5} c_k \cdot \lambda_k \cdot \exp(-\lambda_k \cdot (R_{CSDA} - z)) \cdot \exp(0.25 \cdot \tau^2 \cdot \lambda_k^2) \quad (104)$$

$$\cdot \exp[-((R_{CSDA} - z - 0.5 \cdot \lambda_k \cdot \tau^2)/\tau)^2]\}$$

**(b) Secondary non-reaction protons $S_{sp,n}$**

Abbreviation:

$$Erf\lambda_{ksp} = \tfrac{1}{2} \cdot [1 + erf((R_{CSDA} - z_s - 0.5 \cdot \lambda_k \cdot \tau^2)/\tau)]$$

$$\cdot \exp(-\lambda_k \cdot (R_{CSDA} - z_s)) \cdot \exp(0.25 \cdot \tau^2 \cdot \lambda_k^2)$$

By that, we obtain:

$$S_{sp,n} = -F_{\upsilon'} \cdot z \cdot \sum_{k=1}^{5} c_k \cdot [1 + \lambda_k \cdot (R_{CSDA} - z_s) - 0.5 \cdot \lambda_k^2 \cdot \tau^2] \cdot Erf\lambda_{ksp}$$

$$- F_{\upsilon'} \cdot z \cdot \{\sum_{k=1}^{5} c_k \cdot \lambda_k \cdot \exp(-\lambda_k \cdot (R_{CSDA} - z_s)) \cdot \exp(0.25 \cdot \tau^2 \cdot \lambda_k^2) \cdot$$

$$\cdot \exp[-((R_{CSDA} - z_s - 0.5 \cdot \lambda_k \cdot \tau^2)/\tau)^2]\} \quad (105)$$

**(c) Recoil protons $S_{rp}$**

Abbreviation:

$$Erf\lambda_{krp} = \tfrac{1}{2} \cdot [1 + erf((R_{CSDA} - z_s - 0.5 \cdot \lambda_k \cdot \tau^2)/\tau)]$$

$$\cdot \exp(-\lambda_k \cdot (R_{CSDA} - z_s)) \cdot \exp(0.25 \cdot \tau^2 \cdot \lambda_k^2)$$

By that, we obtain:

$$S_{rp} = -F_{\upsilon} \cdot z \cdot \sum_{k=1}^{5} c_k \cdot [1 + \lambda_k \cdot (R_{CSDA} - z_s) - 0.5 \cdot \lambda_k^2 \cdot \tau^2] \cdot Erf\lambda_{krp}$$

$$- F_{\upsilon} \cdot z \cdot \{\sum_{k=1}^{5} c_k \cdot \lambda_k \cdot \exp(-\lambda_k \cdot (R_{CSDA} - z_s)) \cdot \exp(0.25 \cdot \tau^2 \cdot \lambda_k^2) \quad (106)$$

$$\cdot \exp[-((R_{CSDA} - z_s - 0.5 \cdot \lambda_k \cdot \tau^2)/\tau)^2]\}$$

### 1.8.5 Inclusion of the Landau-tail corrections and their role in the pristine Bragg curves

As, according to Eqs. (89, 96), it is difficult to decide which order of corrections is sufficient, we will provide, in this section, a summary of the theoretical results with GEANT4. It is an interesting feature that the lowest order (i.e., $S_E(l = 0)$) of Eq. (80) yields Bohr's classical formula (Eq. (56)) of fluctuations and energy straggling. Eq. (81) represents the translation in the configuration space. In the following, we keep the terms up to order 2 in Eq. (81), and compare the results with the Monte-Carlo calculations based on GEANT4.

We have not yet accounted for the contributions of the Landau tails. These tails result from the modification of the energy transfer and the stopping power of protons. The theoretical analysis of the preceding section [33] concludes that symmetrical; i.e., Gaussian fluctuations (and the related convolution kernel) of the energy transfer according to the CSDA are only rigorously valid, if the local proton energy and the energy transfer by collisions are non-relativistic. The maximum energy transfer $E_{max}$ has been shown in Figure 10 as a function of the local energy for protons in water. $E_{max}$ has a nonlinear term, which becomes more important with increasing energy. In a similar way, the fluctuations of the energy transfer become less symmetrical; collisions occur significantly less frequently. This behavior can be observed more and more for proton energies $E_0 \gg 100$ MeV. A consequence of this relativistic effect is that protons in the entrance region (e.g., 250 MeV, $E_{max}$ = 617 keV) undergo fewer collisions with environmental electrons than it would be expected from a symmetrical energy transfer. Thus, less energy is locally stored and a contribution to the buildup effect can be seen as long as the symmetrical fluctuation is not yet reached. However, this effect decreases along the proton track, and when the local energy approaches about 100 MeV, the fluctuations of the energy transfer tend to become symmetrical, i.e., the buildup effect is reduced (and vanishes for $E_0 \ll 100$ MeV, since Landau tails and reaction protons become negligible). The preceding relativistic treatment of convolutions provides that the inclusion of the Landau tails (Vavilov distribution function) necessitates generalized Gaussian convolutions, i.e., Gaussian convolutions with relativistic correction terms expressed by two-point Hermite polynomials. However, for proton energies below about 300 MeV, the lower-order corrections are sufficient. The results of the generalized Gaussian convolution are the following contributions:

**(a) Primary protons:**

$$S_{Lan,pp} = \Phi_0 \cdot (I_{Lan1}(z) + I_{Lan2}(z)) \cdot (1 - \tfrac{z}{R_{CSDA}}) \quad (107)$$

**(b) Secondary protons:**

$$S_{Lan,sp} = \Phi_0 \cdot (I_{Lan1}(z_s) + I_{Lan2}(z_s)) \cdot \tfrac{z}{R_{CSDA}} \cdot \upsilon' \quad (108)$$

**(c) Recoil protons:**

$$S_{Lan,rp} = \Phi_0 \cdot (I_{Lan1}(z_s) + I_{Lan2}(z_s)) \cdot (1 - \upsilon) \cdot \tfrac{z}{R_{CSDA}} \quad (109)$$

Concerning the calculation formulas for secondary/recoil protons, the substitutions $z \to z_s$ has to be performed.

$$I_{Lan1} = C_{Lan1} \cdot \left\{ erf\left(\frac{2 \cdot z}{R_{Lan1}}\right) + erf\left(\frac{2 \cdot (R_{Lan1} - z)}{\tau_{Lan1}}\right) \cdot \frac{z}{R_{CSDA}} \cdot \right\}$$
$$\cdot \left(1 + erf\left(\frac{(R_{CSDA} - z)}{\tau}\right)\right) \quad (110)$$

$$\left. \begin{aligned} C_{Lan1} &= -1.427 \cdot 10^{-6} \cdot R_{csda}^{\,3} + 1.439 \cdot 10^{-4} \cdot R_{csda}^{\,2} - 0.002435 \cdot R_{csda} + 0.2545 \\ R_{Lan1} &= R_{csda} \cdot \begin{cases} 0.7 & (\text{if } E_0 < 68 \text{ MeV}) \\ (0.812087912 - 0.001648352 \cdot E_0) & (\text{if } E_0 \geq 68 \text{ MeV}) \end{cases} \\ \tau_{Lan1} &= R_{Lan1} + 0.0492 \cdot \tau_{in} \end{aligned} \right\} \quad (111)$$

$$I_{Lan2} = -\frac{C_{Lan2}}{R_{Lan2}^{\,2}} \left\{ \left(R_{Lan2}^{\,2} - z^2 - \frac{\tau_{Lan2}^{\,2}}{2\sqrt{\pi}}\right) \cdot \left(1 + erf\left(\frac{R_{Lan2} - z}{\tau_{Lan2}}\right)\right) + \frac{(z + R_{Lan2}) \cdot \tau_{Lan2}}{\sqrt{\pi}} \cdot e^{\frac{-(R_{Lan2} - z)^2}{\tau_{Lan2}^{\,2}}} \right\} \quad (112)$$

For proton energies larger than about 120 MeV, the correction term $I_{Lan2}$ is of increasing importance (with the energy). Even below 120 MeV, the contribution $I_{Lan1}$ remains noteworthy. The energy spectrum of the protons has a tail due to the beam-line elements (range-modulator, etc.). It is certainly not sufficient to take account of all these influences on the basis of a half-width parameter $\tau_{in}$ in a Gaussian convolution. Since the influence of the beam-line depends on specific properties of the proton accelerator, an adaptation of the parameters, appearing in $I_{Lan1}$, by a fitting procedure, in addition to the fitting of $\tau_{in}$, is required. As a result, $C_{Lan1}$ (and to a lesser extent $R_{Lan1}$) only represent initial values, which may slightly be modified by fitting procedures, to accommodate machine-specific properties).

$$C_{Lan2} = \begin{cases} 0.000021791 \cdot E_0, & \text{if } E_0 < 120\,\text{MeV} \\ 0.022 \cdot \left(\dfrac{E_0 - 118}{168 - 118}\right)^{0.877}, & \text{if } 120\,\text{MeV} \leq E_0 < 168\,\text{MeV} \\ 0.022, & \text{if } 168\,\text{MeV} \leq E_0 \end{cases} \quad (113)$$

$$\left. \begin{array}{l} R_{Lan2} = (3.19 + 0.00161 \cdot E_0) \cdot \left[1 - \exp\left(-\dfrac{E_0^2}{165.795268^2}\right)\right] \\ \tau_{Lan2} = 2.4 \cdot \sqrt{\tau_{straggle}(z)^2 + \tau_{in}^2} \end{array} \right\} \quad (114)$$

The correction terms introduced to account for the Landau tails, have also been subjected to comparisons with the results of the GEANT4 code in the case of mono-energetic beams; in these calculations, the energy was varied between 2 and 250 MeV, with a step of 2 MeV. The result of these comparisons was that the differences never exceeded 2.2 %, the mean standard deviation amounting to 1.3 %. Concerning the poly-chromaticity of the proton beam, induced by the various beam-line elements, a direct comparison with experimental data was done. It should be noted that the already-described contribution to the buildup effect, induced by the Landau tails, could also be verified by the Monte-Carlo calculations, when the Landau tails were taken into account; additionally, the contribution disappeared when the statistical fluctuations were restricted to a Gaussian kernel. The role of reaction protons and Landau tails to buildup is clearly shown in Figures 14 - 16. Following remarks should be noted: One cannot label the protons in measurements, but in Monte-Carlo calculations we do so. Thus, one can distinguish them (section 1.5). According to Figures 5 - 9, we should expect significant buildup effects of secondary protons between 50 and 150 MeV; thereafter, they should somewhat decrease due to the asymptotic behavior of the nuclear cross-section. However, the controversial fact is observed: an increasing buildup with the incident energy.

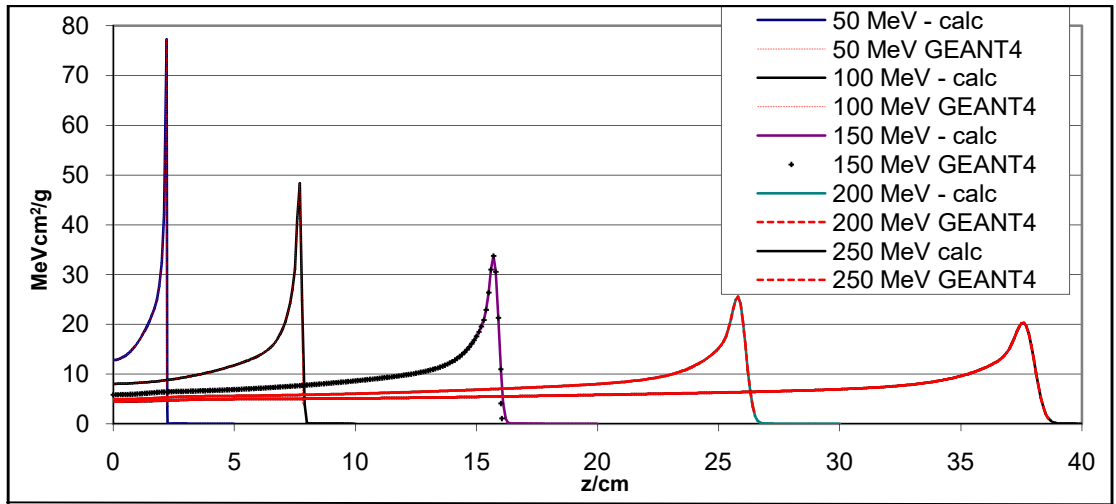

**Figure 14:** Comparison of the Bragg curves between our calculated buildup with the modeling of the Landau tails for primary/polychromatic protons and GEANT4 (for this purpose, the hadronic generator is switched off). Note that these calculations correspond only to primary protons; $\tau_{in}$ is assumed to be $R_{CSDA} \cdot 0.01$. The buildup in the proton depth doses is only partially a result of nuclear interactions or secondary-proton buildup, as suspected by other authors.

According to Figures 5 - 9, the increase of secondary protons is strongly connected to the decrease of primary protons. Since the fluence of secondary protons is zero at the surface and increases significantly along the proton track, this behavior could be connected with the buildup. However, the fluence decrease of the primaries is concurrent with the behavior of the 'secondaries'. The question also arises whether the transport of δ-electrons could be responsible for the buildup (similar to the Compton effect of photons). However, $E_{max}$ of 250 MeV protons amounts to 617 keV (Figure 10) and, with regard to the incident-proton energy of 185 MeV, this energy is lower. The range of these electrons is too small to explain the buildup. In order to produce $E_{max}$ of the order 4 MeV, the proton energy should be of order of 1 GeV.

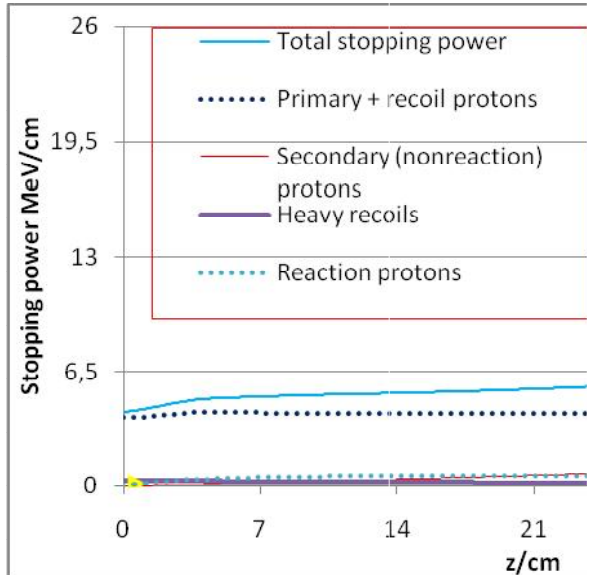

**Figure 15:** Total stopping power and related partial contributions of a Bragg curve of a mono-energetic 250 MeV proton beam.

Contributions from γ-quanta, resulting either from β+-decays of heavy recoils or from the annihilation of positrons, cannot be significant, as these contributions are expected to be isotropic (i.e., there is no preferred direction of these contributions). In agreement with GEANT4, the depth-dose curve of primary protons (250 MeV) shows a valley in the middle part of the plateau, resulting from the corresponding fluence decrease (Figures 5 - 9). If the transport of secondary protons is included, the total depth-dose curve does not show this feature (Figures 14 - 15). A comparison with the results in [20] is noteworthy. If the transport of secondary protons is only partially accounted for or omitted (PTRAN), then this valley can also be observed in the total depth-dose curve.
In addition, it is necessary to include the 'secondaries' in an accurate manner. According to the aforementioned authors, PTRAN leads (for 200-MeV protons) to a dose contribution of about 10 % at depth z = 20 cm, whereas investigations (see, e..g [4, 33]) provided 17 % at the same z value. This is in agreement with Figure 15 and may be calculated on the basis of the formalism developed in section 1.6. In order to be consistent with the total nuclear cross section (Figures 5 - 9), and the classification in section 1.6 we consider all protons as secondary protons, originating from nuclear interactions. The so-called 'reaction protons', which stand in close relation to heavy recoils, are preferably dominant for E > 150 MeV, else they amount only a small percentage. Their depth-dose curve shows certainly a maximum along their track, but, due to the broad spectral distribution, not a typical Bragg peak (Figure 15). If most of the secondary protons are non-reaction protons, the total contribution of the secondary protons yields a Bragg peak, but the whole dose profile is rather different. In various publications on therapeutic protons, this distinction has not been pointed out in a clear manner.

In order to obtain an approximation up to order 2 for the Landau tails according to Eqs. (107 – 115), we have assumed that, in the environment of the impact surface, the first- and second-order terms are the leading ones, whereas the error functions (resulting from first-order corrections) are significant along the entire proton track. All necessary calculation parameters have been determined by fits to the more general theory, as outlined above, and by fits to Monte-Carlo (GEANT4) results, since the option to use a model of Vavilov distributions is available in the code. The terms, which are related to $I_{Lan1}$ given by the usual convolution integrals, are a little bit complicated; therefore we do not go into details here. They contain – besides the pure error functions – also products of error functions with terms proportional to $z/R_{CSDA}$.

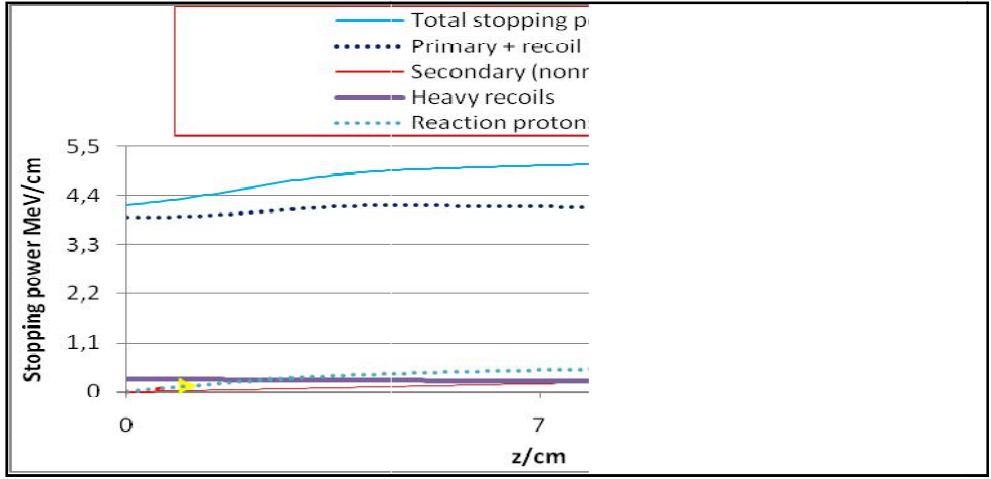

**Figure 16:** Buildup region of Figure 15.

The solution procedure of $I_{Lan2}$ is rather a standard task. Resulting from first- and second-order Hermite polynomials, we have to solve:

$$I_{Lan2} = -C_{lan2} \cdot \frac{2}{\sqrt{\pi} \cdot \tau_{Lan2}} \cdot \int (1 - u^2/R_{Lan2}^2) \cdot \exp[-(u-z)^2/\tau_{Lan2}^2] du \quad (115)$$

In order to carry out the integration, we use the substitution:

$$u' = (u - z)/\tau_{Lan2} \quad (116).$$

Using this substitution, it is easy to solve Eq. (115) via the rule of continued integration, and obtain Eq. (112). It is evident from Figure 15 that the stopping power of primary protons has a valley in the middle part of the pristine Bragg curve. This valley is only removed by the contributions from secondary/recoil protons. Figure 16 provides a clear indication that for the buildup two different contributions are essential, namely reaction protons and Landau tails. For polychromatic protons, the role of asymmetric Landau tails may become an increasing importance. The energy shifts induced by Landau tails have been analyzed in [4].

## 1.9 Calculation procedure of lateral scatter

In the present implementation in Eclipse, the lateral scatter of protons is treated by an approximate version of the multiple-scattering theory [37, 38, 45, 46, 47] . Often only one Gaussian is used for the angular distribution of scattered protons [11, 47]:

$$f(\theta)d\theta \approx \exp(-\theta^2/2\theta_0^2)d\theta \quad (117)$$

The angle $\theta_0$ depends on various physical parameters, e.g., on the radiation length L, the incident energy, etc.

In order to describe accurately the lateral tail of the primary and secondary (non-reaction, *sp,n*) protons, we will make use of two Gaussian kernels; Monte-Carlo calculations with GEANT4 indicated that two Gaussian kernels are sufficient in the case of the primary protons. For secondary (reaction, *sp,r*) protons, we restrict the lateral kernel to one modified Gaussian kernel due to their significantly smaller contribution. Our first Gaussian accounts for the inner part of multiple Molière scatter, which is steeper than some approximations (see e.g. [11, 47]), whereas the second Gaussian has a larger half-width to describe the tail. The Highland approximation [47] assumes a broader half-width to account (partially) for the tail. Thus, our calculation model for the lateral scatter in water is given by:

$$K_{lat,prim}(r, z) = \frac{C_0}{\pi \tau_{lat}(z)^2} \cdot e^{-\frac{r^2}{\tau_{lat}(z)^2}} + \frac{(1-C_0)}{\pi \tau_{lat,LA}(z)^2} \cdot e^{-\frac{r^2}{\tau_{lat,LA}(z)^2}} \quad (118)$$

For the contribution of the main Gaussian, we use a weight coefficient $C_0$ of 0.96. The calculation of $\tau_{lat}(z)$ (inner-part) and $\tau_{lat,LA}(z)$ (large-angle) is carried out as follows:

$$\left. \begin{array}{l} \tau_{lat}(z) = f \cdot 0.626 \cdot \tau_{max} \cdot Q(z) \\ f = 0.9236 \\ \tau_{max} = (E_0 / 176.576)^p \end{array} \right\} \quad (119)$$

$$\left. \begin{array}{l} p = 1.5 + \begin{cases} 0.00150 \cdot (176.576 - E_0), & \text{if } E_0 \leq 176.576 \\ 0.03104 \cdot \sqrt{E_0 - 176.576}, & \text{if } E_0 > 176.576 \end{cases} \\ Q(z) = \frac{e^{\frac{1.61418 \cdot z}{R_{CSDA}}} - 1}{e^{1.61418} - 1} \cdot 0.5 \cdot \left[ erf\left( \frac{R_{CSDA} - z}{\tau} \right) + 1 \right] \end{array} \right\} \quad (119\,a)$$

A good approximation for a model with a single Gaussian for the primary protons can be obtained from the equations above by substituting $C_0$ and f by 1. In agreement with the results in [45, 46], the change from water to other materials is obtained by the scaling of $Q(z)$ on the basis of $R_{CSDA}$. The error-function term models the Gaussian distribution of the stopping distribution due to range straggling. The protons, which have undergone only small-angle scattering, are closer to the central axis of the beam-let and travel further.

A fit to Monte-Carlo results shows that $\tau_{lat,LA}(z)$ can be determined by the kernel:

$$\tau_{lat,LA}(z) = \frac{0.90563}{1 - e^{\frac{-1}{0.252}}} \cdot \tau_{max} \cdot \left[ e^{-\frac{(1-z/R_{CSDA})^2}{0.252}} - e^{-\frac{1}{0.252}} \right] \cdot 0.5 \cdot \left[ erf\left( \frac{R_{CSDA} - z}{\tau} \right) + 1 \right] \quad (120)$$

The scaling properties of $Q(z)$ still hold, since only the ratio $z/R_{CSDA}$ enters Eq. (120). For the secondary protons, one should use different Gaussian kernels for those particles which did or did not undergo nuclear reactions (see $S_{sp}$ and $S_{heavy}$):

$$\left. \begin{array}{l} \tau_{sp}(z) = \tau_{max} \cdot Q(z) \\ \tau_{heavy}(z) = \tau_{heavy}(E(z)) \end{array} \right\} \quad (121)$$

$\tau_{max}$ and $Q(z)$ are the same as defined in Eq. (120); $\tau_{heavy}(E(z))$ is given by a similar equation as above, except that it depends on the local energy $E(z)$ instead of $E_0$. It might appear that Eqs. (119 - 121) are just obtained by fitting methods. However, the spatial behavior of the multiple-scatter theory involves a Gaussian and Hermite polynomials $H_{2n}$:

$$\left. \begin{array}{l} K(r, z, lat) = N \cdot \exp(-r^2 / lat(z)^2) \cdot \\ [a_0 + a_2 \cdot H_2(r/lat(z)) + a_4 \cdot H_4(r/lat(z)) + \\ + a_6 \cdot H_6(r/lat(z)) + 0(higher\ order)] \\ N = 1/(\sqrt{\pi} \cdot lat(z)); \quad r^2 = x^2 + y^2 \end{array} \right\} \quad (122)$$

The parameters of the Hermite-polynomial expansion of multiple-scatter theory are: $a_0 = 0.932$; $a_2 = 0.041$; $a_4 = 0.019$; $a_6 = 0.008$. The Gaussian half-width is the same as assumed for the inner part. The task now is to determine a linear combination of two Gaussians, according to Eq. (122), with different half-widths, on the basis of an optimization problem. By this way, we are able to go beyond the Highland approximation.

The calculation of $\tau_{max}$ is carried out in the following way. The differential cross section is given by:

$$\left.\frac{dlat(z)^2}{dz} = \frac{dlat_E^2}{dE} \cdot \frac{dE}{dz}; \quad \frac{dlat_E^2}{dE} = \frac{\alpha_{Medium}}{E^2}\right\} \quad (123)$$

In our calculations, we have only used $\alpha_{Water}$; $\alpha_{Medium}$ is proportional to $Z_M \cdot \rho_M / A_M$, where $Z_M$, $\rho_M$, and $A_M$ are respectively the nuclear charge, the mass density, and the mass number of the medium. Values for $E(z)$ and $dE/dz$ in Eq. (124) may be obtained by using the BKR:

$$R_{CSDA} = 0.00259 \cdot E_0^p \rightarrow R_{CSDA} - z = 0.00259 \cdot E(z)^p \quad (124)$$

$$E(z) = \left(\frac{R_{CSDA} - z}{0.00259}\right)^{1/p} \quad (125)$$

An accurate application of this rule requires consideration of the dependence of p on $E_0$. The quantity $dE(z)/dz$ can be computed from Eq. (125). Finally, Eqs. (123 - 125) yield:

$$lat_{max} = \sqrt{\int_0^{R_{CSDA}} \left[\frac{\alpha_{Water}}{E(z)^2}\right] \cdot \frac{dE(z)}{dz} \cdot dz} \quad (126)$$

One might expect that the lateral-scatter functions (Eqs. (125 - 126)) continuously increase up to z = $R_{CSDA}$. Actually, this assumption would be valid, if the energy spectrum for the scattered protons would be identical at depth z, independent of the scatter angle and the fluctuations due to the energy/range straggling $\tau_{straggle}$ and $\tau_{in}$. In reality, there are small fluctuations of the lateral-scatter functions along the proton tracks. In particular, from the Bragg peak down to the distal end, there is a significant difference between the protons, which have only undergone small-angle scatter in this domain, and the ones with larger scatter angle. The latter protons have deposited their energy in an oblique path; therefore, they stop earlier and cannot reach the distal end of the Bragg curve. It is clear that the scatter functions for primary and secondary protons ($\tau_{lat}$, $\tau_{lat,LA}$ and $\tau_{sp}$) depend on $\tau_{straggle}$ and $\tau_{in}$, which induce these fluctuations and cause significant changes in the energy spectrum at the end of the proton tracks. In order to describe this behavior by a mathematical model, we prefer to use a Gaussian convolution, which is certainly justified in the domain of Bragg peak (low-energy region of the proton tracks). As an example, we use the function Q(z) in Eq. (127), which fixes both $\tau_{lat}$ and $\tau_{sp}$. We denote the fluctuation parameter by $\tau$; the connection to straggle parameters will be considered thereafter. The crude model assumes:

$$\left.\begin{array}{l} Q_0(z) = \frac{\exp(1.61418 \cdot z / R_{csda}) - 1}{\exp(1.161418) - 1} \quad (if \ z \leq R_{csda}) \\ \qquad = 1 \ (if \ z > R_{csda}) \end{array}\right\} \quad (127)$$

A more realistic model, taking all the arguments with regard to fluctuations into account, is obtained by:

$$Q(z) = \frac{1}{\tau \cdot \sqrt{\pi}} \cdot \int_{-\infty}^{R_{csda}} Q_0(\xi) \exp[-(z-\xi)^2 / \tau^2] d\xi \quad (128)$$

Using standard methods, we obtain the modified Eq. (127) for Q(z), instead of $Q_0(z)$, according to Eq. (128):

$$Q(z) = \frac{\exp(1.61418 \cdot z / R_{csda}) - 1}{\exp(1.161418) - 1} \cdot$$

$$\cdot \tfrac{1}{2} \cdot [1 + erf((R_{csda} - z)/\tau)] \quad (129)$$

This result implies that Q(z) increases exponentially along the proton track, as long as the error function is 1 (or nearly 1). Only in the environment of z = $R_{CSDA}$ Q(z), does it decrease rapidly. However, this behavior actually depends on τ. The connection between τ and the aforementioned convolution parameters for energy/range straggling is a valid question. One might assume that, for proton pencil beams with energy/range straggling, we can set:

$$\tau^2 = \tau_{straggle}^2 + \tau_{in}^2 \quad (130)$$

This assumption might be reasonable, since the proton history, resulting from the beam-line and expressed by $\tau_{in}$, should be accounted for. As already pointed out, we have made use of the GEANT4 results for the adaptation of the scatter functions with regard to mono-energetic protons. It turned out that the best adaptation in the domain from the Bragg peak to the distal end can be obtained, if we put (mono-energetic protons):

$$\tau^2 = \tau_{straggle}^2 + \tau_{Range}^2 \quad (131)$$

The mean standard deviation for protons with $E_0$ = 50 MeV up to 250 MeV in intervals of 25 MeV amounts to 2.6 %. Therefore, it may be justified to modify Eq. (131) by writing:

$$\tau^2 = \tau_{straggle}^2 + \tau_{in}^2 + \tau_{Range}^2 \quad (132).$$

## 1.10 Beam-lines for protons

At several places in this study, we have mentioned that we shall give information on the description of the available beam-lines, which are used in proton treatment. With regard to the pristine Bragg peak, the parameter $\tau_{in}$ takes account of the Gaussian convolution, as well as of the Landau tail. However, a sole Bragg curve does not fulfill the requirement of creating a homogeneous dose distribution in the target, and we have to consider the SOBP via the superposition of different Bragg peaks with modulation of the proton energy. Besides the energy modulation, the lateral distribution of the proton beam is a very important issue. In particular, a sufficient description of the tail, resulting from multiple scattering, goes beyond the single Gaussian approximation. There are three different methods commonly used to solve this task:

1. Active-scanning technique; 2. Passive-scanning technique; 3. Broad-beam technique (double scattering, uniform scattering (wobbling)).

*Active-scanning technique*

The specific proton accelerator is a synchrotron, which provides accurately the required energy of each beam to form an SOBP. Since no further range shifters are necessary, the energy spectrum of the impinging proton beam approaches best the mono-energetic case [44].

*Passive-scanning technique*

In this case, a cyclotron provides only some fixed energies (e.g., 100 MeV, 175 MeV, 250 MeV). The desired energy to form an SOBP is obtained by a range modulator. The energy spectrum of the impinging beam is broadened. Since the past decade revealed that active/passive scanning techniques received the preference modality in proton therapy, we only refer to detailed papers [4, 44] with regard to broad-beam technique.

## 1.11 Role of electron capture in Bragg curves

According to BBE the energy spectrum produced by carbon ions should be the same as that produced by protons, and the only difference between protons and carbon ions should be the intensity of the released collision electrons, i.e. the amplification factor should be 36 for carbon ions. It is well-known that this property is not valid for the following reasons: The average ionization energy for carbon ions turned out to be $E_I = 80$ eV instead of $E_I = 75$ eV for protons [49 - 55]. The second reason is the electron capture of the carbon ion. Thus a carbon ion can capture a free electron, which has been released immediately before (Figure 17). However, only electrons with a slow relative velocity to the carbon ion can account for this process ($v_{relative} \approx 0$). Since the transition time of the capture electron to a lower atomic state of the carbon ion is less than $10^{-10}$ sec with a simultaneous emission of light (UV or visible), it is possible that the captured electrons goes lost again, and only a stripping effect occurs for a short time. If the $C^{6+}$ ions has been finally transferred to a stable $C^{5+}$ ion, the identical process can be repeated until at the end track a neutral carbon atom is obtained having only a thermal energy. In the environment of the Bragg peak the effective charge of the carbon ion is about the same that of a proton, namely $+e_0$. Since the electron capture can only occur for electrons of which the relative velocity is slow, the upper energy limit of the energy exchange $E_{ex}$ is the Fermi edge $E_F$, which is for an electron gas not higher than the thermal energy $k_B T$. If the charge of carbon ion amounts to $+6 \cdot e_0$ and, at least, $> +e_0$, the environmental atomic electrons suffer lowering of the energy levels due to the Coulomb interaction, which leads to an increase of $E_I$. Therefore the stated value of $E_I = 80$ eV represents an average value produced by the fast carbon ion starting with $+6 \cdot e_0$ and ending with an uncharged, neutral carbon atom.

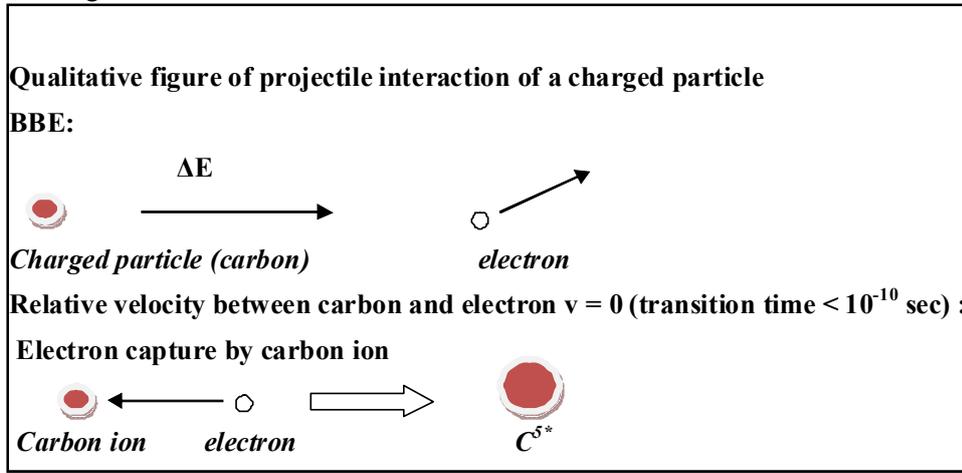

**Figure 17:** Excitation of an atomic electron by the collision interaction of a fast carbon ion with an atomic electron and the reversal process of the electron capture.

As valid in the previous section 1.3, S(z) according to BBE is $\sim q^2$. The Fermi-Dirac operator according to Eqs. (72 - 81) also acts to the square of the projectile charge q; the initial charge now is referred to as $q_0$. For brevity, we only give a hint, that for evaluation of Eq. (133) we have made use of the relationship between the operator function sech($\xi$) and its expansion by a Gaussian and suitable polynomials according to Eq. (77) with $\xi = 0.5 \cdot (H_D - E_F)/E_{exchange}$.

$$\left. \begin{array}{l} [f_{FD}(\hat{H})]^n \cdot S(z) = [\tfrac{1}{2}\exp[-(H_D - E_F)/2E_{exchange}] \cdot \\ \cdot \sec h[-(H_D - E_F)/2E_{exchange}] \cdot [d_s(H_D)^n] \cdot S(z) \end{array} \right\} (133)$$

The application of Eqs. (72 - 81, 133) and the transition to the continuum provides (up to 2$^{nd}$ order):

$$\left. \begin{array}{l} q^2(E) = q_0^{\,2}(E) \cdot [erf(E/s_E) - \\ A \cdot (1 - E/E_0) \cdot (1 - (E/E_0)\exp(-E^2/s_E^{\,2}))] \\ s_E^{\,2} = q_0^{\,3} \cdot \pi \cdot m^2 \cdot c^4 \cdot (1 + m^2 c^4 / M^2 c^4) \\ A = q_0^{\,2} \cdot \pi^2 \cdot \tfrac{mc^2}{Mc^2} \end{array} \right\} (134)$$

The effective charge $q_{eff}$ is calculated by the integral:

$$q_{eff}^2 = \int_0^{E_0} q(E)^2 \cdot dE \quad (134a)$$

The $R_{CSDA}$ formula (Eqs. (34 - 35)) now becomes:

$$\left.\begin{array}{l} R_{csda} = \alpha \cdot (E_0 \cdot A_N / q_{eff}^2) + \beta \cdot (E_0 \cdot A_N / q_{eff}^2)^2 + \\ \gamma \cdot (E_0 \cdot A_N / q_{eff}^2)^3 + \delta \cdot (E_0 \cdot A_N / q_{eff}^2)^4 \end{array}\right\} \quad (135)$$

Now $E_0$ refers to the initial energy per nucleon. For further detail, the publication [45] should be consulted. The parameters have slightly to be modified: α = 0.0069465598, β = 0.0008132157, γ = -0.00000121069, δ = 0.000000001051. In the following we present results of calculations of $q^2(E)$ for protons, He ions and carbon ions; the initial energy amounts to 400 MeV/nucleon (Figures 18 - 20). This appears to be a reasonable restriction with regard to therapeutic conditions. Figures 19 - 20 show that at the end of the projectile track all charged ions nearly behave in the same manner. In the following, we present results of calculations for protons, He ions and carbon ions; the initial energy amounts to 400 MeV/nucleon. This appears to be a reasonable restriction with regard to therapeutic conditions. Thus Figure 18 shows that at the end of the projectile track all charged ions nearly behave in the same manner.

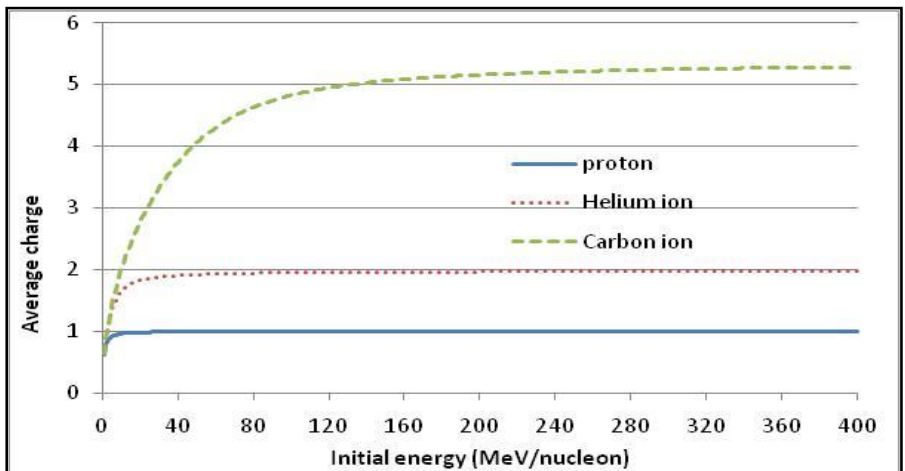

**Figure 18:** Effective charge of protons, He- and C-ions as a function of the initial energy $E_0$ (for protons, we have to assume $q_{eff}$ = 0.995)
.

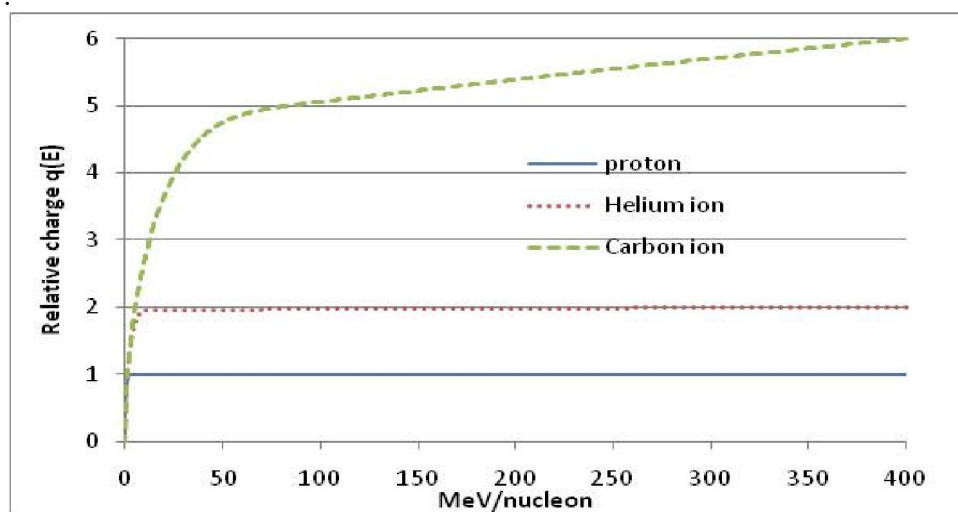

**Figure 19:** Actual charge of protons, Helium and Carbon ions in dependence of the residual energy /MeV/nucleon).

With regard to the therapeutic efficacy the behavior of the LET in the environment of the Bragg peak is very significant. For a comparison, we first regard a previous result [4, 33] referring to the LET of protons. According to Figure 21 the stopping power of protons at the end track depends significantly on the initial energy $E_0$ and on the beam-line (energy spectrum at the impinging plane). The electron capture of the proton at the end track is ignored. However, Figure 21 clearly shows that with regard to protons the electron capture only becomes more and more significant, when the actual proton energy is smaller than E = 2 MeV. The electron capture of protons at the end track would make the LET of protons zero independent of the initial energy.

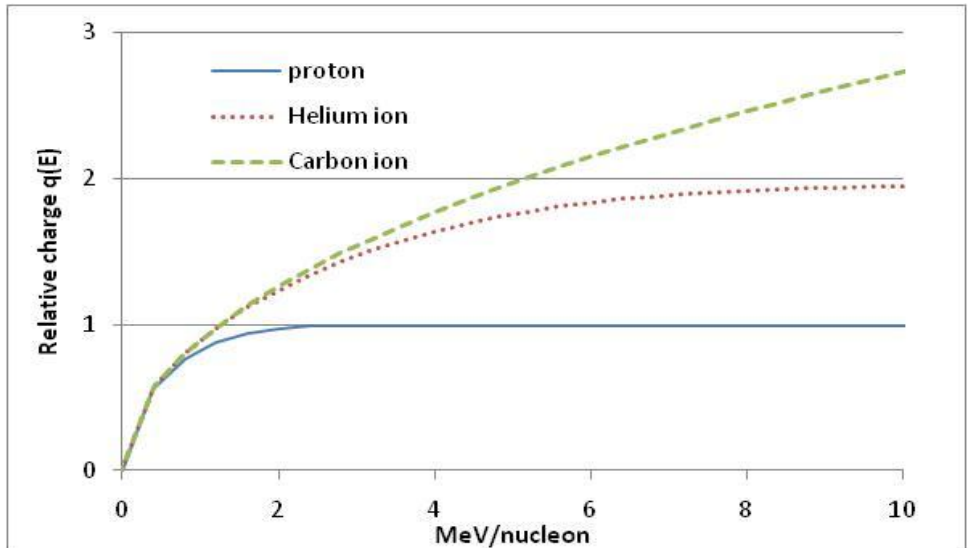

**Figure 20:** Effective charge q(E) of carbon ions in dependence of the initial energy for the cases $E_0$ = 200, 300 and 400 MeV/nucleon.

The succeeding Figure 22 presents E(z) and S(z) = dE(z)/dz of protons and S(z) of carbon ions with taking account for electron capture. The initial proton energy amounts to 270 MeV, whereas the initial carbon ion energy is 400 MeV/nucleon. Most significant is the height of the Bragg peak, which is resulting from the electron capture only a factor 1.7 higher than that of protons. In both cases CSDA is assumed. Since protons are much more influenced by energy straggling and scatter, their peak height are reduced again, whereas for carbon ions scatter and energy straggling do not play a very significant role due to the mass factor 12.

A rigorous consideration of the LET of carbon ions is given the following Figure 24, whereas the different ranges of protons and carbon ions is demonstrated in Figure 23. It makes only sense to consider the total energy of 4800 MeV of the carbon ions (Figure 25). Due to this order of magnitude E(z) of the carbon ion has not been presented in Figure 22. Energy straggling and scatter have been ignored in Figure 22, which is justified for heavy carbons. On the other side, this figure makes also apparent the well-known disadvantage of carbon ions, namely the enormous amount of energy of carbon ions in order to reach an acceptable dose distribution in the domain of the target, where a SOBP is required.

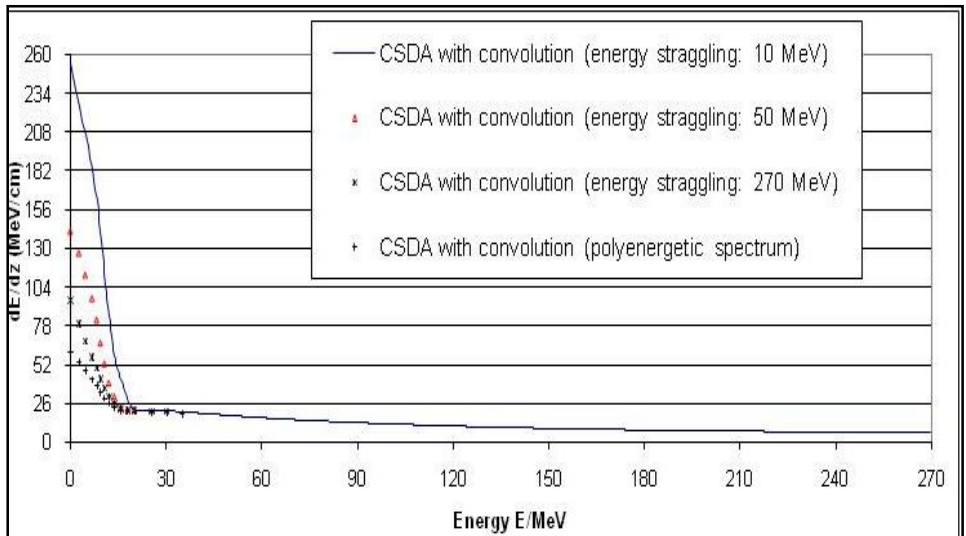

**Figure 21:** Calculation of the LET (mono-energetic proton beams with energy straggling of 10 MeV, 50 MeV, and 270 MeV and polychromatic proton beam with energy/range straggling corresponding to 270 MeV; $\tau^2 = \tau_{straggle}^2 + \tau_{in}^2$). Stopping power of protons in dependence of the energy straggling (without taking account of electron capture).

With the help of GEANT4 a real depth dose curve (HIMAC, 290 MeV/nucleon [50 – 55]) has been determined (Figure 25). The role of GEANT4 was only to account for the nuclear reactions, which are based in this Monte-Carlo code on an evaporation model. The electronic stopping power S(z) has been determined by the tools worked out in this communication, the electron capture effect has been accounted for. Further parameters for a calculation of S(z) have been used based on the proton calculation model [4, 33] by appropriate modifications. The Gaussian convolution kernels for energy straggling and lateral scatter have been rescaled according to the corresponding mass properties.

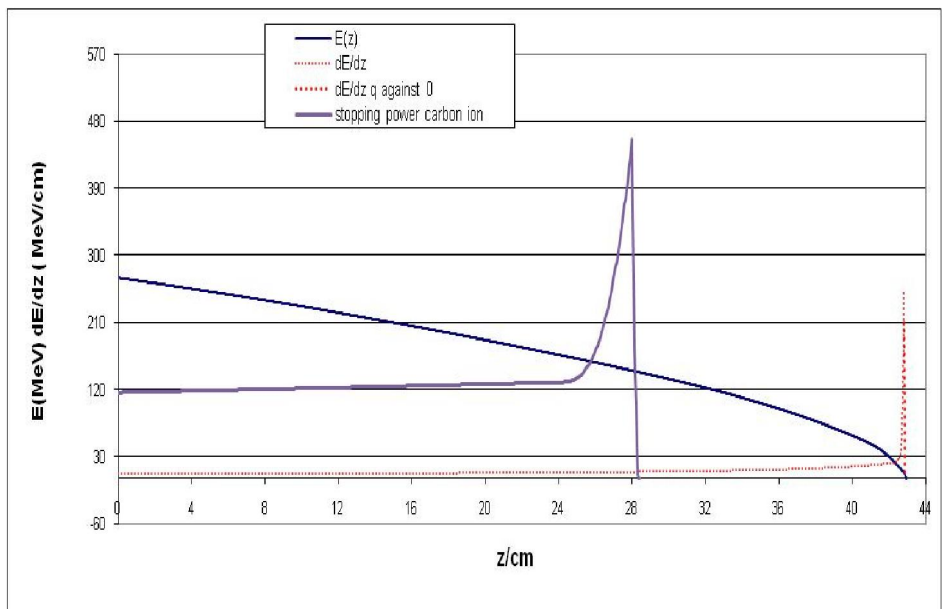

**Figure 22:** LET for mono-energetic protons (dots) and overall stopping power S(z) of carbon ions 400 MeV/nucleon.

With regard to the decrease of fluence of primary carbon ions we have derived some modifications of the corresponding decrease curves for protons. However, it appears not to be appropriate to go into further details. A further aspect is the use of the code GEANT4. Since this Monte-Carlo code represents an open programming package, some suitable additional reaction channels have been introduced with reference to nuclear reaction detected by HIMAC [53].

The main purpose of this communication was the derivation of a systematic theory of electron capture of charged particles and the role for the LET. There are purely empirical trials to include charge capture in Monte-Carlo codes. However, it appears that a profound basis for the calculation of $q^2(E)$, E(z), S(z) and $R_{CSDA}(E_0)$ depending besides the initial energy $E_0$ also on the nuclear mass number N is required to account for further influences of Bragg curves such as the density of the medium and its nuclear mass/charge $A_N$ and Z.

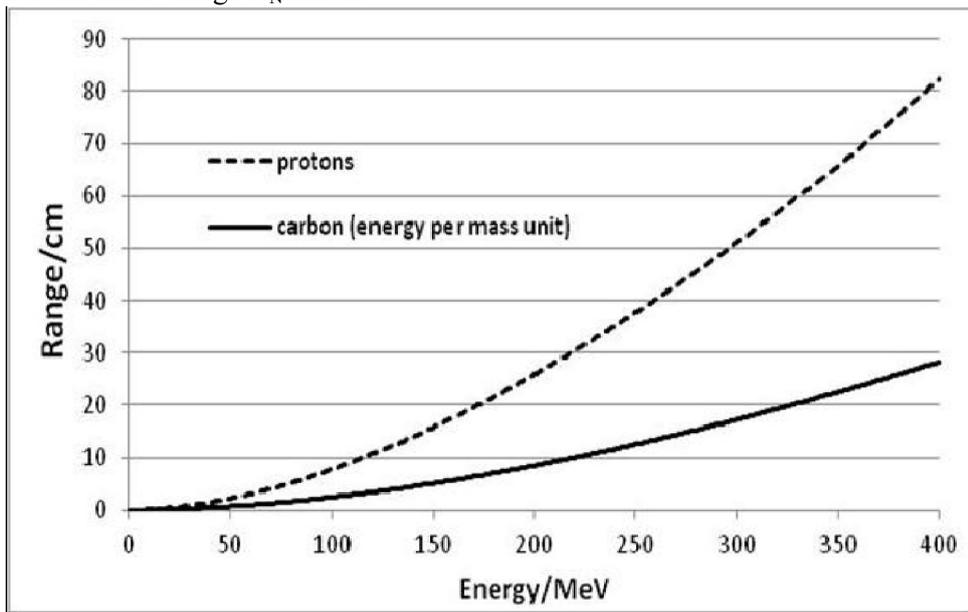

**Figure 23.** Comparison between proton range calculated by Eq. (37) and range of the carbon ion determined by Eq. (147)

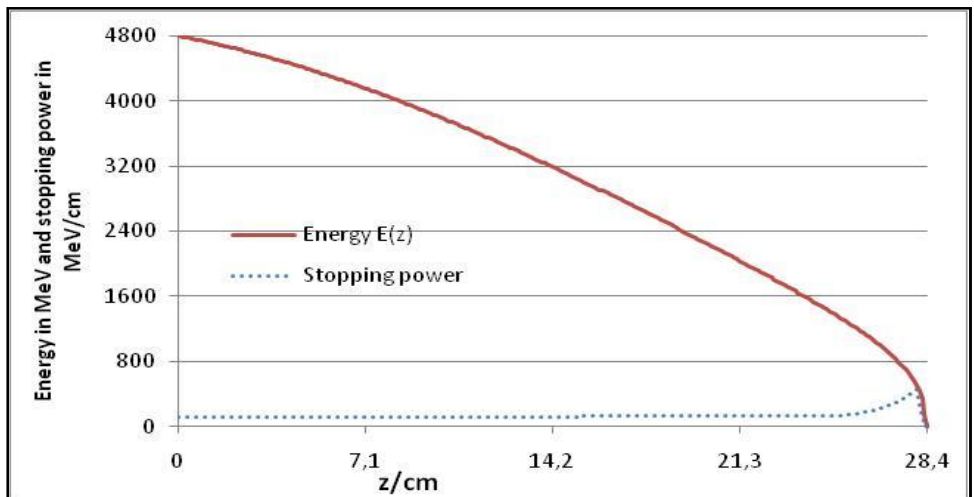

**Figure 24:** LET of carbon ions (400 MeV/nucleon).

The unmodified use of BBE leads to wrong results and the Barkas correction, which does not affect the factor $q^2$ of BBE, only works for protons or antiprotons, whereas for projectile particles like He or carbon ions this correction cannot be considered as small. The presented theory includes the *Barkas* effect without any correction model. In view of the enormous effort with reference to the acceleration of $C^6$ ions, the application of this therapy modality should be critically reviewed.

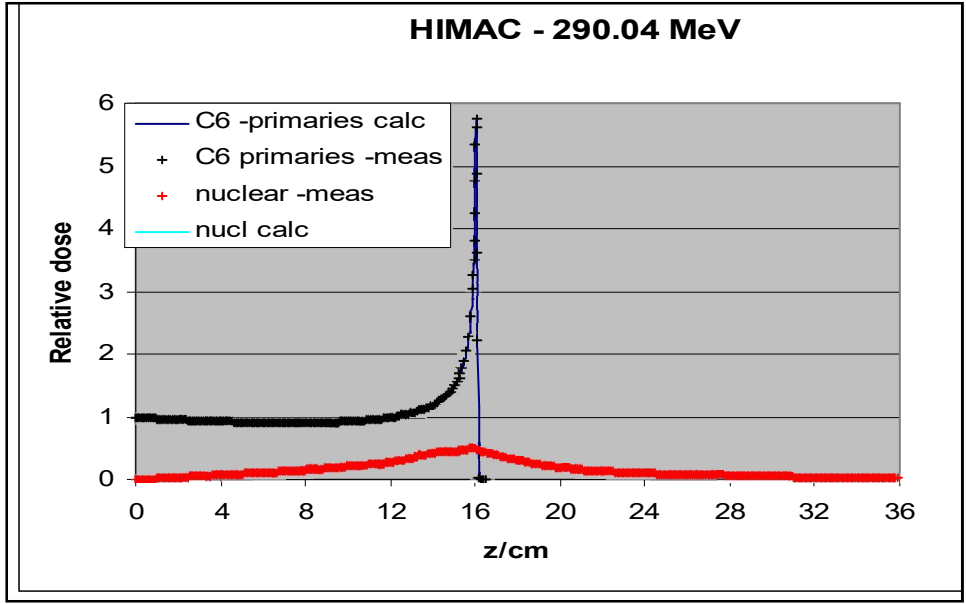

**Figure 25:** Measurement (HIMAC) and theoretical calculation of the Bragg curve of carbon ions (290 MeV/nucleon).

Figure 22 shows relative depth-dose curves for mono-energetic protons in water. The agreement between GEANT4 and the analytical calculation is excellent up to the distal end. In this domain, we could only verify very small deviations of the order of 0.01 %; they are probably due to numerical inaccuracy. In the distal end, the calculated values exceed the GEANT4 results by about 1 - 8 %. We assume that this difference originates from the energy cutoff in GEANT4.

## 2. Applications and comparisons with measurements

### 2.1 Bragg curves of protons (based on BBE: model M2)

In [3] we have pointed out that the measured Bragg curves can be adapted optimally, if a certain approach is followed. The starting point is the energy $E_0$ of the incident proton beam and the assumption of an incident spectral distribution described by $\tau_{in}$. The rough estimate of $C_{Lan1}$, according to Eq. (111), can also be considered as a starting value. Small variations of $E_0$ and $\tau_{in}$ lead to minimum standard deviation. If the Landau tails are included, the mean standard deviation did not exceed 0.1 - 0.3 %. The source-surface-distance (SSD) may also be subjected to a small variation; in the scanning technique, SSD $\rightarrow \infty$. We shall next show that the aforementioned parameters can be estimated on the basis of the beam-line characteristics of each proton-treatment machine. As already mentioned, we have accounted for all the terms relating to the stopping power of the BBE according to the recommendations in [9]. Since the numerical simulation procedure in GEANT4 has to account for all these terms separately, a compensation of the logarithmic term by the other terms in the low-energy region cannot occur, and a cutoff in the proton energy at 1 MeV has been imposed. Figures 26 - 27 pristine Bragg curves, obtained at PSI by degrading a 600 MeV proton beam. In contrast to Figure 28 (Varian-Accel), a buildup effect could not be verified in this beam-line, and it should be mentioned that only the contribution $I_{Lan1}$ plays a certain role with regard to the calculation procedure, since it slightly modifies the energy spectrum in the middle part of a Bragg curve. The comparably high values for $\tau_{in}$ result from the degrading. Figures 26 - 27 present pristine Bragg curves, obtained at PSI by degrading a 600 MeV beam. Unfortunately, a buildup effect could not be verified in this beam-line, and it should be mentioned that only the contribution $I_{Lan1}$ plays a certain role with regard to the calculation procedure, since it slightly modifies the energy spectrum in the middle part of a Bragg curve. The comparably high values for $\tau_{in}$ result from the degrading.

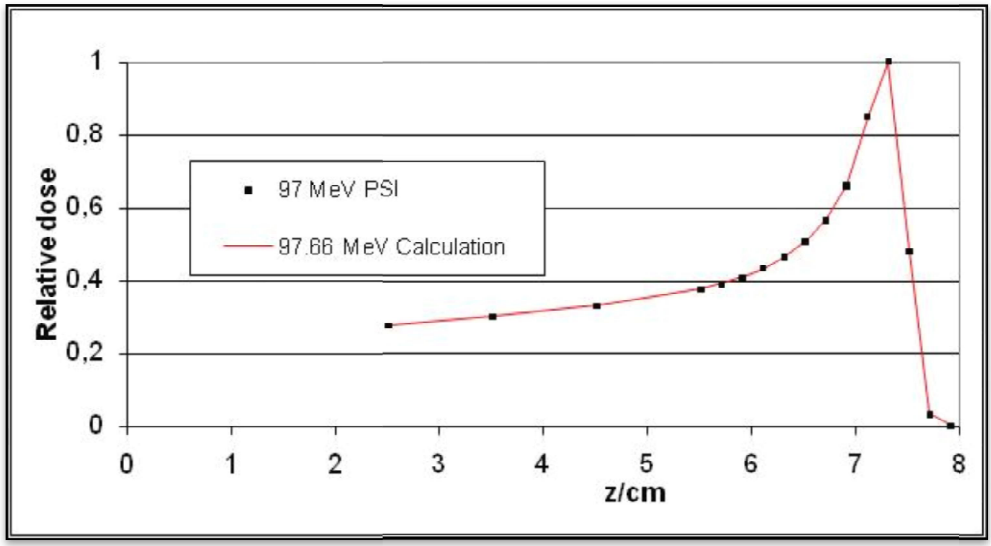

**Figure 26:** 97 MeV – PSI (discrete points: measurements): dev=0.14 %, $\tau_{in}$ = 0.159 cm.

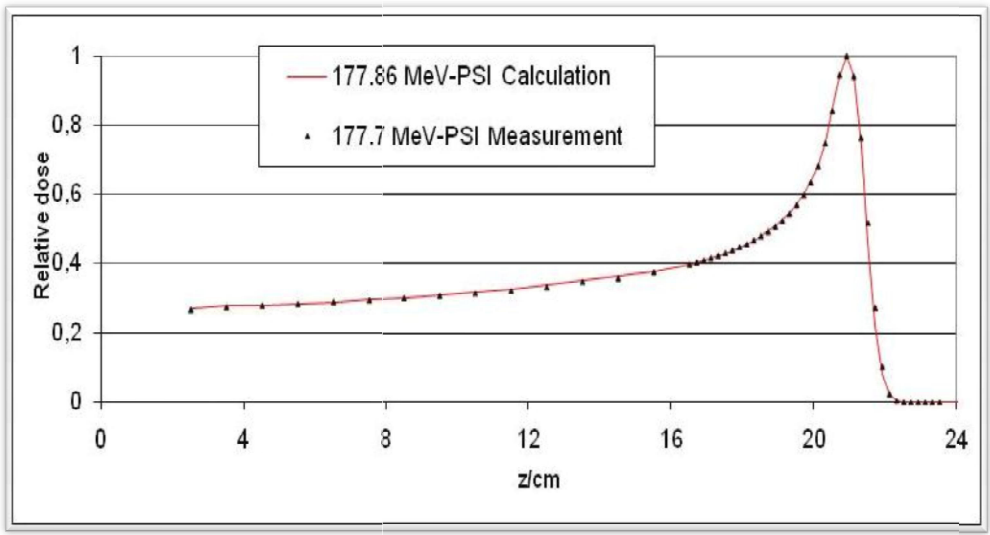

**Figure 27:** 177 MeV – PSI (discrete points: measurements): dev=0.16 %, $\tau_{in}$ = 0.347 cm.

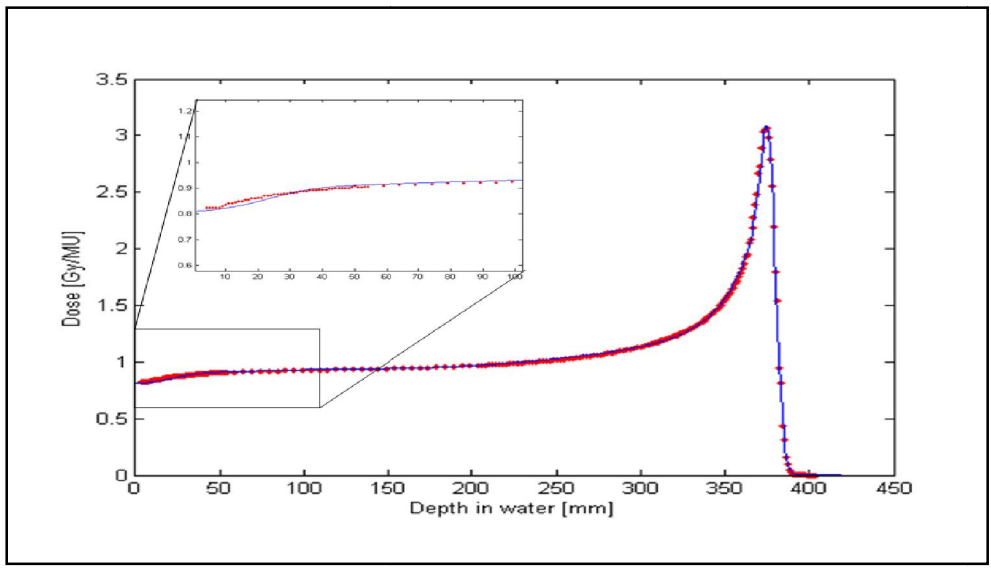

**Figure 28:** Comparison between a measured and a calculated pristine Bragg peak. The dose is measured with a large parallel plate chamber across a single 250-MeV pencil beam delivered by the Varian-Accel machine. The buildup is most visible for high-energy beams without absorbing material in the beam-line; it is well described by the model.

## 2.2 Some features of lateral scattering

Since lateral scattering and the determination of the parameters $\tau_{in}$ and $C_{Lan1}$ are reasons that the proposed calculation procedure might not provide sufficiently accurate results;, it is necessary to introduce a per-case (for each machine, separately) procedure to obtain their values. The experimental verification of the lateral-scattering calculation is very challenging due to the small contributions of the large-angle scattered primary protons and of the secondary protons. Furthermore, there are other contributions to the lateral distribution of protons (initial phase space of a scanned beam or effective-source size and block-scattering effects in a scattered beam), which affect the measurement. The following plots show comparisons between calculated and fitted beam width (Figures 29 - 33).

Figure 29 shows how the beam width is reduced when passing from a single-Gaussian model (dashed line, x) to a model with higher-order contributions (thick solid line and o for the main Gaussian). Figure 30 provides the magnitude of the contributions from higher-order scattering. Despite of the fact that these contributions may be negligible in the situation treated here, they are nevertheless expected to play an important role at higher energies, as well as when a range shifter is present in the planning. Especially in the latter case, an impact is also expected on the estimation of the MU factor [22]. Inspection of the results for the fitted parameter leads to the conclusion that the difference between our fitted-energy values and those claimed by the machine manufacturer is kept below 1 MeV. Interesting observations can be made by plotting the initial range spectrum $\tau_{in}$ as a function of the nominal energy for a number of different machines and techniques (Figures 31 - 32). The Hitachi machine is a synchrotron, which produces naturally a narrow Bragg peak. The other machines are cyclotrons; the lower energies are obtained by degrading the initial beam. This process creates a broad energy spectrum, which needs to be narrowed through energy-selection slits. The setting of the energy-selection slits is a compromise between the width of the Bragg peak and the beam intensity. It seems that this compromise leads to similar width values for the three machines by Varian-Accel, PSI, and IBA. The width of the PSI beam is somewhat broader, probably due to the fact that the data from the PSI beam-line originates from the 600-MeV physics-research accelerator and, therefore, has to be degraded more than the beams of the other machines.

The data points from the Hitachi machine show that the range spectrum seems to increase with the amount of high-density material in the beam-line (i.e., the large-field configuration (□) has more Pb than the medium field size (◊); the scanning-beam line (∇) has no extra material). This demonstrates convincingly that the range straggling in higher-density materials is larger than in the same (water-equivalent) amount of low-density materials. Since our model does not distinguish between different compositions of the beam-line (i.e., $\tau_{straggle}$ is not changed), the additional straggling component is accounted for by the fitting procedure in beam configuration through an increase in $\tau_{in}$. We found that the resulting range spectra, depending on the nominal energy, can be fitted well by a third-degree polynomial for both machine types, i.e., synchrotron and cyclotron; this enables us to model the pristine Bragg curves for intermediate (not configured) energies. Due to a restriction to horizontal beam geometry on the other double-scattering beam-lines, the pristine Bragg measurements could not be measured up to the surface of the water phantom; this implies that the most-relevant data points for the determination of $C_{Lan1}$ are missing. The results for the pristine Bragg measurements from the modulated-scanning beam-lines show a much clearer trend for $C_{Lan1}$ as a function of the residual range (Figure 33). We usually substitute $C_{Lan1}$ from Eqs. (111 - 112) by another third-degree polynomial with parameters obtained from fits to the data points of each machine – as indicated by the straight lines in Figure 33.

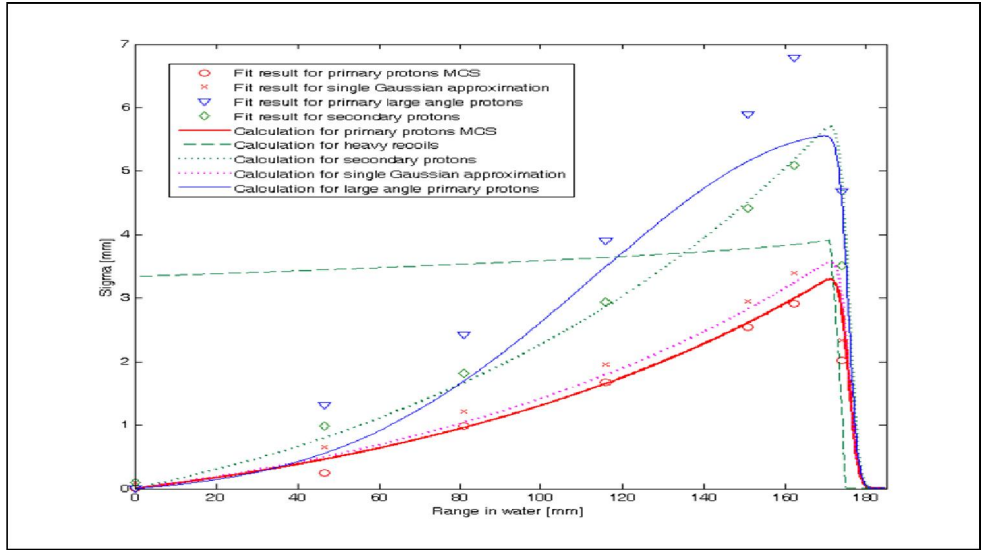

**Figure 29:** Comparison of the theoretical beam width for a 160-MeV scanning beam (Varian) and the results obtained by fitting a single Gaussian (x) or our theory (open symbols) to the measured profiles. Due to the smallness of the majority of the contributions, the fit results are extremely sensitive to the starting values. Furthermore, limits had to be defined in order to keep the secondary-proton and the large-angle contributions to reasonable values. The calculated width for a single Gaussian was obtained from Eq. (130) by setting $C_0$ to 1. The quantity shown in this figure is defined as $\tau_{lat}/\sqrt{2}$.

The fitting of the Landau parameter $C_{Lan1}$ for double scattering (Figure 33) shows quite a bit of scatter between different machines and also for different hardware configurations of the beam-line (often called 'options') within the same machine. However, the magnitude of $C_{Lan1}$ is the same for all machines and there is a trend towards a minimum at residual-range values of 100-150 mm. The effect of these variations of $C_{Lan1}$ on the total depth-dose calculation is very small and we normally use the nominal $C_{Lan1}$ according to Eqs. (111 - 112) for scattering- and uniform-scanning dose calculations. In a recent paper [17] it has been pointed out that, in the domain of the Bragg peak, the Gaussian solution (one Gaussian) is sufficient for both longitudinal energy straggling and lateral scatter. The corresponding arguments are based on the transition of the more general Boltzmann transport theory to the Fermi-Eyges theory in the low-energy limit. However, it appears that the conclusion is only partially true, since the history of the proton track has an influence on the behavior in the low-energy domain (see above results referring to the Landau tails of the energy transfer), and, all types of transport equations also have more general solutions than given by one Gaussian in the diffusion limit. Yet the linear combination of two Gaussians with different half-widths, as used in the present study, is not a solution of the Fermi-Eyges theory, but a corresponding one of a nonlocal Boltzmann equation. This is an integro-differential equation with different transition probabilities for the local and nonlocal part (long-range interaction). In the diffusion limit, the nonlocal part provides, at least, one additional Gaussian (e.g., see the adaptation of multiple-scatter theory by two Gaussians).

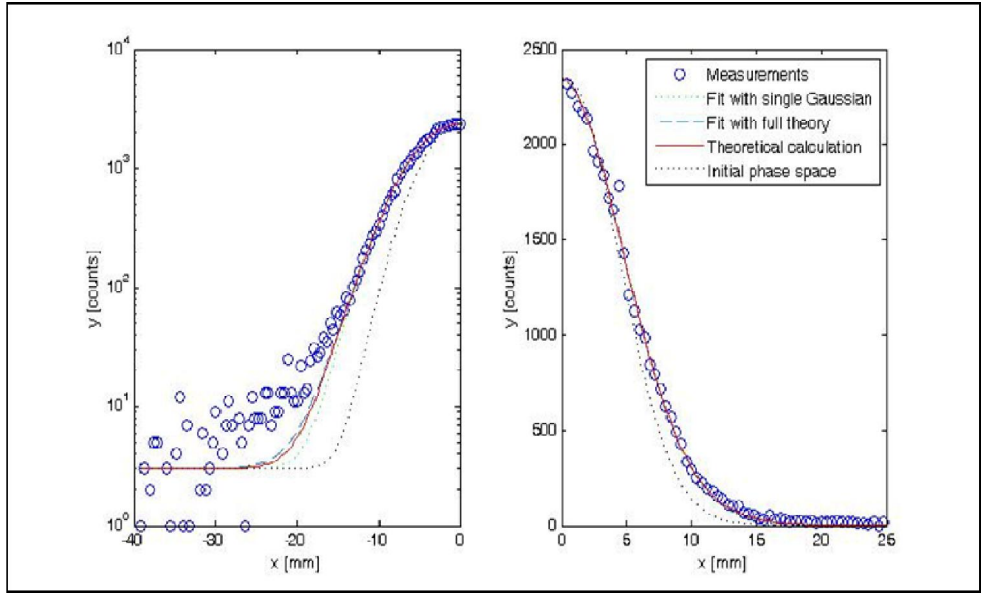

**Figure 30**: Logarithmic and linear plots of a spot-size measurement of a 160-MeV beam at the water-equivalent depth of 151mm for an Accel machine. The black dotted line shows the initial phase space, which is subtracted quadratic from the other contributions.

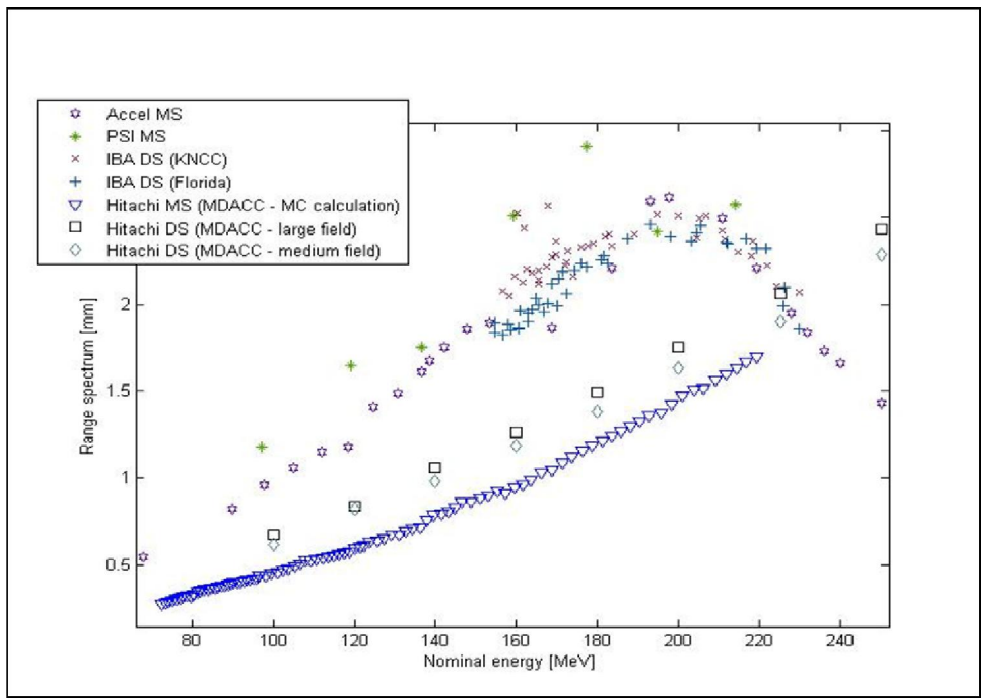

**Figure 31:** The range spectra ($\tau_{in}$) obtained from our depth-dose model for a variety of treatment machines and techniques (MS: modulated scanning, DS: double scattering). It is well known that a synchrotron (Hitachi) usually creates a smaller initial range spectrum than a cyclotron (all other machines). It is interesting to observe that all cyclotrons produce similar range spectra. It has to be commented that, as measured data for MS in the Hitachi machine has not yet been available to us, the data, shown in this figure, has been obtained via a Monte-Carlo simulation of the pristine Bragg curves.

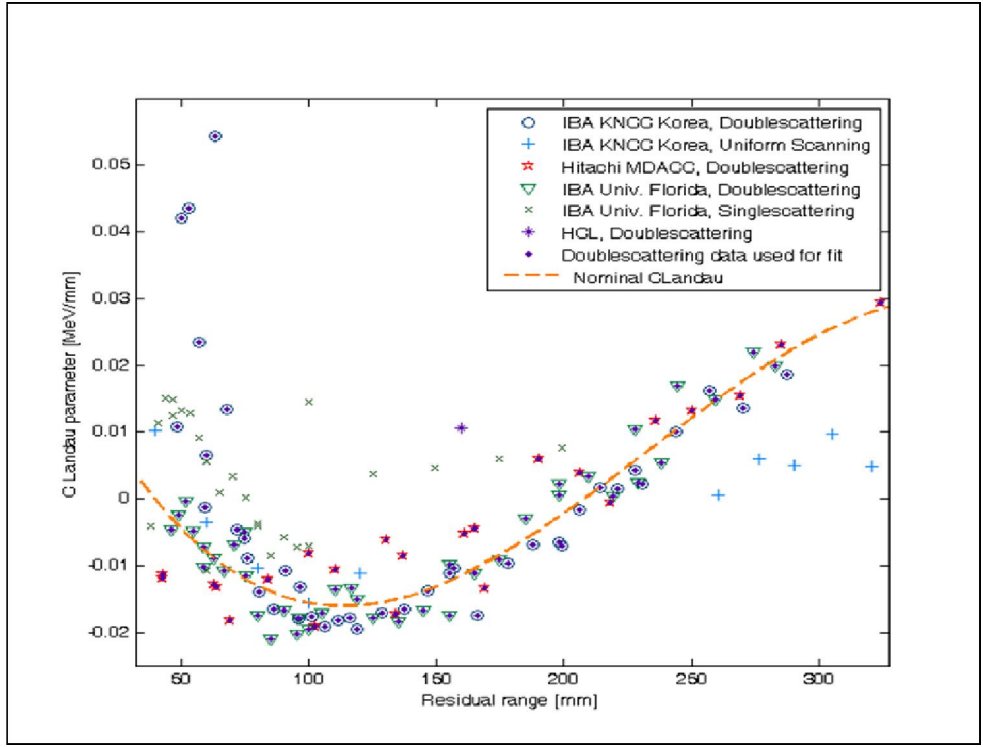

**Figure 32**: Fit results for $C_{Lan1}$ for a number of different double-scattering beam-lines, as well as for one uniform-scanning beam-line. The line corresponds to $C_{Lan1}$ according to Eq. (111); it has been obtained from a polynomial fit to the data sets of Hitachi MDACC, IBA NCC (double scattering) and IBA Florida. The other data sets are plotted only for comparison.

We have developed analytical models for the depth-dose distribution of a proton beam – the pristine Bragg peak. The models depend on a few beam-line-specific parameters (nominal energy, energy/range spread, Landau parameter, contribution of secondary protons), which need to be obtained by fitting the model to the measured pristine Bragg curves. We have shown that the models can reproduce the pristine Bragg curves for different accelerator and beam-line designs. An interpolation of the key parameters permits the accurate calculation of any intermediate pristine Bragg peaks; this is particularly important for delivery machines which feature an analog energy tuning. The lateral distribution of the protons is modeled separately for primary and secondary protons; in order to describe better the large-angle scattering, the lateral distribution of the primary protons is modeled by a sum of two Gaussians. However, it has been shown [22, 52] that a correct modeling of the large-angle scattered primary protons and the scattering of the secondary protons has an impact on the determination of the MU factor. Figures 30 - 33 make apparent that with regard to accurate 3D dose calculations and the determination of the MU values (absolute dosimetry) the flexibility of the presented methods represents the outstanding feature developed by the theoretical means of the preceding sections. This flexibility is also reached by the simplified version of the model M2, which is implemented in the VARIAN therapy planning system Eclipse [3, 21] in order to speed up the calculation procedure without significant loss of accuracy.

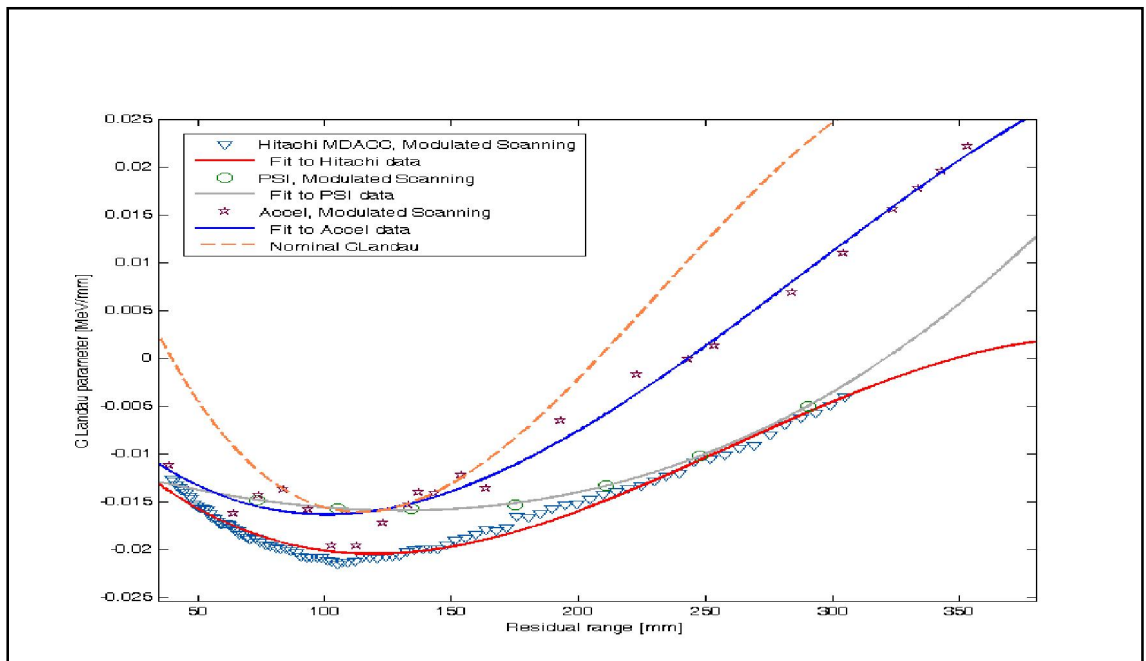

**Figure 33**: The value of $C_{Lan1}$ for different modulated-scanning beam-lines; the Hitachi MDACC values originate from Monte-Carlo calculated pristine Bragg peak. The fitted lines are third-degree polynomials; they replace the nominal $C_{Lan1}$ from Eqs. (111 - 112) in the dose calculation for the respective machines.

## 2.3 Results: applications in therapy planning and comparison with VMAT (Rapidarc)

The most outstanding feature represents a comparison between therapy planning with proton beams and γ-irradiation with VMAT-technique. Figure 34 provides the proton irradiation of a liver carcinoma at PSI (Villigen, Switzerland - a courtesy of A. Lomax) , whereas Figure 35 (irradiation with 6 MV γ-beam, VMAT, Radio-Oncology Hirslanden, Switzerland - a courtesy of U. Schneider) results from a calculation with Eclipse [21, 48]. The principal differences are: 1. Maximum dose of PTV amounts to 107.2 % and mean value 98.9 % (VMAT) and 107.8 % and mean value 100 % (protons). This is not a striking difference, except the higher RBE, which is assumed to be 1.10 - 1.17 for protons. 2. More important are the large iso-dose values of VMAT in the normal liver tissue, which include large ranges from 80 % to 10 %; even the spinal cord is charged with ca. 15 % (≈ 9 Gy). Thus the normal liver tissue is charged in noteworthy regions with ca. 42 Gy or slightly more, whereas irradiation with protons does not imply such a problem due to the properties of Bragg curves.

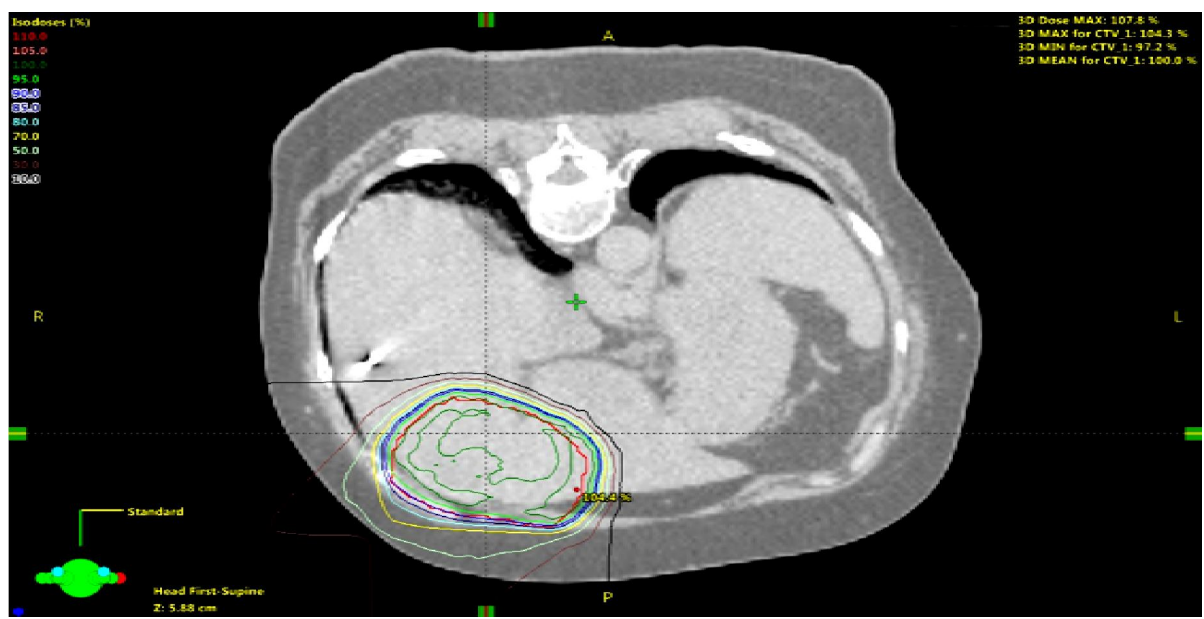

**Figure 34:** Proton treatment plan for a tumour in the liver. The patient was planned using three proton fields from 0°, 45°

and 90 ° (note: images are displayed upside down) to a dose of 60 Gy.

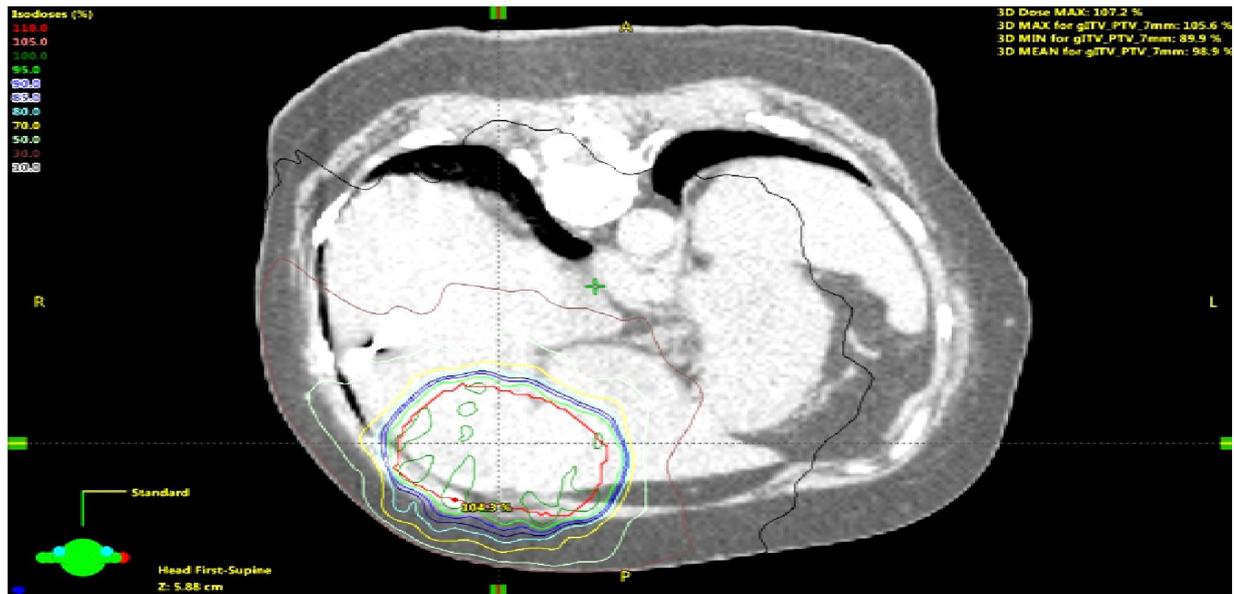

**Figure 35:** Volumetric arc photon treatment plan for a tumour in the liver. The patient was planned using one arc of 6 MV photons. The delivered dose was 60 Gy.

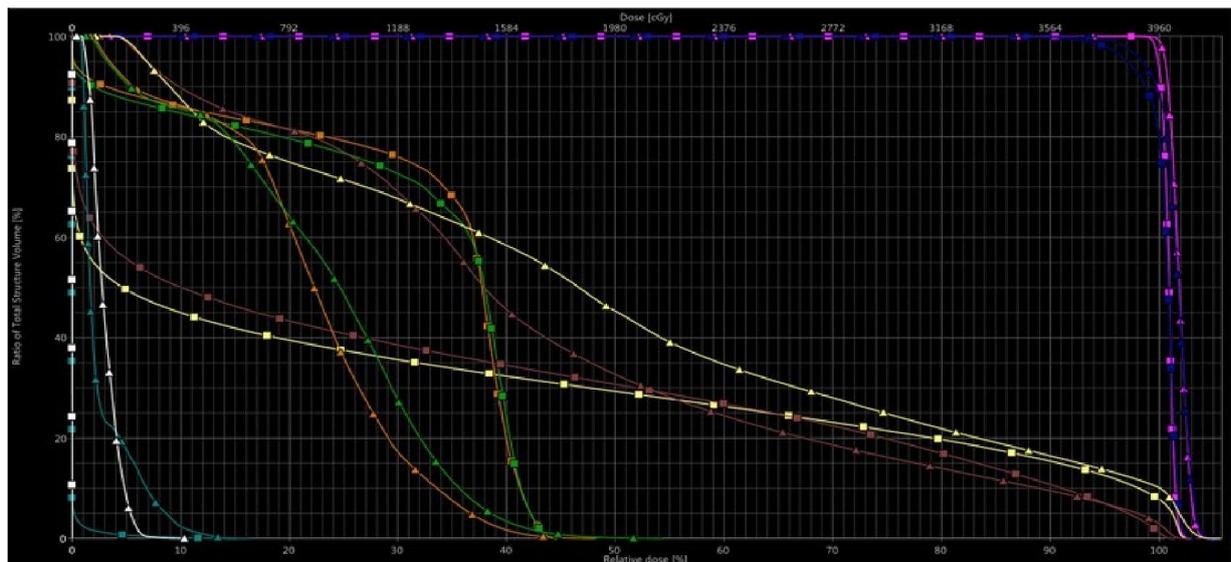

**Figure 36:** Explanation of Figures 37 - 38 referring to DVH of both Figures. The squares in this figure refer to protons and the triangles to photons; the colours of the figure are associated with the following regions: small bowel (white), rectum (brown), yellow (bladder), dark blue (PTV 79.20 Gy), red (CTV 79.20 Gy), green (femoral head L), orange (femoral head R).

The comparison therapy plans of prostate irradiation (Figures 36 - 38) have been worked out at the Centre of Radiation Oncology of the University of Pennsylvania (Philadelphia, USA). In this case, both irradiation plans are based on calculations via Eclipse implementation [21, 48]. The benefits of proton therapy can be verified at the DVH (Figure 36). Thus in Figure 36 the PTV maximum dose (VMA) amounts to 105.7 % (mean value: 101.5 %) and in Figure 38 (protons) the maximum amounts to 103.5 % (mean value: 100.1 %). The significantly smaller charge of normal tissue irradiation is - besides RBE - the most outstanding feature of protons.

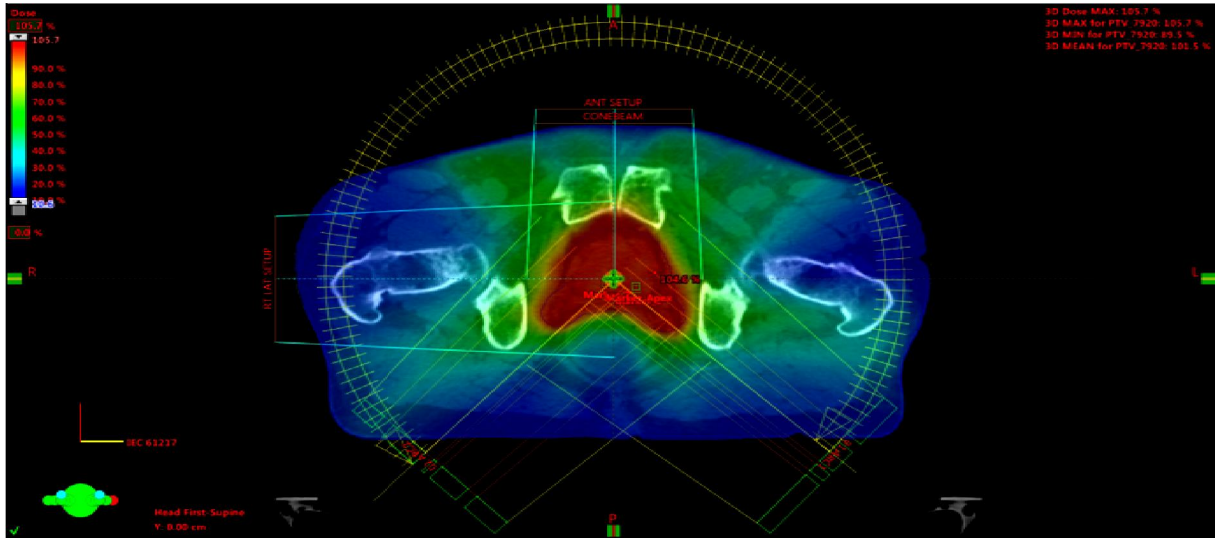

**Figure 37:** Treatment of a prostate tumour with γ-irradiation (VMAT-technique). The arc of beam entrance starts at 220° and stops at 140°.

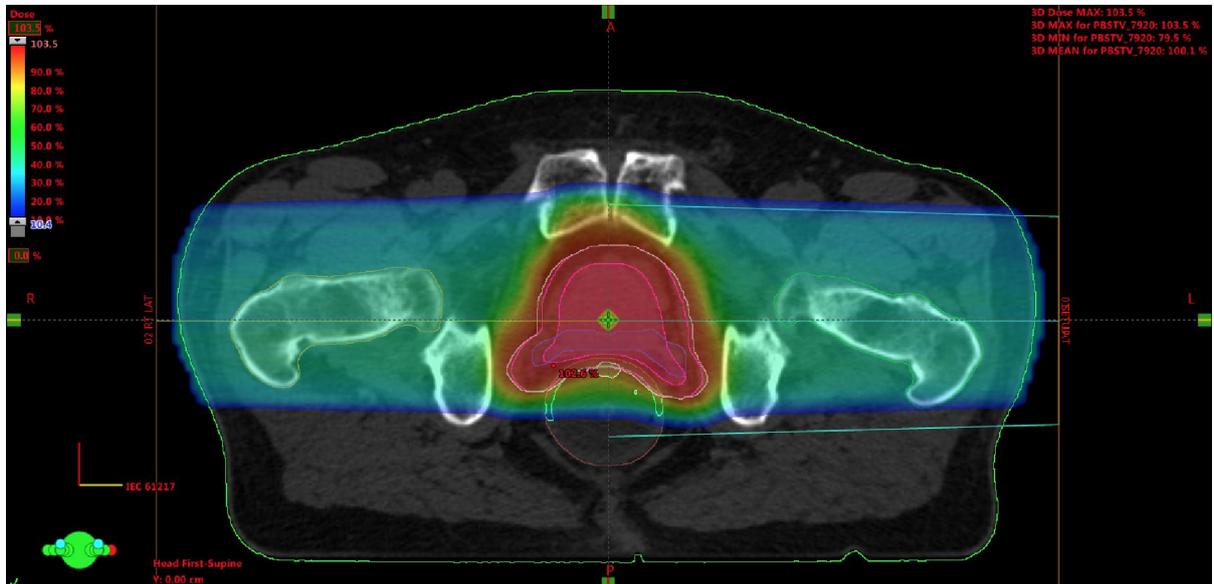

**Figure 38:** Treatment of the target as shown in Figure 37 with two opposing proton beam (beam entrances 270° and 90°)

## 3. APPENDIX: Nuclear interaction channels of protons

The calculation of the total nuclear cross section requires some information acquired by fitting the Los-Alamos data, results of extended nuclear shell theory, and empirical rules. Let us at first recall Eqs. (46 - 51, 55 – 56). It is now the task to correlate the theoretical results of this section with the parameters Z and $A_N$. We assume an iso-scalar nucleus, i.e., one in which the numbers of protons and neutrons are equal ($A_N = 2 \cdot Z$). This assumption holds for almost all light nuclei. For nuclei with spherical symmetry, the nuclear radius $R_{strong}$ is given by:

$$R_{strong} = 1.2 \cdot 10^{-13} \cdot \sqrt[3]{A_N} \quad (in \ cm) \quad (136)$$

With regard to the determination of the total nuclear cross section $Q^{tot}$, we have borrowed some elements from the collective model of nuclear interactions. First of all, we have to know $Q^{tot}_{max}$ required for the calculation of other quantities. We start with the following 'ansatz', and thereafter we give some explanations:

$$Q^{tot}{}_{max} = a \cdot A_N + b \cdot A_N^{2/3} + c \cdot A_N^{1/3} + d \cdot Z^{\kappa} / A_N^{1/3}$$
$$(Q^{tot}{}_{max} \quad in \quad mb) \quad (137)$$

What is the physical interpretation? With the aid of Eq. (137), we obtain the following properties:

Term a: Connection of $Q^{tot}{}_{max}$ to the complete volume of the nucleus. It is important in the resonance domain; it includes resonance scatter via nuclear deformations (vibrations) of the whole nucleus, resonance excitations by changing the spin multiplicity (all effects are inelastic), and transformation of a nuclear neutron (see section 3.4).

Term b: Proportional to the area of the geometric cross section. It contains potential scatter (major part, elastic), rotations induced by Coulomb repulsion/strong-interaction attraction (elastic and inelastic), and nuclear reactions by changing the iso-spin multiplicity (inelastic).

Term c: Proportional to the nuclear radius $R_{strong}$. Excitations by spin-orbit coupling, when the whole nucleus changes its angular momentum, inelastic resonance effect, and elastic spin-spin scatter.

Term d: Proportional to $Z^k/R_{strong}$. Excitation of nuclear vibrations by Coulomb repulsion (resonance effect, inelastic) and elastic scatter.

The asymptotic behavior $Q^{tot}{}_{as}$ of $Q^{tot}$ is given by the relation:

$$Q^{tot}{}_{as} \cong C_{as} \cdot A_N^{2/3} \quad (C_{as} = 85.203266 \quad mb) \quad (138)$$

This connection mainly contains the *term b* above. We have verified the validity of this property for nuclei up to Zn. The order of magnitude of the term b is a clear indication that elastic potential scatter of the nucleus via the strong interaction is the main contribution of the total nuclear cross section. However, Eq. (138) can also be used for the determination of $Q^{tot}{}_c$ and $Q^{tot}{}_{as}$, and only the four coefficients are different. Therefore we write Eq. (49) in a modified form (the parameters are given in Table 8):

$$Q^{tot}{}_{type} = a \cdot A_N + b \cdot A_N^{2/3} + c \cdot A_N^{1/3} + d \cdot Z^{\kappa} / A_N^{1/3} \quad (139)$$

| $Q^{tot}{}_{(type)}$ | a | b | c | d |
|---|---|---|---|---|
| $Q^{tot}{}_{max}$ | 2.616961 | 81.292396789 | 2.942205176087 | - 1.952388200516 |
| $Q^{tot}{}_c$ | 2.613238 | 76.416450001 | 2.405500581216 | - 1.262097902713 |
| $Q^{tot}{}_{as}$ | 0.262441 | 46.681178969 | 0.377143799339 | - 0.141664052734 |

**Table 8:** Parameters a, b, c and d for some different types of $Q^{tot}$.

$E_{res}$ is given by $E_{res} = E_m + E_{Th}$. Results obtained by using the extended nuclear theory and Los-Alamos data indicate the following connection:

$$E_m = 11.94 + 0.29 \cdot (A_N - 12) \quad (if \quad A_N \geq 12)$$
$$E_m = [(A_N - 1/11)]^{2/3} \cdot 11.94 \quad (if \quad A_N < 12) \quad (140)$$

The parameter $\sigma_{res}$, according to Eq. (48), results from a fitting procedure of calculated and measured data. $I_c$ is defined by the continuity condition for the Gaussian (Eq. (47)) and the hyperbolic-tangent function (Eq.(49)). It is known that the nuclear cross section can be described by a series of exponential functions, before the asymptotic behavior is reached. We have verified that the hyperbolic-tangent function, which can be expanded in terms of exponential functions, provides optimal results, and it easily accommodates the continuity conditions at $E = E_c$ and $Q^{tot}_c = Q^{tot}_{max} \cdot I_c$. The Gaussian behavior in the resonance domain is due to the numerous resonance excitations occurring at $E \approx E_{res}$ according the generalized Breit-Wigner formula [41].

## 3.1 Harmonic oscillator 3D models

At first, we consider the 3D harmonic oscillator, described by the Hamiltonian $H_{osc}$:

$$H_{osc} = \frac{1}{2M} \cdot \sum_{k=1}^{3} p_k^2 + \frac{M}{2} \cdot \omega_0^2 \cdot \sum_{k=1}^{3} q_k^2 \quad (141)$$

In this equation, iso-spin symmetry is assumed to hold, i.e., $M_p = M_{neutron} = M$. There are three ways to obtain the general solution of this equation, well-known from standard textbooks of quantum mechanics and nuclear physics; herein, we only present the results:
1. Use of creation and annihilation operators (algebraic method) based on the commutation relation:

$$p_k \cdot q_l - q_l \cdot p_k = \frac{\hbar}{i} \cdot \delta_{kl} \quad (142)$$

2. Replacement of $p_k$ by $-i \cdot \hbar \cdot \partial/\partial q_k$ in the Hamiltonian and solving the resulting Schrödinger equation by a Gaussian function multiplied with Hermite polynomials. Solving the Schrödinger equation in terms of spherical harmonics and Laguerre polynomials. The methods of this point 2 are not discussed here, see e.g. [4, 5 and textbooks of quantum mechanics].

We rewrite the Hamiltonian of the 3D oscillator as:

$$\left.\begin{array}{l} b_k = (M \cdot \omega_0 / 2\hbar)^{1/2} \cdot q_k + (i/(2 \cdot M \cdot \omega_0 \cdot \hbar)) \cdot p_k \\ b^+_k = (M \cdot \omega_0 / 2\hbar)^{1/2} \cdot q_k - (i/(2 \cdot M \cdot \omega_0 \cdot \hbar)) \cdot p_k \end{array}\right\} (k=1,..,3) \quad (143)$$

These operators obey the commutation relations for bosons:

$$\left.\begin{array}{l} [b_k, b^+_l] = \delta_{kl} \quad (k,l = 1,...3) \\ [b^+_k, b^+_l] = [b_k, b_l] = 0 \end{array}\right\} \quad (144)$$

With the help of these operators, the Hamiltonian assumes the shape:

$$H = \tfrac{1}{2} \cdot \hbar \cdot \omega_0 \cdot \sum_{k=1}^{3}(b^+_k \cdot b_k + b_k \cdot b^+_k) = \hbar \cdot \omega_0 \cdot (\sum_{k=1}^{3} b^+_k \cdot b_k + \tfrac{3}{2}) \quad (145)$$

The operator of the angular momentum is given by:

$$\Im_{kl} = q_k \cdot p_l - p_k \cdot q_l = i \cdot (b_k b^+_l - b_l b^+_k) \quad (146)$$

The angular-momentum operator commutes with the Hamiltonian H and, therefore, it only connects degenerate states of H, by transforming a quantum state k to the state l and vice versa. The operator $b^+_k$ (absorption operator) and $b_k$ (emission operator) modify (increase and decrease, respectively) the energy $\hbar \cdot \omega_0$. There are nine independent types of bilinear products $b^+_k \cdot b_l$ (i.e., k = 1, …, 3 and l = 1, ..., 3), which implies that they can be the generators of $SU_3$ in the configuration space. This means that there is a correspondence between $SO_3$ (rotational symmetry in the configuration space) and $SU_3$, in analogy to the one between the group $SO_2$ (x/y – plane) and $SU_2$ for the two-dimensional harmonic oscillator. In nuclear physics, the group $SU_2$ is connected to the iso-spin, referring to both nucleons obeying anti-commutation rules. Although bilinear products of Fermion-operators satisfy the above commutation rules, physical differences exist. Applied to a whole nucleus, the oscillator model is rather a collective description of physical properties as oscillations/vibrations via deformation or creation of rotational bands (quanta of the angular momentum of the whole nucleus) due to interactions with comparably low-energy protons (e.g., see resonance scatter of $Q^{tot}$, in particular the first term of Eq. (137)). A further critical aspect is that the oscillator potential has a minimum for $E = 0$, not for $E \ll 0$; bound states exist for arbitrarily high energies. Nevertheless, with the help of some modifications this model will become suitable for practical problems.

The concept of spin can be introduced to the 3D harmonic oscillator by the commutation rule [4]:

$$\sigma \otimes p \cdot q_l - q_l \cdot \sigma \otimes p = \tfrac{\hbar}{i} \cdot \sigma_l \quad (l=1,2,3) \quad (147)$$

In this case, σ represents the three Pauli spin matrices and the unit matrix. The result is the Pauli equation of a 3D harmonic oscillator. Similarly, we introduce the iso-spin τ by the substitution:

$$\sigma \Rightarrow \sigma \otimes \tau \quad (148)$$

By that, we obtain the Pauli equation for a 3D oscillator for a proton and neutron (in nuclear many-particle theory, the Pauli principle is generalized, i.e., the total wave-function has to be anti-symmetric with regard to spin and iso-spin; however, the property $g_p/g_n = -3/2$ does not follow from the scope of the present theory and further principles have to be introduced):

$$\left. \begin{array}{l} \tfrac{1}{2M} \cdot ((\tfrac{\hbar}{i}\nabla - \tfrac{e_0}{c}A)^2 \psi_p + +(V_{osc} - \mu_p \cdot B) \cdot \psi_P = -i\hbar \tfrac{\partial}{\partial t}\psi_p \\ -\tfrac{\hbar^2}{2M} \cdot \Delta\psi_n + (V_{osc} - \mu_n \cdot B) \cdot \psi_n = -i\hbar \tfrac{\partial}{\partial t}\psi_n \\ \mu_p = g_p \cdot \mu_0 \cdot s; \quad \mu_n = g_n \cdot \mu_0 \cdot s; \quad B = \nabla \times A \end{array} \right\} (149)$$

Even by extending Eq. (149) to a many-particle equation (including the spin-orbit coupling) and to a Slater determinant (Hartree-Fock: ground state), the problem of nuclear reactions, due to the properties of the harmonic oscillator potential $V_{osc}$, cannot be solved. The problem of simple oscillator models can be verified in the following Figure 39, which shows the effective nuclear potential energy for O. The abscissa is expressed in units of $r = 1.2 \cdot 10^{-13} \cdot A_N^{1/3}$ cm ($A_N = 16$).

We approximate the potential of Figure 39 (the complete potential function can be expressed as a linear combination of two Gaussians):

$$\left. \begin{array}{l} \varphi(r) = V_0 \cdot \exp(-r^2/\sigma_0^2) + V_1 \cdot \exp(-r^2/\sigma_1^2) \\ \sigma_1 = 3.37402; \; V_0 = -27.75925; \; V_1 = 7.75935 \end{array} \right\} (150)$$

$$\left. \begin{array}{l} \varphi(r) = V_0(1 - r^2/\sigma_0^2) \\ \sigma_0 \approx 0.423901 \end{array} \right\} (151)$$

A property of the Gaussian function is that its curvature changes sign at $r = r_c$. For a single Gaussian, as the first one in Eq. (150), $r_c$ is given by:

$$r_c = \sigma_0 / \sqrt{2} = 0.29974 \quad (152)$$

Only for $r \leq r_c$, is a harmonic-oscillator approach useful, and the deviation to a Gaussian in this domain small. This is, however, not true for $r > r_c$. In the case of the linear combination of two Gaussians, $r_c$ is broader:

$$r_c \approx \tfrac{1}{\sqrt{2}} \cdot \sigma_0 \cdot \sigma_1 \cdot \sqrt{(-\sigma_1^{\,2} \cdot V_0 - \sigma_0^{\,2} \cdot V_1 \cdot A_\sigma)/(-\sigma_1^{\,4} \cdot V_0 - \sigma_0^{\,4} \cdot V_1 \cdot A_\sigma)}$$
$$r_c \approx 0.4582, \qquad A_\sigma = \exp[(\sigma_1^{\,2} - \sigma_0^{\,2})/2 \cdot \sigma_1^{\,2})] \quad (153)$$

Figure 39 shows that, for r < 0.4982, strong interactions are dominant (all other interactions are negligible). If 0.4682 ≤ r < 1, strong interactions are still present with decreasing tendency, whereas Coulomb repulsion is increasing; finally, for r = 1, strong interactions are negligible. We will come back to these results in the next section.

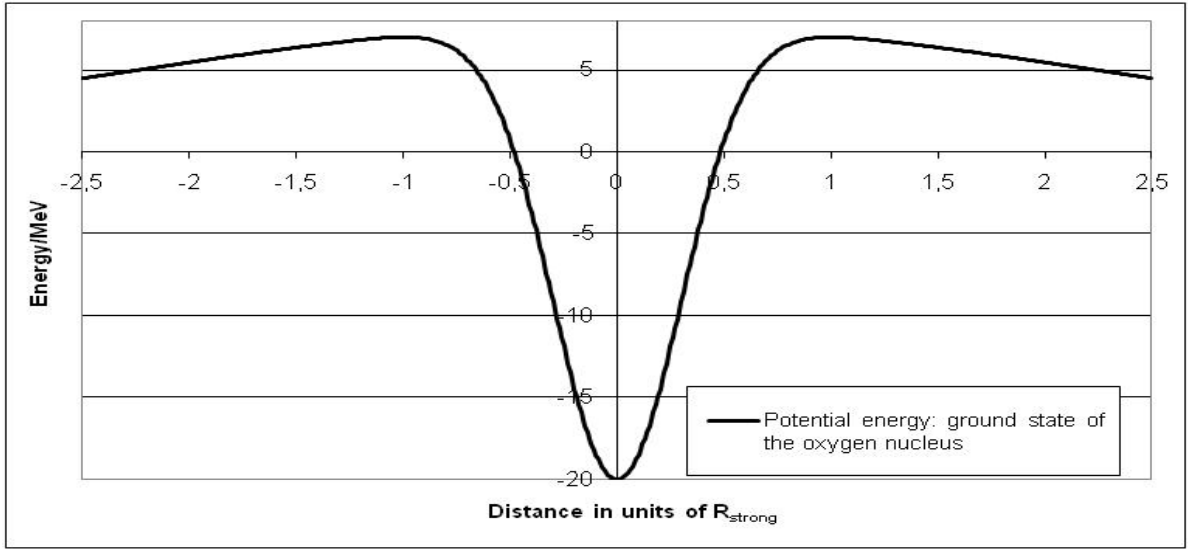

**Figure 39:** Total (effective nuclear potential plus Coulomb repulsion) for the oxygen nucleus.

## 3.2 Nonlinear/nonlocal Schrödinger equation, anharmonic oscillators with self-interaction and Hartree-Fock (HF) method inclusive configuration interaction via open shells

Let us first consider the usual Schrödinger equation for a bound system:

$$E \cdot \psi + \tfrac{\hbar^2}{2 \cdot M} \cdot \Delta \psi = \varphi(x, y, z) \cdot \psi$$
$$\varphi \le 0 \; (x, y, z \in R_3); \Delta : 3D - Laplace\,operator \quad (154)$$

A nonlinear Schrödinger equation is obtained by introducing the potential φ, proportional to the density of solutions:

$$\varphi = \lambda \cdot |\psi(x, y, z)|^2 =$$
$$\lambda \cdot \int \delta(x - x') \cdot \delta(y - y') \cdot \delta(x - x') \cdot |\psi(x', y', z')|^2 d^3 x' \quad (155)$$

During the past decades, this type has been encountered in many fields of physics, such as superconductivity, nuclear and plasma physics [4, 5, 55, and references therein]. The coupling constant λ is negative (in which case, the solutions are bound states with E < 0); Eq. (155) can be interpreted as a contact interaction. It is known from many-particle problems (e.g., quantum electrodynamics, HF method, etc.) that the mutual interactions between the particles in configuration space lead to nonlinear equations in quantum mechanics. However, in these cases, there are not at all contact interactions; the nonlinear Schrödinger equation above is an idealistic case. By taking ε → 0, the Gaussian kernel is transformed into a δ kernel:

$$\varphi = \lambda \cdot \int (1/(\sqrt{\pi}^3 \cdot \varepsilon^3)) \cdot \exp[-((x-x')^2 + (y-y')^2 + (z-z')^2)/\varepsilon^2] \cdot$$
$$\cdot |\psi(x', y', z')|^2 d^3 x' \quad (156)$$

The nonlinear/nonlocal Schrödinger equation can be interpreted as a self-interaction of a many-particle system with internal structure, and it is possible to generalize this type by incorporation of additional internal symmetries (e.g., the introduction of the spin to obtain spin-orbit coupling, $SU_2$, $SU_3$, and also discrete-point groups). According to the principles developed in this work, we are able to write Eq. (156) in the form of an operator equation (the Gaussian kernel is Green's function):

$$\varphi = \lambda \cdot \exp(\tfrac{1}{4} \cdot \varepsilon^2 \cdot \Delta) \cdot |\psi(x, y, z)|^2 \quad (157)$$

Expanding this operator in terms of a Lie series and keeping only the terms up to Δ, Eq. (157) becomes a stationary Klein-Gordon equation, which describes the interaction between the particles obeying the Ψ-field:

$$\left. \begin{array}{l} \exp(\tfrac{1}{4} \cdot \varepsilon^2 \cdot \Delta) = 1 + \tfrac{1}{4} \cdot \varepsilon^2 \cdot \Delta + 0(higher-order\ terms) \\ (1 + \tfrac{1}{4} \cdot \varepsilon^2 \cdot \Delta) \cdot \varphi = \lambda \cdot |\psi|^2 \end{array} \right\} (158)$$

By rescaling the Klein-Gordon equation, we obtain the more familiar form: $1 + 0.25\ \varepsilon^2 \Delta \to k^2 + \Delta$; Green's function is of the form:

$$\left. \begin{array}{l} G(r, r') = N \cdot \exp(-k \cdot |r - r'|) \cdot \tfrac{1}{|r-r'|} \\ N: normalization\ factor \end{array} \right\} (159)$$

By setting k → 0, the Poisson equation of electrostatics is obtained, if $|\psi|^2$ is interpreted as a charge density. The Gaussian kernel K also represents the exchange of virtual particles between the nucleons. In view of this fact, we point out that we have incorporated a many-particle system from the beginning. Which information now does this nonlinear/nonlocal Schrödinger equation provide? In order to obtain a connection of the combined Eqs. (156, 157) with the oscillator model of nuclear shell theory, we analyze the kernel K in detail. In the propagator method [4, 32], the expansion of K in terms of generating functions is an important tool:

$$\left. \begin{array}{l} K(\varepsilon, x'-x, y'-y, z'-z) = (1/\sqrt{\pi}^3 \cdot \varepsilon^3) \sum_{n1=0}^{\infty} \exp(-x'^2/\varepsilon^2) \cdot \\ \cdot H_{n1}(x'/\varepsilon) \cdot x^{n1}/(\varepsilon^{n1} \cdot n_1!) \cdot \sum_{n2=0}^{\infty} \exp(-y'^2/\varepsilon^2) \\ \cdot H_{n2}(y'/\varepsilon) \cdot y^{n2}/(\varepsilon^{n2} \cdot n_2!) \cdot \\ \quad \cdot \sum_{n3=0}^{\infty} \exp(-z'^2/\varepsilon^2) \cdot H_{n3}(z'/\varepsilon) \cdot z^{n3}/(\varepsilon^{n3} \cdot n_3!) \end{array} \right\} (160)$$

Inserting Eq. (160) into the nonlinear/nonlocal Schrödinger equation, we obtain:

$$E \cdot \psi + \tfrac{\hbar^2}{2 \cdot M} \cdot \Delta \psi = \varphi(x,y,z) \cdot \psi =$$
$$\lambda \cdot (1/\sqrt{\pi}^3 \cdot \varepsilon^3) \sum_{n1=0}^{\infty} \sum_{n2=0}^{\infty} \sum_{n3=0}^{\infty} \Phi_{n1,n2,n3} \cdot x^{n1} \cdot y^{n2} \cdot z^{n3} \cdot \psi$$
$$\Phi_{n1,n2,n3} = \tfrac{1}{n1!} \cdot \tfrac{1}{n2!} \cdot \tfrac{1}{n3!} \cdot \tfrac{1}{\varepsilon^{n1+n2+n3}} \cdot \int |\psi(x',y',z')|^2$$
$$\cdot \exp(-(x'^2+y'^2+z'^2)/\varepsilon^2) \cdot H_{n1}(x'/\varepsilon) \cdot H_{n2}(y'/\varepsilon) \cdot H_{n3}(z'/\varepsilon) d^3 x' \quad (161)$$

Eq. (161) represents a highly anharmonic oscillator equation of a self-interacting field. Since the square of the wave-function is always positive definite, all terms with odd numbers of n1, n2, and n3 vanish due to the anti-symmetric properties of those Hermite polynomials. For $r_c \leq \varepsilon/\sqrt{2}$ (domain with positive curvature), the whole equation is reduced to a harmonic oscillator with self-interaction; the higher-order terms are small perturbations. We summarize the results and refer to previous publications [4, 5]:

$$E \cdot \psi + \tfrac{\hbar^2}{2 \cdot M} \cdot \Delta \psi = \varphi(x,y,z) \cdot \psi =$$
$$\lambda \cdot (1/\sqrt{\pi}^3 \cdot \varepsilon^3) \cdot [\Phi_{0,0,2}(x^2+y^2+z^2) + \Phi_{0,0,0}] \cdot \psi \quad (162)$$

The solutions of this equation are those of a 3D harmonic oscillator; the classification of the states by $SU_3$ and all previously developed statements with regard to the angular momentum are still valid. The only difference is that the energy levels are not equidistant; this property can easily be verified in one dimension. The usual ground state energy is $\hbar\omega_0/2$. This energy level is lowered by the term $\sim\lambda \cdot \Phi_{0,0,0}$, depending on the ground-state wave-function. The energy difference between the ground and the first excited state amounts to $\hbar\omega_0$; this is not true in the case above, since the energy level of the excited states depends on the corresponding eigen-functions (these are still the oscillator eigen-functions!). Next, we will include the terms of the next order, which are of the form $\sim \lambda \cdot (\Phi_{0,2,2}, \Phi_{2,2,0}, \Phi_{2,0,2})$:

$$E \cdot \psi + \tfrac{\hbar^2}{2 \cdot M} \cdot \Delta \psi = \varphi(x,y,z) \cdot \psi$$
$$= \lambda \cdot (1/\sqrt{\pi}^3 \cdot \varepsilon^3) \cdot [\Phi_{0,0,0} + \Phi_{0,0,2}(x^2+y^2+z^2) + T] \cdot \psi \quad (163)$$
$$T = \Phi_{2,2,0} \cdot x^2 \cdot y^2 + \Phi_{2,0,2} \cdot x^2 \cdot z^2 + \Phi_{0,2,2} \cdot y^2 \cdot z^2$$

The additional term T represents tensor forces. The whole problem is still exact soluble. In further extensions of the nonlinear/nonlocal Schrödinger equation, we are able to account for spin, iso-spin, and spin-orbit coupling.

The spin-orbit coupling, as an effect of an internal field with nonlocal self-interaction, is plausible, since the extended nucleonic particle has internal structure; consequently, we have to add $H_{so}$ to the nonlinear term:

$$H_{so} \cdot \psi = g_\tau \cdot \frac{\hbar \cdot \sigma}{4 \cdot M \cdot c^2} \cdot \nabla \varphi \times p \cdot \psi \quad (164)$$

$\Psi$ is now (at least) a Pauli spinor (i.e., a two-component wave-function), and together with $H_{so}$ the $SU_3$ symmetry is broken. We should like to point out that the operation $\nabla \varphi$ acts on the Gaussian kernel K:

$$\nabla \varphi = -\frac{2}{\varepsilon^2} \cdot [H_1((x-x')/\varepsilon), H_1((y-y')/\varepsilon),$$
$$H_1((z-z')/\varepsilon)] \cdot \varphi \qquad (165)$$

The expression in the bracket of the previous equation represents a vector, and p (p → -i$\hbar\nabla$) acts on the wave-function. Since the neutron is not a charged particle, the spin-orbit coupling of a neutron can only involve the angular momentum of a proton. In nuclear physics, these nonlinear fields are adequate for the analysis of clusters (deuteron, He, etc.). In [53] the theory has been extended to describe nuclei with odd spin. The complete wave-function $\Psi_c$ is now given by the product of a function in configuration space $\Psi$ multiplied with the total spin and iso-spin functions.

We should like to add that an extended harmonic oscillator model with tensor forces has been regarded in [57 - 60]. Feynman and Schwinger, (see [32]) have verified that the use of Gaussians in the description of meson fields provides many advantages compared to the Yukawa potential (Green's function according to Eq. (159)). In a final step, we consider the generalized HF method to solve the many-particle problem. In order to derive all required formulas, it is convenient to use second quantization. The method of second quantization is only suitable to derive the calculation procedure: extension of the Pauli principle to iso-spin besides spin, inclusion of spin-orbit coupling, and exchange interactions. This is the consequence of dealing with identical particles, in which case every state can only occupy one quantum number. In order to get numerical results (i.e., the minimum of the total energy of an ensemble of nucleons, the extraction of the excited states, the scatter amplitudes, etc.), we have to use representations of the wave-function by at least one determinant in the configuration space. In the 'language' of second quantization of fermions, we would have to regard expressions like:

$$\left. \begin{array}{l} a^+_{k,\sigma,\tau} \cdot a_{l,\sigma',\tau'} + a_{l,\sigma',\tau'} \cdot a^+_{k,\sigma,\tau} = \delta_{kl} \cdot \delta_{\sigma\sigma'} \cdot \delta_{\tau\tau'} \\ a_{k,\sigma,\tau} \cdot a_{l,\sigma',\tau'} + a_{l,\sigma',\tau'} \cdot a_{k,\sigma,\tau} = 0 \\ a^+_{k,\sigma,\tau} \cdot a^+_{l,\sigma',\tau'} + a^+_{l,\sigma',\tau'} \cdot a^+_{k,\sigma,\tau} = 0 \end{array} \right\} \qquad (166)$$

The operators of the form $a_k^+$ and $a_k$ (k being a set of quantum numbers) are creation and destruction operators in the state space. The nonlinear/nonlocal Schrödinger equation with Gaussian kernel for the description of the strong interaction, including the spin-orbit coupling, can be written by these operators, leading from an extended particle with internal structure to a many-particle theory. Before we start to explain the calculations by including one or more configurations, we recall that, according to Figure 39, we have an increasing contribution of the Coulomb repulsion for r > $r_c$, though in the domain r < $r_c$, the contributions of the Coulomb interactions are negligible. Since all basis elements of the calculation procedures, i.e., the calculation of eigen-functions in the configuration space, two-point kernels of strong interactions between nucleons, and the spin-orbit coupling can be expressed in terms of Gaussians and Hermite polynomials, we want to proceed in the same fashion with regard to the Coulomb part. According to results of elementary-particle models [59], the charge of the proton is located in an extremely small sphere with radius $r_p = 10^{-14}$ cm, not at one 'point'. Therefore, we write the decrease of the proton Coulomb potential by 1/(r+$r_p$); for r = 0, we then obtain $10^{14}$ cm$^{-1}$, not infinite. In a sufficiently small distance of r = 2.4·$10^{-13}$ cm, we can approximate the Coulomb potential with high precision by:

$$\frac{1}{r+r_p} = c_0 \cdot \exp(-r^2/r_0^2) + c_1 \cdot \exp(-r^2/r_1^2)$$
$$+ c_2 \cdot \exp(-r^2/r_2^2) \qquad (167)$$

The mean standard deviation amounts to $10^{-5}$, if the parameters of Eq. (167) are chosen as:

$$c_0 = 0.5146 \cdot 10^{14}; \ c_1 = 0.3910 \cdot 10^{14}; \ c_2 = 0.0944 \cdot 10^{14}$$
$$r_0 = 0.392 \cdot 10^{-13} \ cm; \ r_1 = 0.478 \cdot 10^{-13} \ cm; \ r_2 = 2.5901 \cdot 10^{-13} \ cm \quad (168)$$

If necessary, it is possible to rescale $r_0$, $r_1$, and $r_2$ by dividing by $(A_N)^{1/3}$. The contribution with $c_2$ incorporates a long-range correction. In the absence of an external electromagnetic field, the Hamiltonian reads as:

$$H = \sum_j -\frac{\hbar^2}{2M} \Delta_j + H_{so} + H_{Coul} + H_{strong}$$

$$H_{Coul} = e_0^2 \cdot \sum_{j,proton} \sum_{l,proton \neq j} \sum_{k=0}^{2} c_k \cdot \exp(-(r_j - r_l)^2 / r_k^2) \quad (169)$$

$$H_{strong} = -g_s \cdot \sum_j \sum_{l \neq j} \exp(-(r_j - r_l)^2 / \sigma_s^2)$$

Note that it is possible to distinguish between the proton and the neutron masses by indexing M; the $\square$, previously used in Eq. (169), has been replaced by $\sigma_s$. The coupling constant of $g_s$ is 1, if the Coulomb interaction is scaled to:

$$g_s = 1; \ e_0^2 / (\hbar \cdot c) = 1/137 \quad (170)$$

Thus, in theoretical units with $e_0 = c = h/2\pi = 1$, the coupling constant $g_s$ assumes 137. This relation can be best seen in the Dirac equation containing a Coulomb repulsion potential $\sim e_0^2$ and a strong interaction term $\sim -g_s$. The aforementioned relation is obtained by dividing the kinetic-energy operator $c \cdot \alpha \cdot p \rightarrow -c \cdot \alpha \cdot \hbar \cdot \nabla$ and $\beta \cdot mc^2$ by $(c \cdot \hbar)$. In the calculations for deuteron, $He^3$, and He, we have assumed the range length $\sigma_s$:

$$\sigma_s = \sigma_{sp} = \frac{\hbar}{m_p \cdot c} \approx 10^{-13} \ cm$$
$$m_p : \ mass \ of \ \pi-meson \quad (171)$$

This assumption turned out to be not sufficient; a replacement of $\sigma_s$ was justified to distinguish between the range length $\sigma_{sp}$ ($\pi$-mesons) and $\sigma_{sk}$ (K-mesons):

$$-g_S \cdot \exp(-(r_j - r_l)^2 / \sigma_s^2)$$
$$\Rightarrow -g_S \cdot [c_{sp} \cdot \exp(-(r_j - r_l)^2 / \sigma_{sp}^2)$$
$$+ c_{sk} \cdot \exp(-(r_j - r_l)^2 / \sigma_{sk}^2)] \quad (172)$$
$$c_{sp} = 1 - (\sigma_{sk} / \sigma_{sp})^2; \ c_{sk} = (\sigma_{sk} / \sigma_{sp})^2;$$
$$\sigma_{sp} = 1.02 \cdot 10^{-13} \ cm; \ \sigma_{sk} = 0.29 \cdot 10^{-13} \ cm$$

The range length $\sigma_{sk}$ is proportional to $1/m_k$ ($m_k$: mass of the K-meson). The HF-method provides the best one-particle approximation of the closed-shell case.

$$\Phi = \frac{1}{\sqrt{N!}} \begin{vmatrix} \varphi_{k1}(1) \ldots \ldots \varphi_{k1}(N) \\ \varphi_{k2}(1) \ldots \ldots \varphi_{k2}(N) \\ \ldots \ldots \ldots \ldots \ldots \ldots \ldots \\ \varphi_{kN}(1) \ldots \ldots \varphi_{kN}(N) \end{vmatrix} \quad (173)$$

The one-particle functions $\varphi_{k1}(1), \ldots, \varphi_{kN}(N)$ contain all variables (configuration space of position coordinates, spin, and iso-spin). By using a complete system of trial functions, e.g., a Gaussian multiplied with Hermite polynomials, the HF limit is obtained. In view of our question to calculate the S-matrix and the cross section of the proton-nucleon interactions (elastic, inelastic, resonance scatter, and nuclear reactions), this restriction is insufficient. In particular, we have to add excited configurations and virtually-excited configurations. The role of excited states is clear. As an example, we regard the oxygen nucleus, where the total spin is 0. If a proton or neutron of the highest-occupied shell is excited, then the spin may change, and both, highest-occupied and lowest-unoccupied shell, are occupied by one nucleon. The emitted nucleon may be regarded as a 'hole'. This procedure can be repeated to higher-unoccupied states and to linear combinations of configurations with different nucleon numbers. A virtually-excited state is produced, if the configuration of the excited state only formally exists for the calculation procedure, but cannot be reached physically. An example of this case is already the deuteron with iso-spin 0 and spin 1. An excited state with spin 1 or 0, where proton and neutron occupy different energy levels (shells), does not exist. In spite of this situation, the HF method does not provide the correct ground state, and linear combinations of determinants with different spin states (S = 1, -1, 0) and 'holes' have to be included. These virtual states also enter the calculation of the S-matrix and of the cross section.

We have performed HF configuration-interaction calculations (HF - CI) for the nuclei: deuteron, He$^3$, He, Be, C, Si, O, Al, Cu, and Zn. The set of basic functions comprises $2 \cdot (A_N + 13)$ functions with the following properties:

$$\left. \begin{array}{l} \varphi(x) = \sum_{j=0}^{N} [A_j \cdot H_j(\alpha_1 \cdot x) \cdot \exp(-\tfrac{1}{2} \cdot \alpha_1^2 \cdot x^2) \\ + B_j \cdot H_j(\alpha_2 \cdot x) \cdot \exp(-\tfrac{1}{2} \cdot \alpha_2^2 \cdot x^2)] \end{array} \right\} \quad (174)$$

$$\varphi(x,y,z) = \varphi(x) \cdot \varphi(y) \cdot \varphi(z); \quad N = 2 \cdot (A_N + 13) \quad (175)$$

Both $\alpha_1$-functions and $\alpha_2$-functions are chosen such that the number of functions is $A_N + 13$. The different range parameters $\alpha_1$ and $\alpha_2$ are useful, since different ranges can be accounted for. If $\alpha \gg \beta$, the related wave-functions decrease much more rapidly (central part of the nucleus), whereas the $\beta$-contributions preferably describe the behavior in the domain $r \geq r_c$. With the help of this set of trial functions[1] (Ritz's variation principle), we obtain the best approximation of the total energy E by $E_{app}$ and the nuclear shell energies (for occupied and unoccupied shells). For bound states, $E_{app} > E$ is always fulfilled. It should be noted that for computational reasons it is useful to replace the set of functions (174) by the non-orthogonal set:

$$\varphi_{n1,n2,n3} = x^{n1} \cdot y^{n2} \cdot z^{n3} \cdot [A_{n1,n2,n3}(\alpha_1) \cdot \exp(-\tfrac{1}{2}\alpha_1^2 \cdot r^2)$$
$$+ B_{n1,n2,n3}(\alpha_2) \cdot \exp(-\tfrac{1}{2}\alpha_2^2 \cdot r^2)] \quad (176)$$

By forming arbitrary linear combinations depending on $\alpha_1$ and $\alpha_2$ we obtain the same results as by the expansion (174). The exploding coefficients of the Hermite polynomials are an obstacle in numerical calculations and can be avoided by the expansion (176). The minimal basis set for the calculation of deuteron would be one single trial function, i.e. a Gaussian without further polynomials. This is, however, a crude approximation and already far from the HF limit. Using this simple approximation, we obtain the result that the ground state $E_g$ depends solely on $\alpha_1$. The best approximation exceeds the HF limit by about 15 %. Various tasks, such as resonance scatter, nuclear reactions, and spin-orbit coupling cannot be described; the cross section of the pure potential scatter is also 12 % too low.

Using 13 $\alpha_1$-dependent and 13 $\alpha_2$-dependent functions, we have obtained the HF limit and virtually-excited states (a bound excited state does not exist). The HF wave-function had to be subjected to virtually-excited configurations, i.e., all possible singlet and triplet states. This calculation had to be completed by introducing a further proton (interaction proton) and including all virtual configurations (besides a configuration with three independent nucleons, a configuration of a virtual $He^3$ state). Thus, for low proton energies (slightly above $E_{Th}$), the $He^3$ formation is possible. The exceeding energy can be transferred to the total system and/or to rotations/vibrations of $He^3$. In the same fashion, we have to proceed to the calculations for other nuclei: the configurations of all possible fragments have also to be taken into account (see e.g., the final section). In order to keep these considerations short, we now only give a skeleton of the calculation procedures, which are necessary to evaluate the cross sections. When – besides the ground state – all excited states (including virtually excited states and configurations of fragments) are determined (wave-functions and related energy levels), then Green's function is readily determined by taking the sum over all states. This function contains all coordinates in the configuration space (including the spin), quantum numbers of oscillations, and rotational bands:

$$G(r_n, r'_n) = \sum_{j=0}^{N} \psi^*_j(r'_n) \cdot \psi_j(r_n) \quad (177)$$

The S-matrix is given by:

$$S_{kl} = \int \psi^*_k(r'_n) \cdot G(r_n, r'_n) \cdot$$
$$\cdot \psi_l(r_n) d^3 r_1 \ldots d^3 r_{2N} \cdot d^3 r'_1 \ldots d^3 r'_{2N} \quad (178)$$

The transition matrix $T_{kl}$ is defined by all transitions with $k \neq l$:

$$T_{kl} = S_{kl} - \delta_{kl} \quad (179)$$

In order to determine the differential cross section, we need the transition probability. For this purpose, we assume that, before the interaction of the proton with the nucleus, this nucleus is in the ground state. Thus, it might be possible that a proton produces excited states of the nucleus by resonance scatter (inelastic), and a second proton hits the excited nucleus before the transition to the ground state (by emission of a γ quantum) has occurred. The second proton would require a lower energy to release either a nucleon or to induce a much higher excited state of the nucleus. However, due to the nuclear cross section, the probability for an inelastic nuclear reaction is very small and would require a very high proton density to yield a noteworthy effect. Therefore, we have calculated the transition probability using the assumption that the occupation probability of the ground state $P_0$ is 1, i.e., $P_0 = 1$ and $P_k = 0$ ($k > 1$). (This is very special case of the Pauli master equation). The differential cross section is obtained by the transition probability divided by the incoming proton flux:

$$dq/d\Omega = \frac{\text{Transition probability}}{\text{Incoming proton current}} \quad (180)$$

At lower energies, this flux could be calculated by the current given by the Schrödinger equation. To be consistent, we have always used the Dirac equation, since proton energies E > 200 MeV show a significant relativistic effect. With regard to the incoming proton current, we have to point out an important feature:

- The Breit-Wigner formula only considers S states and the incoming current is along the z direction.
- The generalization of this formula [41] includes P states, but the incoming beam is also restricted to the z direction.

Since for our purpose it is necessary to take account for the x/y/z direction by $k_x$, $k_y$, $k_z$ in the Dirac equation, we have not yet succeeded in obtaining a compact and simple analytical form.

We have already pointed out that the main purpose for calculations with the extended nuclear shell theory incorporate nuclear reaction contributions of protons, neutrons and further small nuclei to the total nuclear cross sections of nuclei discussed in this presentation. We should also mention that the default calculation procedure of nuclear reactions in GEANT4 is an evaporation/cascade model, which has been developed on the basis of statistical thermodynamics.

Figures 5 - 6, 14 do not yet provide final information about the contributions $S_{sp,n}$ and $S_{sp,r}$. The first case of non-reaction protons has already treated. According to Figure 15 the contribution of reaction protons is particular important for E > 150 MeV with increasing energy. We now present the calculation formulas for this case. Thus $S_{sp,r}$ is proportional to $\Phi_0 \cdot 2 \cdot \upsilon \cdot C_{heavy}$ and a function $F_r$, depending on some further parameters. It should be mentioned that the parameters of Eq. (182) exclusively refer to the oxygen nucleus. However, from Figure 40 the corresponding parameters of some further nuclei can be verified, e.g., $E_{Th}$, $E_{res}$, and some necessary information on the total cross section. We use the following definitions and abbreviations:

$$\left. \begin{array}{l} z_R = R_{CSDA}/\pi;\, \tau_s = 0.55411;\, \tau_f = \tau_s - 0.000585437 \cdot (E - E_{res}) \\ \arg 1 = z/\sqrt{\tau_s^2 + \tau_{in}^2 + (R_{CSDA}/4\pi)^2} \\ \arg 2 = (R_{CSDA} - z - \sqrt{2}\cdot\pi\cdot z_{shift}) \\ /\sqrt{\tau_f^2 + (\tau_{straggle}/7.07)^2 + (R_{CSDA}/\sqrt{3}\cdot 4\pi)^2} \\ \arg = (R_{CSDA} - 0.5\cdot z_{shift}\cdot\sqrt{\pi} - z)/(\sqrt{\pi}\cdot z_{shift}) \end{array} \right\} \quad (181)$$

Eqs. (181 – 182) can only partially derived, and the adaptation to computed data with the help of the extended nuclear-shell theory is also needed. The transport of secondary reaction protons resulting from the spectral distribution of these protons has to be taken into account; and the spectral distributions rather obey a Landau than a Gaussian distribution (Figure 40).

The tails at $z \geq R_{CSDA}$ result from tertiary protons induced by neutrons and the resonance interaction via meson exchange as pointed out in a previous section. Some consequences of these contributions with regard to buildup have been thoroughly discussed in this work, since the role skew symmetric energy transfer (Landau distributions) and energy transfer from reaction protons along the proton track represents a principal question in understanding the physical foundation of Bragg curves. It appears that the interpretation [20] too simplified with regard to the contributions of the so-called secondary protons. With regard to Figure 40 and the proton - oxygen nucleus interaction, we should finally point out that by a suitable modification a similar listing for nuclear reactions will be obtained, if the oxygen nucleus is replaced by another one such as calcium or copper. All formulas necessary for the calculation of $Q^{tot}$ are applicable, as long as the rotational symmetry of the nuclei approximately holds.

The result is the following connection:

$$\begin{aligned}
&S_{sp,r} = (E_0/N_{abs}) \cdot \Phi_0 \cdot [2 \cdot \upsilon \cdot C_{heavy} \cdot F_r + G] \\
&\cdot Q^{tot}_{as}(medium) \cdot A_N^{1/3} / (Q^{tot}_{as}(oxygen) \cdot A_{oxygen}^{1/3}) \\
&F_r = \varphi \cdot [\varphi_1 + \varphi_2 \cdot \theta] \\
&\varphi = 0.5 \cdot (1 + erf(\arg)); \quad \varphi_1 = erf(\arg 1); \quad \varphi_2 = erf(\arg 2) \\
&\theta = \begin{pmatrix} e^{-1} - \exp(-(z-z_R)^2/z_R^2)/(e^{-1}-1) & (if \; z \leq z_R) \\ 1 & (else) \end{pmatrix} \\
&G = (c_1 \cdot z_{shift} \cdot \sqrt{\pi}/R_{CSDA}) \cdot \exp(-(\tfrac{1}{\sqrt{\pi}}R_{CSDA} - z)^2/z_R^2)
\end{aligned} \quad (182)$$

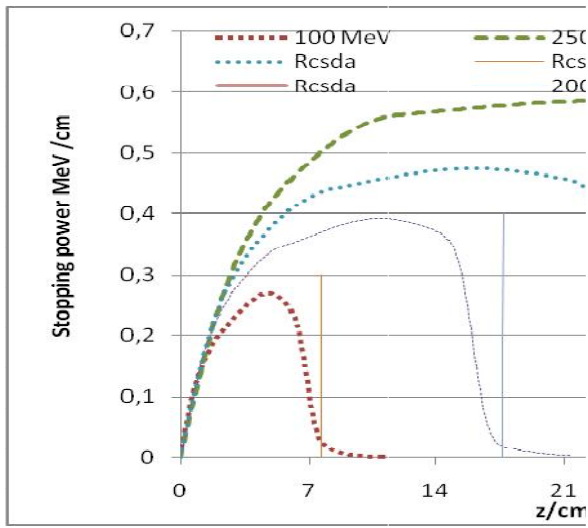

**Figure 40:** Stopping power of secondary/tertiary protons (+ further charged particles) induced by nuclear reactions of protons with oxygen.

## 3.3 Some consequences with respect to the determination of $Q^{tot}(E)$

The wave-functions applied in the previous section to determine $Q^{tot}(E)$ suggest to adapt Eqs. (46 - 49) in an analytical way by three Gaussian distributions and an error-function with regard to the asymptotic behavior [5 - 6]. This procedure enables to simplify the determination of the fluence decrease [28] with the help of Figures (5 - 9).

The presented results show that a suitable combination of the collective model with extended nuclear shell theory can be adequate to solve problems, which are rather outstanding in many practical problems. Besides the radiotherapy with protons it should be mentioned that the cross-sections of those nuclei/isotopes are important to reduce the half-times of the corresponding isotopes significantly. The storage of long-existing isotopes should be avoided. However, a thorough discussion of technical details of these processes goes beyond the scope of this study.

The total nuclear cross-section $Q^{tot}(E)$ of the proton - nucleus interaction is described by Eq. (183):

$$Q^{tot} = w_0 \cdot Q^{tot}{}_{as} \cdot erf(\tfrac{E-E_{Th}}{\sigma_{as}}) + w_{g0} \cdot \exp(-(\tfrac{E+\delta-E_{res}}{\sigma_{res}})^2)$$
$$+ w_{g1} \cdot A_m \cdot \exp(-(\tfrac{E-E_{res}}{\sigma_1})^2);$$
$$w_{g1} \cdot (1-A_m) \cdot \exp(-(\tfrac{E-\gamma-E_{res}}{\sigma_2})^2) + A_{boundary} \quad (183)$$

The parameters of Eq. (183) are given by Eqs. (184, 185) and Tables 9 and 10.

| Nucleus | $E_{Th}$/MeV | $E_{res}$/MeV | $Q^{tot}{}_{max}$/mb | $Q^{tot}{}_c$/mb | $Q^{tot}{}_{as}$/mb |
|---|---|---|---|---|---|
| C | 5.7433 | | 447.86 | | 247.64 |
| | | 17.5033 | | 426.91 | |
| O | 6.9999 | | 541.06 | | 299.79 |
| | | 20.1202 | | 517.31 | |
| Ca | 7.7096 | | 984.86 | | 552.56 |
| | | 25.2128 | | 954.82 | |
| Cu | 8.2911 | | 1341.94 | | 752.03 |
| | | 33.4733 | | 1308.07 | |
| Cs | 8.5620 | | 2009.71 | | 1254.93 |
| | | 51.2022 | | 1726.48 | |

**Table 9:** Some essential data of the energy E and $Q^{tot}(E)$ necessary in Eqs (183) and (184).

$$A_{boundary} = (1-A_m) \cdot wg_0 - A_m \cdot wg_{00} + (1-A_m) \cdot$$
$$(wg_0 - wg_{f_0}) \cdot erf(\tfrac{E-E_{Th}}{\sigma_{as}}) +$$
$$A_m \cdot erf(\tfrac{E-E_{Th}}{\sigma_{as}}) \cdot (wg_{00} - wg_{f_1}) + wg_{f_3} \cdot w_{Gauss} +$$
$$+ w_{Gauss} \cdot wg_0 - A_m \cdot wg_{00} + (1-A_m) \cdot (wg_0 - wg_{f_0}) \cdot erf(\tfrac{E-E_{Th}}{\sigma_{as}}) +$$
$$A_m \cdot erf(\tfrac{E-E_{Th}}{\sigma_{as}}) \cdot (wg_{00} - wg_{f_1}) + wg_{f_3} \cdot w_{Gauss} + w_{Gauss} \cdot (wg_{f_3} - wg_{f_2}) \quad (184)$$
$$wg_0 = \exp(-(\tfrac{E_{Th}-\gamma-E_{res}}{\sigma_2})^2); \; wg_{00} = \exp(-(\tfrac{E_{Th}-E_{res}}{\sigma_1})^2);$$
$$wg_{f_0} = \exp(-(\tfrac{E_{res}-E_f-\gamma}{\sigma_2})^2); \; E_f = 270\,MeV$$
$$wg_{f_1} = \exp(-(\tfrac{E_{res}-E_f}{\sigma_1})^2); \; wg_{f_2} = \exp(-(\tfrac{E_{res}+\delta-E_f}{\sigma_{res}})^2);$$
$$wg_{f_3} = \exp(-(\tfrac{E_{Th}-E_{res}+\delta}{\sigma_{res}})^2)$$

The parameters of Table 9 can be calculated by the parameters of Table 10 used in a function of Z and $A_N$ according to Eq. (185):

$$P_p = C_p \cdot Z^p / A_N^q \quad (185)$$

| Parameters $P_p$ in formula (194) | $C_p$ | p | Q |
|---|---|---|---|
| $w_{Gauss}$ | 36.05 | 1.421 | 1.811 |
| δ | 0.09335 | -1.621 | -0.405 |
| γ | -9.155 | 2.396 | 1.763 |
| $\sigma_{res}$ | 0.925 | -1.232 | -1.595 |
| $\sigma_1$ | 17.215 | 0.6375 | 0.31 |
| $\sigma_2$ | 11.575 | 1.13 | 0.38 |
| $\sigma_{as}$ | 1.074 | 1.745 | 2.102 |
| $A_m$ | 0.06257 | -1.102 | -1.335 |
| $E_{res}$ | 4.2064 | -0.7932 | 1.1561 |
| $w_{as}$ | 47.354 | 0.0055 | -0.6616 |

**Table 10:** Parameters $P_p$ of the cross-section formula (185).

Deconvolutions of cross-section Eq. (183) to determine $Q^{tot}(E)$ can be performed with regard to the elastic and to an inelastic part (change of the nuclear quantum numbers) of nuclear interactions (proton scatter), i.e. if $A_N$ and $Z$ remain constant! This means that if a proton hits the domain of strong interaction a part of the energy is transferred to the whole nucleus. Thus it is possible that the elastic scatter is the only interaction, but usually beside this part other excitations can occur: vibrations by deformations of the nucleus, rotations, and nuclear excited states. The proper nuclear reactions result from extremely high excited and/or virtual states. With reference to the complete expression (183), yielding $Q^{tot}(E)$, the operators $O^1(E)$ and $O^{-1}(E)$ are the tool to calculate $Q^{red}(E)$. However, this situation is in fact more difficult, since we need linear combinations of the operators with different half-widths. Therefore, we only give an outline in this section. Since the skeleton of Eq. (183) consists of three Gaussians and one error function, the action of $O^1$ and $O^{-1}$ can be summarized by the following way:

*(a) Gaussian distribution function*

$$O^{\pm 1}(\varepsilon, E) \cdot \exp(-[(E-\lambda_E)^2/\sigma^2]$$
$$= [\sigma/(\sigma^2 \pm(-\varepsilon^2))^{1/2}] \cdot \exp(-[(E-\lambda_E)^2/(\sigma^2 \pm(-\varepsilon^2))]) \quad (186)$$

Thus $\lambda_E$ represents an arbitrary parameter (e.g. a shift or threshold energy $E_{Th}$) in Eq. (195), and for $\varepsilon = \sigma$ the deconvolution operator $O^1(\varepsilon, E)$ converts the Gaussian distribution (196) to a δ-distribution function. Therefore we can summarize that the convolution operator $O^{-1}$ broadens a Gaussian distribution, whereas $O^1$ (deconvolution operator) affects an increased steepness of a Gaussian with decreasing half-width. It should be added that the 3 Gaussian terms appearing in $Q^{tot}(E)$ of Eq. (192) act with different σ-values.

*(b) Error function erf(ξ)*

As the error function erf(ξ) plays a significant role with regard to the asymptotic behavior of $Q^{tot}(E)$ it appears that the statement of the complete deconvolution term resulting from the expression (193) is justified. Now the above argument ξ refers to the substitution *ξ = (E-$E_{Th}$)/$\sigma_{as}$*, and the following terms result from a Lie series expansion:

$$O^1(\varepsilon, E) \cdot \text{erf}(E) = \text{erf}(E)$$

$$+ \tfrac{2}{\sqrt{\pi}} \cdot \sum_{n=0}^{\infty} (-1)^n \cdot H_{2n+1}(E) \cdot \exp(-E^2) \cdot (\varepsilon/\sigma_{as})^{2n} /(n! \cdot 4^n) \quad (187)$$

The polynomials of odd order $H_{2n+1}(E)$ refer to as Hermite polynomials, of which the recurrence formula is stated in [8, 31]. The result (187) yields an interesting property, known already from quantum mechanics, namely oscillations along the asymptotic behavior of $Q^{red}(E)$, which may be regarded as characteristic resonances.

Since deconvolutions are rather sensitive and in order to avoid artifacts, we have restricted ourselves with regard to parameter fixations in Eqs. (186 - 190). By that, $Q^{red}(E)$ may still contain 5 % - 10 % contributions of elastic and, above all, inelastic scatter of protons at nuclei. The principal difficulty comes from proper nuclear reactions changing $A_N$ or $Z$, because these processes always are closely connected with radiation transitions.

With regard to the deconvolution of $Q^{tot}(E)$ providing $Q^{red}(E)$, we have to account for different values for ε, and therefore we need a linear combination of operators $O^1$. Denoting this linear combination by the operator $A^1$, which reads:

$$\left. \begin{array}{l} A^1 = c \cdot O^1(\varepsilon, E) + c_1 \cdot O_1^{\,1}(\varepsilon_1, E) \\ + c_2 \cdot O_2^{\,1}(\varepsilon_2, E) + c_3 \cdot O_3^{\,1}(\varepsilon_3, E) \end{array} \right\} \quad (188)$$

The deconvolution calculation applied to $Q^{tot}(E)$ now becomes:

$$A^1 \cdot Q^{tot}(E) = Q^{red}(E) \quad (189)$$

The necessary parameters of $A^1$ are given by:

$$\left. \begin{array}{l} c = 0.4194 \; ; \; c_1 = 0.3211 \; ; \; c_2 = 0.1862 \; ; \\ c_3 = 0.0733 \; ; \; \varepsilon = 0.56 \cdot s_{as} \; ; \; \varepsilon_1 = 0.41 \cdot \varepsilon \; ; \\ \varepsilon_2 = 1.95 \cdot \varepsilon \; ; \; \varepsilon_3 = 0.59 \cdot \varepsilon \end{array} \right\} \quad (190)$$

### 3.4. Application of deconvolution methods to $Q^{tot}(E)$ of oxygen, carbon. calcium, copper, cesium and determination of $Q^{red}(E)$

A general aspect of all applications to nuclei stated above represents the tunneling effect of protons determined previously [61] via quantum mechanical methods. This effect consists of the property that protons with kinetic energy $E_{kin} < E_{Th}$ can pass through the potential barrier (Figure 41) and Table 9, where $E_{Th}$ is stated for the considered nuclei. However, this is only a noteworthy effect, if $E_{kin}$ of protons amounts to 90 % - 95 % of $E_{Th}$, otherwise the effect is very small and for $E_{kin} < 0.4 \cdot E_{Th}$ negligible. Figure 41 indicates that due the tunneling effect there is a small probability for a proton to surpass the potential barrier, which is 7 MeV for oxygen, but rather similar for other nuclei. Therefore Figure 41 may stand for all other nuclei considered in this study. What may happen for protons $E_{kin} < E_{Th}$? The smallest probability is that the proton can leave the strong interaction area by a further tunneling effect. Since the probability of the proton having undergone some reflections within $R_{strong}$ is extremely small to leave again the domain of strong interaction, then either *a. transitions to lower excited states with emission of γ-quanta* or *b. conversion of the proton to a neutron by exchange of a meson is possible (quantum mechanical exchange interaction due to the Pauli principle)*. The latter case b is the most probable one, since the neutron can pass through the potential wall and leave the strong interaction domain, if $E_{kin} > 0$. Then in the case of oxygen the residual nucleus is the isotope $F_9^{16}$, which undergoes ß$^+$ decay to yield again an oxygen nucleus. The behavior of the other nuclei under consideration is rather equivalent. Therefore we can summarize: If *$E_{kin} < E_{Th}$*, the most important reaction channel is stated for the nuclei studied in this communication.

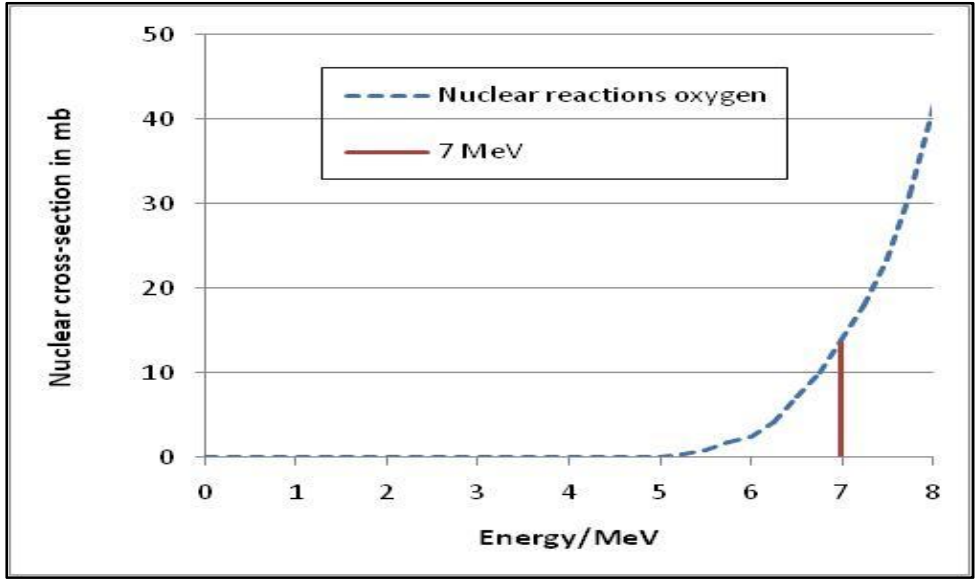

**Figure 41:** Nuclear cross-section $Q^{tot}(E)$ at $E_{kin} < 8$ MeV.

Figure 41 can make evident some possible nuclear reaction channels indicating the importance of the impinging angle of the proton and its history by preceding scatter processes (this is referred to as Molière multiple scatter at distances $R > R_N$ or $R \gg R_N$). For this purpose, we have split the cross-section in separated zones Z1, Z2, Z3 of a spherical nuclei. As examples we look at the positions $P_1$, $P_2$, $P_3$. In order to explain Figure 5 we use the abbreviations: $E_{pot}$ represents the excitation energy of nucleon to the state of positive energy $E = 0$ (a neutron release with $E_{kin,n} \geq 0$ must receive the energy $E > E_{pot}$); $E_{pot,wall}$ represents the potential energy at $R_{strong}$ (this is the minimum excitation energy of a proton release, i.e., $E_{pot,wall} = E_{pot} + E_{Th}$ and $E_{kin,p} \geq E_{pot,wall}$). In the following, we shall use the abbreviations: $D_1^2$ (deuterium) and $T_1^3$ (tritium); all other nuclei and their isotopes follow the usual denominations of the periodic system.

By regarding position $P_1$ of a proton we can verify that a nucleus can be hit in different zones and the distance from $P_1$ to a nucleus is rather far. By that, the angles between the three arrows are very small. If a nucleus exhibits approximately spherical symmetry, i.e. number of neutrons $N_n \approx Z$, the highest probability to hit the some nucleus is given by the outermost zone Z1, while it decreases significantly in direction to the central domain (Z1 → Z2 → Z3). With increasing energy of the impinging proton ($E > E_{pot}$), at first, a neutron $n$ can be released. The release of $2 \cdot n$, $p + n$, $D_1^2$, $T_1^3$ requires increasing energy again. If $N_n > Z$ (e.g. with regard to Cu) or $N_n \gg Z$ (e.g. $C_{55}^{137}$) the preference for neutron release growths. A hit at the next inner zone Z2 is only interesting for impinging protons with $E \gg E_{kin,p}$ to push out nucleonic fragments (clusters) such as α-particles or an isotope of Li. Otherwise ($E > E_{pot}$) a nucleus absorbs the impinging proton and after reorganization of the nucleonic states the behavior is equivalent to the outermost zone with a preference release of neutrons. The same fact is true for a hit of the impinging proton in the central domain (Z3), where the lowest differential cross-section has to be accounted for. If $E \gg E_{pot}$ a nuclear fission may occur with two fragments of about the same number of nucleons. Otherwise the behavior is comparable to the already discussed cases of the other zones. The positions $P_2$ and $P_3$ of the impinging protons have to be associated with the same probability behavior as for $P_1$. These both positions indicate that the impinging protons may have undergone further scatter characterized by the Molière scatter theory before they enter the domain $R < R_{Nucleus}$. However, $E_{proton}$ might already be lower than for $P_1$. The consequence of the diminished energy of impinging protons may result in a reduced kinetic energy of secondary particles, and since neutrons require the lowest energy transfer for their release, the $N_n$ may increase, whereas the release of fragments growths less.

With the help of Figures 14, 21 - 22 we are able to associate the energy for the possible nuclear reactions. Although both figures refer to the medium *'water'*, they can be transferred to other media/elements by their knowledge of $A_N$, Z and mass density $\rho$ due to the properties of BBE. However, we should like to point out that the simple substitution rule for molecules, namely $Z \rightarrow Z_{sub}$ and $A_N \rightarrow A_{N,sub}$ is rather insufficient. Thus for water we would obtain $Z_{sub} = 10$ and $A_{N,sub} = 18$, which would be identical to an isotope of the noble gas $Ne_{10}^{18}$. A quantum chemical calculation of the center of mass and center of charge for water provided $Z_{sub} = 9.02$ and $A_{N,sub} = 17.73$ [3, 4] with the ratio 0.5087422, whereas 10/18 yields 0,555556. This results implies that $A_N$ for water with $A_{N,sub} = 17.73$ is a good approach due to the mass number $2 \cdot Z$ of oxygen, but for the effective charge $Z_{sub} = 10$ is rather unrealistic. By use of the correct ratio for $Z_{sub}/A_{N,sub}$ for the reference system *'water'* conventional in therapy planning algorithms the transition to different media can be carried out more precisely. Taking these transitions into account we are now able to connect the zones (Figure 42) with the corresponding areas of Bragg curve (Figures 15, 21 - 24).

Thus the very small tunneling effect with release of a neutron for $E_{proton} < E_{Th}$ occurs at the Bragg peak with growing less in direction to the distal end. The emission of a neutron, and with decreasing probability the release of *2·n, n + p, $D_1^2$, $T_1^3$* takes place in the area of exponentially ascending slope. However, all other emissions of fragment, clusters, simultaneous emissions of some single neutrons or protons with clusters and/or fragments can only occur in the initial plateau!

With regard to the zones (Figure 42) analyzing the differential cross-section we can summarize: Figures 14, 21 - 22 serve for consecution of energy domains of specific nuclear reactions:

a. *Thus for $E \approx E_{Th}$ the proton has reached the Bragg peak and travels to distal end ($E \rightarrow 0$). Nuclear reactions* in *this energy domain referring to neutron release are classified by (10a - 10e) via exchange of mesons, if $R < R_{strong}$, but, in addition, $R \geq R_{Nucleus}$.*

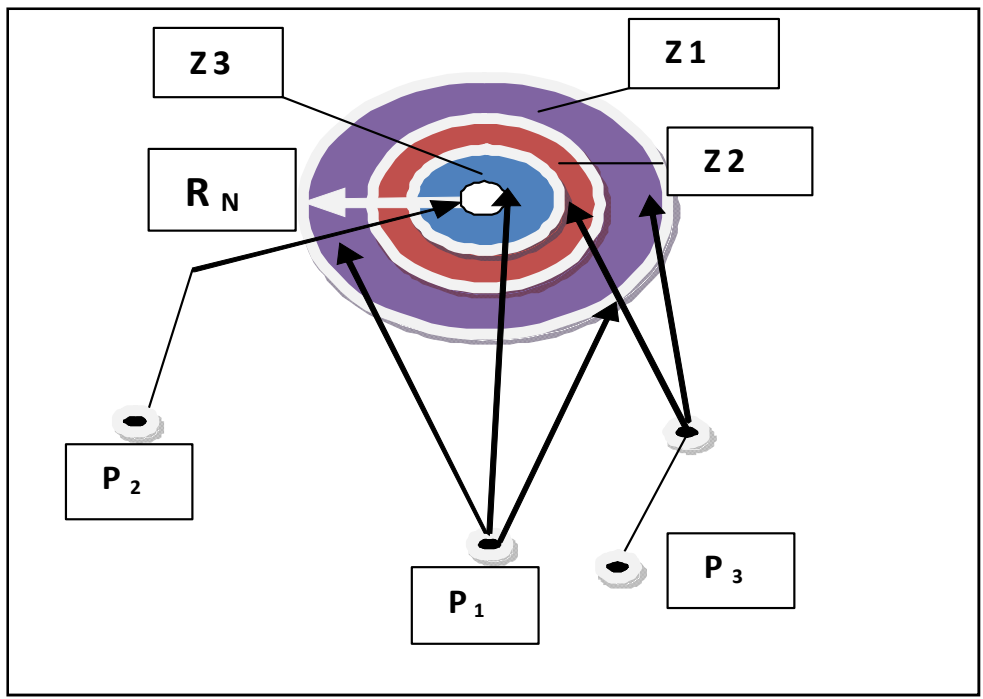

**Figure 42:** Some possible paths of protons from positions $P_1$, $P_2$, $P_3$ to hit a nucleus within different zones Z1, Z2, Z3 ($R_N = R_{Nucleus}$ according to Eq. (148)). The area of Z1 amounts to $\pi \cdot R_N^2 - 4 \cdot \pi \cdot R_N^2/9 = 5 \cdot \pi \cdot R_N^2/9$, the area of Z2 to $\pi \cdot R_N^2/3$, and that of Z3 to $\pi \cdot R_N^2/9$.

b. *The release of nucleons (preferably neutrons) from the outermost zone Z1 of Figure 42 represent the dominant part of secondary particles in dependence of the kinetic energy $E_{proton}$ of the impinging protons.*

c. *The inner zone Z2 and the central area Z3 may also emit bigger nuclear fragments and fissions besides the release of further nucleons.*

The total nuclear cross-section $Q^{red}(E)$ resulting from Figure 42 is given by:

$$'Q^{red}(E) = \int\int q_{Nuclear}(E, \theta, \varphi) \cdot \sin\theta \cdot d\theta\, d\varphi \quad (191)$$

For nuclei with spherical symmetry the integration over φ can be omitted. The total nuclear cross-section $Q^{tot}(E)$ accounts for all kinds of nuclear interactions (elastic scatter and inelastic scatter by changing some quantum numbers). The differences between $Q^{red}(E)$ and $Q^{tot}(E)$ are presented in Figures 43 - 47. In general, it can be mentioned that $Q^{red}(E)$ shows some characteristic resonance energies for nuclei with $N_n = Z$ with growing less tendency for $N_n > Z$ or $N_n \gg Z$ as true for $Cs_{55}^{137}$.

Figure 43 is most important for nuclear reactions of protons in water with regard to therapy planning algorithms, since $Q^{tot}(E)$ determines the behavior of Figure 43 (decrease if fluence of primary protons), whereas Figure 44 (carbon) basicly refers to organic molecules. A general feature of all figures is that the absolute maximum of $Q^{red}(E)$ is slightly shifted to a lower energy compared to the overall maximum of $Q^{tot}(E)$.

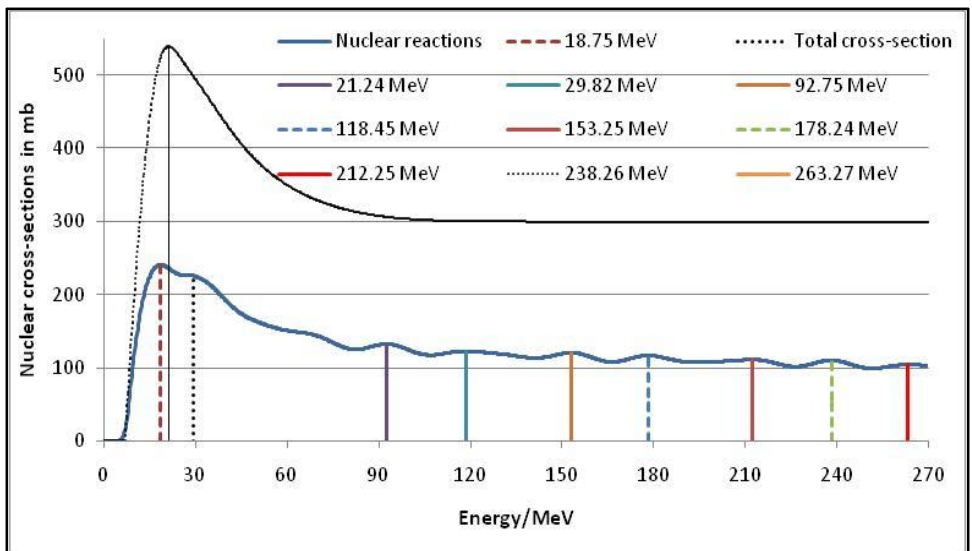

**Figure 43:** $Q^{tot}$ (dots, one maximum at 21.24 MeV)) and $Q^{red}$ (solid, 9 maxima from 18.75 MeV up to 263.27 MeV) of *oxygen*. For E > 270 MeV the existence of maxima is unimportant, since the asymptotic behavior is reached (the therapeutic proton energies basicly remain below 270 MeV).

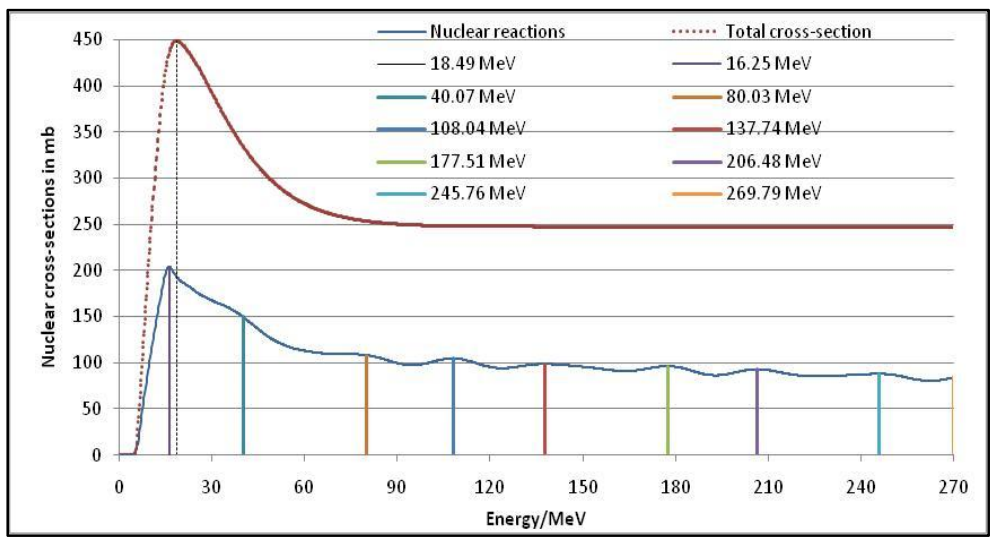

**Figure 44:** $Q^{tot}$ (dots, one maximum at 18.49 MeV) and $Q^{red}$ (solid, maxima at 16.25 MeV up 269.79 MeV) of *carbon*. If E > 270 MeV the asymptotic behavior is reached.

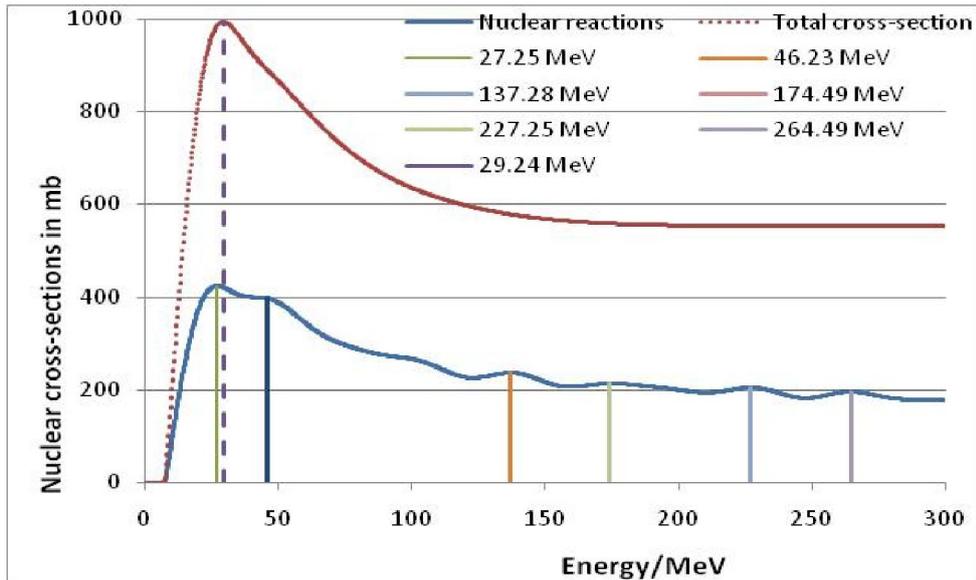

**Figure 45:** $Q^{tot}$ (dots, one maximum at 29.24 MeV) and $Q^{red}$ (solid, maxima at 27.25 MeV up to 264.49 MeV) of *calcium*.

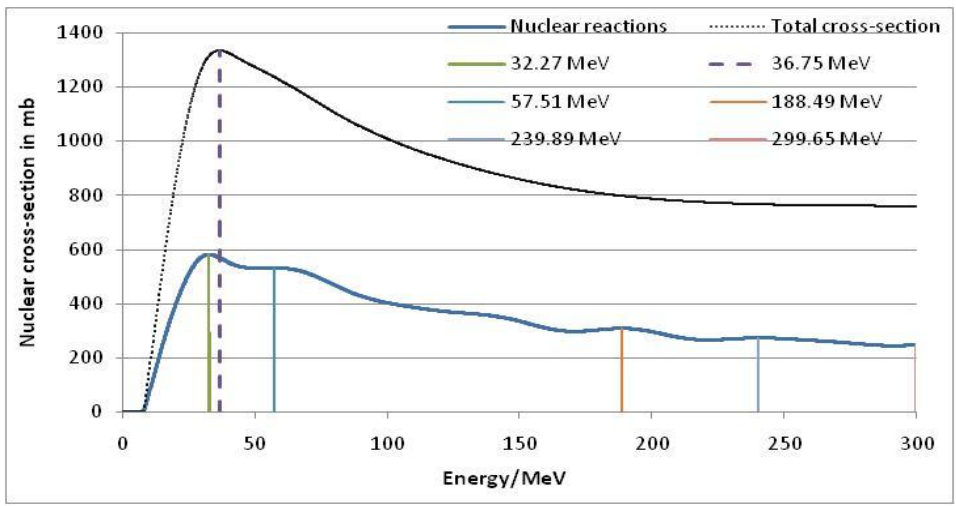

**Figure 46:** $Q^{tot}$ (dots, one maximum at 36.75 MeV) and $Q^{red}$ (solid, maxima at 32.27 MeV up to 299.65 MeV) of *copper*.

The behavior of secondary protons is clear: they all are able to undergo nuclear reactions with lower energies after traveling along the related medium. The reaction products (isotopes) of protons with nuclei (here oxygen) are denoted by *'heavy recoils'*, which undergo $\beta^+/\beta^-$ decay or electron capture with further emission of γ-quanta. However the neutrons show a rather different behavior, since their threshold energy does not exist. Some essential nuclear reactions have already previously described. Thus for the neutron energy E < 1 MeV which can be considered as 'thermal neutrons' the formation of $D_1^2$ with emission of γ-quanta is rather probable, whereas for higher energies preferably the collisions with hydrogen protons represent the main effect of neutron stopping and the formation of the oxygen isotope $O_8^{17}$ is less significant.

The interaction of neutrons with environmental electrons via magnetic dipole coupling of the related spins is nearly negligible with regard to neutron stopping, since the effect is only of higher order. With regard to Figures 43 - 47 and their maxima of the reaction channels we are able to state essential reaction types occurring in various energy domains:

$$p + X_m^{m'} \rightarrow n + X_{m+1}^{m'} + \gamma; \quad (192)$$

$$p + X_m^{m'} \rightarrow n + p + X_m^{m'-1} + \gamma; (192a)$$

$$p + X_m^{m'} \rightarrow 2 \cdot n + X_{m+1}^{m'-1} + \gamma; (192b)$$

$$p + X_m^{m'} \rightarrow D_1^2 + X_m^{m'-1} + \gamma. \quad (192c)$$

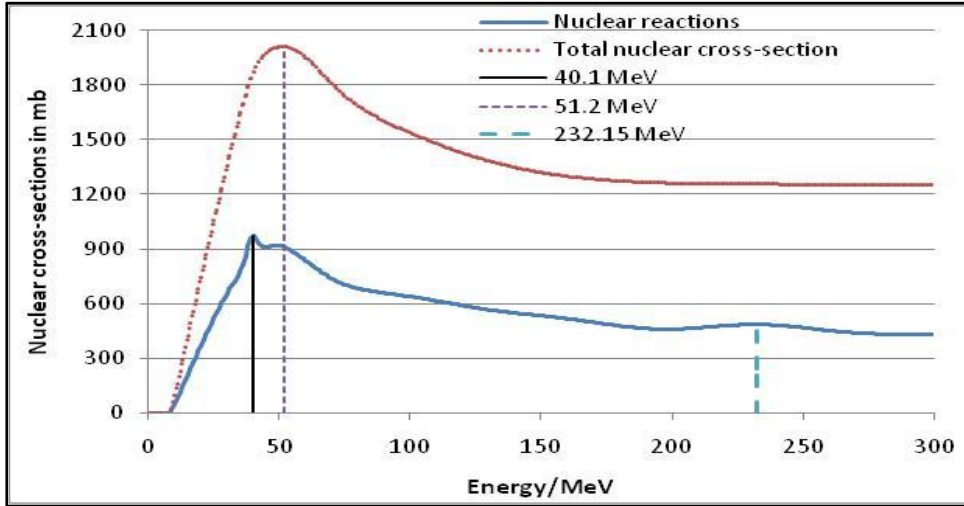

**Figure 47:** $Q^{tot}$ (dots, one maximum at 51.2 MeV) and $Q^{red}$ (solid, maxima at 40.103 MeV, 51.2 MeV, 232.15 MeV) of **cesium**.

In the survey of this appendix we shall abbreviate the reaction channel (192) with $R_0$ and the channels (192a) - (192c) with $R_{1a}$, $R_{1b}$, $R_{1c}$. Thus the reaction channel (192) is only possible for lower initial proton energies, whereas the cases (192a) - (192c) are very probable in the outermost zone Z1 of Figure 42 and they are able to occur even at proton energies E > 200 MeV. Consequently these cases form a certain underground, when other reaction channels exhibit a maximum according to Figures 43 - 47. Now we present the most probable nuclear reactions in dependence of the actual proton energy.

We wish point out two aspects referring to the role of nuclear reactions:

1. The release of neutrons preferably is of essential importance either in the domain of the Bragg peak and behind in direction of the distal end or in the area of exponential growing of the Bragg curve. Thus the algorithm of proton therapy planning systems and even Monte-Carlo calculations underrate the release of neutrons [3, 4, 33, 46, 60].

2. The presented method for the determination of $Q^{tot}(E)$ and $Q^{red}(E)$ only require Z and $A_N$ as basic parameters. Therefore the handling of nuclear scatter processes and nuclear reactions can be carried out via voxel data based on 3D CT information using the calibrations/substitutions:

$$Z \rightarrow Z_{effective} \text{ and } A_N \rightarrow A_{N,effective}. \quad (193)$$

It should be noted that the information given by the substitution (203) is already required by handling the interaction of protons with shell electrons due to BBE.

3. The reaction channels occurring with regard to $Cs_{55}^{137}$ certainly do not exhibit any therapeutic relevance, but it should be demonstrated that the calculation methods developed in this study are also applicable to technical problems such as *'transmutations'* of reactor products with long half-times. In such a situation it is superfluous to deflect the proton beam before the nozzle by suitable magnets to obtain a narrow energy spectrum of the impinging protons and a rather broad energy spectrum of the proton beam is desired. Since the nuclear cross-sections $Q^{tot}(E)$ and $Q^{red}(E)$ refer to single hit processes, the yield of nuclear reactions can be strengthened by high intensity of projectile proton beams, e.g. if the first target hit leads to nucleus vibrations or excitations, and an immediate following second target hit can be able to transfer proton energy to an excited nuclear state, where the range $R_{Nuclear}$ is somewhat increased. In order to reach an optimal efficacy of the transmutations, it is necessary to perform these reactions under extremely high density concentrations with the aim to operate beyond single hit events.

### 3.5 Nuclear reactions channels in dependence of the actual proton energy and the role of resonances in Figures 43 - 47

Owing to the deconvolution procedure applied in this study we have been able to separate the energy transfer of the projectile proton to elastic/inelastic scatter and that part being in connection with nuclear reaction channels. The above mentioned figures indicate that the number of maxima decreases with increasing Z and $A_N$, and, above all, the property $N_n > Z$ and $N_n \gg Z$ seems to be decisive. Thus the cross-sections according to Figures 43 and 44 are important in irradiation of soft tissue targets. Figure 45 is important, when either a proton passes bone tissue or the target is incorporated by a bone. Figures 46 and 47 basically are of interest in technical applications (beam-guide of proton beams and transmutations). A general feature of rather low proton energies, i.e. $E_{prot} < E_{pot}$, is the exchange interaction between **proton** and a nucleon (denoted by X) due to meson exchange with a particular role of Pauli principle of the overall system:

$$p + X_m^{m'} : \left| \begin{array}{c} proton \Leftrightarrow nucleus \\ meson \ exchange \end{array} \right\rangle \to n + X_{m+1}^{m'}. \quad (194)$$

The resulting 'heavy recoil' $X_{m+1}^{m'}$ can either undergo a ß$^+$ decay (194) or become immediately a stable isotope as true in the case (196a). This reaction type always occurs with regard to every nucleus, if the available proton energy is $E_{prot} < E_{pot,wall}$ for protons and $E_{prot} < E_{pot}$ for
neutrons. In Figures 43 and 44 the first maxima ($C_6^{12}$ at 16.25 MeV and $O_8^{16}$ at 18.75 MeV) are exclusively determined by (201), whereas for $Ca_{20}^{40}$ (Figure 45, 27.25 MeV) and the cases of Figure 46 ($Cu_{29}^{63,64}$, 36.75 MeV) and Figure 47 ($Cs_{55}^{137}$, 40.1 MeV) the first maxima include further reactions. Therefore only the first maximum of carbon related to (202a) is peculiar, since a second maximum of $Q^{red}$ in the immediate environ does not exist and $Q^{red}$ continuously is growing down to 80.03 MeV. This behavior results from specific properties of this nucleus with the smallest $R_{Nucleus}$ considered here favoring higher fractions of elastic/inelastic scatter. The very small second maximum of $O_8^{16}$ (29.82 MeV) enables the reactions (201a) and (196). All other cases (Figures 43 - 47) contain the reactions within the first maxima (192a) and (192b), and the second maximum also includes the reaction (192c). In the following, we use the normalization value *'1'* with regard to the weight of all possible nuclear reactions up to proton energies of 270 MeV/300 MeV, and we recall that $\boldsymbol{R_0}$ refers to the channel (194) and $\boldsymbol{R_{1a}, R_{1b}, R_{1c}}$ to (201a - 201c). The types $\boldsymbol{R_{1a,b,c}}$ occur with regard to the total energy domain, and we attribute them rather higher relevance. Tables 11 - 15 present the nuclear reaction channels of the reaction p + X → secondary particles + *heavy recoils* and their possible decay products depending of the actual proton energy $E_{proton}$ ($E_p$), which is stated within boundaries. If the weight of a channel fraction is ≤ 0.001, then it is ignored and neighboring fraction is corresponding rounded or added to a suitable channel; the extremely small fractions of $R_0$ and $R_1$ in Z3 is added to Z1. The specific weight of each channel is stated in the related parenthesis, if more than one channel is possible. In the following tables we also shall use the abbreviations:

$p + X_m^{m'} \to D_1^2 + p + X_{m-1}^{m'-2} + \gamma$ (*$R_{1d}$*); $p + X_m^{m'} \to D_1^2 + n + X_m^{m'-2} + \gamma$ (*$R_{1e}$*); $p + X_m^{m'} \to T_1^3 + X_m^{m'-2} + \gamma$ (*$R_{2a}$*); $p + X_m^{m'} \to T_1^3 + p + X_{m-1}^{m'-3} + \gamma$ (*$R_{2b}$*); $p + X_m^{m'} \to T_1^3 + n + X_m^{m'-3} + \gamma$ (*$R_{2c}$*); $p + X_m^{m'} \to \alpha + X_{m-1}^{m'-3} + \gamma$ (*$R_{2d}$*); $p + X_m^{m'} \to \alpha + n + X_{m-1}^{m'-4} + \gamma$ (*$R_{2e}$*); $p + X_m^{m'} \to \alpha + p + n + X_{m-2}^{m'-5} + \gamma$ (*$R_{2f}$*); $p + X_m^{m'} \to \alpha + 2n + X_{m-1}^{m'-5} + \gamma$ (*$R_{2g}$*); $p + X_m^{m'} \to \alpha + p + 2n + X_{m-2}^{m'-6} + \gamma$ (*$R_{2h}$*); $p + X_m^{m'} \to Li_3^5 + X_{m-2}^{m'-4} + \gamma$ (*$R_{3a}$*); $p + X_m^{m'} \to Li_3^6 + X_{m-2}^{m'-5} + \gamma$ (*$R_{3b}$*); $p + X_m^{m'} \to Li_3^7 + X_{m-2}^{m'-6} + \gamma$ (*$R_{3c}$*); $p + X_m^{m'} \to Li_3^5 + n + X_{m-2}^{m'-5} + \gamma$ (*$R_{3d}$*); $p + X_m^{m'} \to Be_4^7 + X_{m-3}^{m'-6} + \gamma$ (*$R_{4a}$*); $p + X_m^{m'} \to Be_4^8 + X_{m-3}^{m'-7} + \gamma$ (*$R_{4b}$*); $p + X_m^{m'} \to Be_4^9 + X_{m-3}^{m'-8} + \gamma$ (*$R_{4c}$*).

| Z1: $E_{proton}$/MeV lower/upper boundaries | $w_1$ | channel type |
|---|---|---|
| $0 > E_p \leq 22.95$ | 0.08 | $R_0$ |
| $22.95 > E_p \leq 65.03$ | 0.07 | $R_0$ (0.02), $R_{1a}$ (0.02), $R_{1b}$ (0.02), $R_{1c}$ (0.01) |
| $65.03 > E_p \leq 105.17$ | 0.07 | $R_0$ (0.01), $R_{1a}$ (0.02), $R_{1b}$ (0.02), $R_{1c}$ (0.01), $R_{1d}$ (0.01) |
| $105.17 > E_p \leq 138.21$ | 0.07 | $R_{1a}$ (0.01), $R_{1b}$ (0.01), $R_{1c}$ (0.01), $R_{1d}$ (0.02), $R_{1e}$ (0.02) |
| $138.21 > E_p \leq 176.34$ | 0.06 | $R_{1c}$ (0.01), $R_{1d}$ (0.01), $R_{1e}$ (0.01), $R_{2a}$ (0.03) |
| $176.34 > E_p \leq 191.58$ | 0.05 | $R_{1e}$ (0.01), $R_{2a}$ (0.02), $R_{2b}$ (0.02) |
| $191.58 > E_p \leq 206.72$ | 0.04 | $R_{1e}$ (0.01), $R_{2a}$ (0.01), $R_{2b}$ (0.01), $R_{2c}$ (0.01) |
| $206.72 > E_p \leq 222.30$ | 0.04 | $R_{2a}$ (0.01), $R_{2b}$ (0.01), $R_{2c}$ (0.02) |
| $235.64 > E_p \leq 254.92$ | 0.04 | $R_{2a}$ (0.01), $R_{2b}$ (0.01), $R_{2c}$ (0.01), $R_{2d}$ (0.01) |
| $254.92 > E_p \leq 270.00$ | 0.04 | $R_{2b}$ (0.01), $R_{2c}$ (0.01), $R_{2d}$ (0.01), $R_{2e}$ (0.01) |
| Z2: $E_{proton}$/MeV lower/upper boundaries | $w_2$ | channel type |
| $0 > E_p \leq 22.95$ | - | - |
| $22.95 > E_p \leq 65.03$ | 0.05 | $R_{1a}$ (0.01), $R_{1b}$ (0.02), $R_{1c}$ (0.02) |
| $65.03 > E_p \leq 105.17$ | 0.05 | $R_{1b}$ (0.01), $R_{1c}$ (0.02), $R_{1e}$ (0.02) |
| $105.17 > E_p \leq 138.21$ | 0.05 | $R_{1b}$ (0.01), $R_{1c}$ (0.01), $R_{2a}$ (0.01), $R_{2b}$ (0.02) |
| $138.21 > E_p \leq 176.34$ | 0.05 | $R_{2a}$ (0.01), $R_{2b}$ (0.01), $R_{2c}$ (0.01), $R_{2d}$ (0.02) |
| $176.34 > E_p \leq 191.58$ | 0.04 | $R_{2b}$ (0.01), $R_{2c}$ (0.01), $R_{2d}$ (0.01), $R_{2e}$ (0.01) |
| $191.58 > E_p \leq 206.72$ | 0.03 | $R_{2b}$ (0.005), $R_{2c}$ (0.005), $R_{2d}$ (0.005), $R_{2e}$ (0.005), $R_{2f}$ (0.01) |
| $206.72 > E_p \leq 222.30$ | 0.02 | $R_{2b}$ (0.004), $R_{2c}$ (0.004), $R_{2d}$ (0.004), $R_{2e}$ (0.004), $R_{2f}$ (0.002), $R_{2g}$ (0.002) |
| $235.64 > E_p \leq 254.92$ | 0.02 | $R_{2b}$ (0.002), $R_{2c}$ (0.002), $R_{2d}$ (0.003), $R_{2e}$ (0.003), $R_{2f}$ (0.004), $R_{2g}$ (0.004), $R_{2h}$ (0.002) |
| $254.92 > E_p \leq 270.00$ | 0.02 | $R_{2b}$ (0.002), $R_{2c}$ (0.002), $R_{2d}$ (0.002), $R_{2e}$ (0.002), $R_{2f}$ (0.003), $R_{2g}$ (0.003), $R_{2h}$ (0.003). $R_{3a}$ (0.003) |
| Z3: $E_{proton}$/MeV lower/upper boundaries | $w_3$ | channel type |
| $0 > E_p \leq 22.95$ | - | - |
| $22.95 > E_p \leq 65.03$ | 0.010 | $R_{1b}$ (0.004), $R_{1c}$ (0.004), $R_{1d}$ (0.002) |
| $65.03 > E_p \leq 105.17$ | 0.010 | $R_{1b}$ (0.003), $R_{1c}$ (0.003), $R_{1d}$ (0.002), $R_{1e}$ (0.002) |
| $105.17 > E_p \leq 138.21$ | 0.010 | $R_{1b}$ (0.002), $R_{1c}$ (0.002), $R_{1d}$ (0.002), $R_{1e}$ (0.002), $R_{2a}$ (0.002) |
| $138.21 > E_p \leq 176.34$ | 0.010 | $R_{1d}$ (0.002), $R_{1e}$ (0.002), $R_{2a}$ (0.002), $R_{2b}$ (0.002), $R_{2c}$ (0.002) |
| $176.34 > E_p \leq 191.58$ | 0.010 | $R_{2d}$ (0.002), $R_{2e}$ (0.002), $R_{2f}$ (0.002), $R_{2g}$ (0.002), $R_{2h}$ (0.002) |
| $191.58 > E_p \leq 206.72$ | 0.010 | $R_{2d}$ (0.001), $R_{2e}$ (0.001), $R_{2f}$ (0.001), $R_{2g}$ (0.001), $R_{2h}$ (0.002), $R_{3a}$ (0.002), $R_{3b}$ (0.001), $R_{3c}$ (0.001) |
| $206.72 > E_p \leq 222.30$ | 0.015 | $R_{2g}$ (0.002), $R_{2h}$ (0.002), $R_{3a}$ (0.002), $R_{3b}$ (0.003), $R_{3c}$ (0.003), $R_{3d}$ (0.003) |
| $235.64 > E_p \leq 254.92$ | 0.015 | $R_{3a}$ (0.002), $R_{3b}$ (0.002), $R_{3c}$ (0.002), $R_{3d}$ (0.003), $R_{4a}$ (0.003), $R_{4b}$ (0.003) |
| $254.92 > E_p \leq 270.00$ | 0.020 | $R_{3a}$ (0.003), $R_{3b}$ (0.003), $R_{3c}$ (0.003), $R_{3d}$ (0.003), $R_{4a}$ (0.003), $R_{4b}$ (0.003), $R_{4c}$ (0.002) |

**Table 11:** List of nuclear reactions of the proton - oxygen interaction with $m = 8$ and $m' = 16$ yielding $p + O_8^{16} \rightarrow X_{m-q}^{m'-r}$ + *secondary particles* and their weights. The overall weights of the zones Z1, Z2. Z3 amount to Z1: w = 0.56; Z2: w = 0.33; Z3: w = 0.11.

As already pointed out, the heavy recoils of oxygen ($O_{8-q}^{16-r}$) usually undergo ß$^+$-decay; a rather important case is the decay reaction (204) $F_9^{16} \rightarrow O_8^{16} + e_0^+$ with $T_{1/2} = 22$ sec. With regard to the recoils of oxygen, all decay processes are rather fast (i.e. less than 10 minutes). Since the heavy recoils stated Table11, their decay products and half-times can be taken from web, we do not intend to present this subject of matter here. The noteworthy neutron release may also imply reactions of neutrons with oxygen, and an important case is the channel:

$$n + O_8^{16} \rightarrow p + N_7^{16} + \gamma; \; (\beta^- : N_7^{16} \rightarrow O_8^{16} + e_0^- + \gamma, T_{1/2} = 120 \text{ sec} \quad (195)$$

Reactions of the type (195) occur by many modifications in those heave recoils resulting from nuclei with $Z < N_n$, e.g. $Cu_{29}^{63}$ or $Cs_{55}^{137}$.

| Z1: $E_{proton}$/MeV lower/upper boundaries | $w_1$ | channel type |
|---|---|---|
| $0 > E_p / \leq 20.92$ | 0.10 | $R_0$ |
| $20.92 > E_p \leq 42.47$ | 0.09 | $R_0$ (0.03), $R_{1a}$ (0.03), $R_{1b}$ (0.02), $R_{1c}$ (0.01) |
| $42.47 > E_p \leq 92.65$ | 0.08 | $R_0$ (0.02), $R_{1a}$ (0.02), $R_{1b}$ (0.02), $R_{1c}$ (0.01), $R_{1d}$ (0.01) |
| $92.65 > E_p / \leq 122.28$ | 0.07 | $R_{1a}$ (0.02), $R_{1b}$ (0.02), $R_{1c}$ (0.01), $R_{1d}$ (0.01), $R_{1e}$ (0.01) |
| $122.28 > E_p \leq 164.17$ | 0.06 | $R_{1c}$ (0.02), $R_{1d}$ (0.02), $R_{1e}$ (0.01), $R_{2a}$ (0.01) |
| $164.17 > E_p \leq 192.53$ | 0.05 | $R_{1e}$ (0.01), $R_{2a}$ (0.02), $R_{2b}$ (0.01), $R_{2c}$ (0.01) |
| $192.53 > E_p \leq 224.65$ | 0.04 | $R_{1e}$ (0.01), $R_{2a}$ (0.01), $R_{2b}$ (0.01), $R_{2c}$ (0.01) |
| $224.65 > E_p \leq 255.37$ | 0.04 | $R_{2b}$ (0.01), $R_{2c}$ (0.01), $R_{2d}$ (0.01), $R_{2e}$ (0.01) |
| $255.37 > E_p \leq 270.00$ | 0.03 | $R_{2b}$ (0.01), $R_{2c}$ (0.01), $R_{2d}$ (0.005), $R_{2e}$ (0.005) |
| **Z2: $E_{proton}$/MeV lower/upper boundaries** | $w_2$ | channel type |
| $0 > E_p \leq 20.92$ | - | - |
| $20.92 > E_p \leq 42.47$ | 0.05 | $R_0$ (0.01), $R_{1a}$ (0.01), $R_{1b}$ (0.015), $R_{1c}$ (0.015) |
| $42.47 > E_p \leq 92.65$ | 0.04 | $R_{1b}$ (0.01), $R_{1c}$ (0.02), $R_{1d}$ (0.01) |
| $92.65 > E_p \leq 122.28$ | 0.04 | $R_{1b}$ (0.01), $R_{1c}$ (0.01), $R_{1d}$ (0.01), $R_{1e}$ (0.01) |
| $122.28 > E_p \leq 164.17$ | 0.04 | $R_{2a}$ (0.01), $R_{2b}$ (0.01), $R_{2c}$ (0.01), $R_{2d}$ (0.01) |
| $164.17 > E_p \leq 192.53$ | 0.04 | $R_{2b}$ (0.005), $R_{2c}$ (0.005), $R_{2d}$ (0.01), $R_{2e}$ (0.02) |
| $192.53 > E_p \leq 224.65$ | 0.04 | $R_{2b}$ (0.005), $R_{2c}$ (0.005), $R_{2d}$ (0.005), $R_{2e}$ (0.005), $R_{2f}$ (0.01), $R_{2g}$ (0.01) |
| $224.65 > E_p \leq 255.37$ | 0.04 | $R_{2b}$ (0.005), $R_{2c}$ (0.005), $R_{2d}$ (0.005), $R_{2e}$ (0.005), $R_{2f}$ (0.005), $R_{2g}$ (0.005), $R_{2h}$ (0.01) |
| $255.37 > E_p \leq 270.00$ | 0.04 | $R_{2b}$ (0.004), $R_{2c}$ (0.004), $R_{2d}$ (0.004), $R_{2e}$ (0.004), $R_{2f}$ (0.004), $R_{2g}$ (0.005), $R_{2h}$ (0.005). $R_{3a}$ (0.01) |
| **Z3: $E_{proton}$/MeV lower/upper boundaries** | $w_3$ | channel type |
| $0 > E_p \leq 20.92$ | - | - |
| $20.92 > E_p \leq 42.47$ | 0.015 | $R_{1b}$ (0.003), $R_{1c}$ (0.003), $R_{1d}$ (0.003), $R_{1e}$ (0.003), $R_{2a}$ (0.003) |
| $42.47 > E_p / \leq 92.65$ | 0.015 | $R_{1b}$ (0.003), $R_{1c}$ (0.003), $R_{1d}$ (0.003), $R_{1e}$ (0.002), $R_{2a}$ (0.002), $R_{2b}$ (0.002) |
| $92.65 > E_p \leq 122.28$ | 0.015 | $R_{1b}$ (0.002), $R_{1c}$ (0.002), $R_{1d}$ (0.002), $R_{1e}$ (0.003), $R_{2a}$ (0.002), $R_{2b}$ (0.002), $R_{2c}$ (0.002) |
| $122.28 > E_p / \leq 164.17$ | 0.015 | $R_{1d}$ (0.002), $R_{1e}$ (0.002), $R_{2a}$ (0.002), $R_{2b}$ (0.003), $R_{2c}$ (0.003), $R_{2d}$ (0.002) |
| $164.17 > E_p \leq 192.53$ | 0.015 | $R_{2d}$ (0.002), $R_{2e}$ (0.002), $R_{2f}$ (0.002), $R_{2g}$ (0.003), $R_{2h}$ (0.003), $R_{3a}$ (0.002) |
| $192.53 > E_p / \leq 224.65$ | 0.015 | $R_{2g}$ (0.002), $R_{2h}$ (0.002), $R_{3a}$ (0.002), $R_{3b}$ (0.003), $R_{3c}$ (0.003), $R_{3d}$ (0.002) |
| $224.65 > E_p \leq 255.37$ | 0.01 | $R_{3a}$ (0.002), $R_{3b}$ (0.002), $R_{3c}$ (0.003), $R_{3d}$ (0.003) |
| $255.37 > E_p \leq 270.00$ | 0.01 | $R_{3a}$ (0.0015), $R_{3b}$ (0.0015), $R_{3c}$ (0.003), $R_{3d}$ (0.004) |

**Table 12:** List of nuclear reactions of the proton - carbon interaction with *m =6 and m' = 12* yielding ***p + $C_6^{12}$ → $X_{6-q}^{12-r}$ +** *secondary particles*, weights and decay products. The overall weights of the zones Z1, Z2. Z3 amount to Z1: $w_1 = 0.56$; Z2: $w_2 = 0.33$; Z3: $w_3 = 0.11$.

With regard to the following nuclei with an increased number of Z and, above all, $N_n$ we need additional abbreviations for nuclear reaction channels in order to be able to represent all additional possibilities by a reasonable way. The purpose of the reaction channels ($R_{5a}$) to ($R_{6k}$) is to account for fissions of the nuclei $Ca_{20}^{40}$, $Cu_{29}^{63}$ and $Cs_{55}^{137}$ to two nuclei with about the half of Z and the half of $N_n$. With regard to the comparably light nuclei $C_6^{12}$ and $O_8^{16}$ this problem is already accounted for by considering the isotopes of *lithium* and *oxygen* according to the reaction channels ($R_{3a}$) - ($R_{4d}$):

$p + X_m^{m'} \rightarrow Be_4^7 + n + X'_{m-3}^{m'-7} + \gamma$ ($R_{4d}$); $p + X_m^{m'} \rightarrow Be_4^7 + 2n + X'_{m-3}^{m'-8} + \gamma$ ($R_{4e}$);

$p + X_m^{m'} \rightarrow Be_4^8 + n + X'_{m-3}^{m'-8} + \gamma$ ($R_{4f}$); $p + X_m^{m'} \rightarrow Be_4^8 + 2n + X'_{m-3}^{m'-9} + \gamma$ ($R_{4g}$);

$p + X_m^{m'} \rightarrow Be_4^9 + n + X'_{m-3}^{m'-9} + \gamma$ ($R_{4h}$); $p + X_m^{m'} \rightarrow Be_4^9 + 2n + X'_{m-3}^{m'-10} + \gamma$ ($R_{4i}$).

**If m and m' even ($Ca_{20}^{40}$):**

$p + X_m^{m'} \rightarrow \gamma + 2 \cdot X_{m/2}^{m'/2} + p$ ($R_{5a}$); $p + X_m^{m'} \rightarrow \gamma + 2 \cdot X_{m/2}^{(m'/2)-1} + p + n$ ($R_{5b}$);

$p + X_m^{m'} \rightarrow \gamma + 2 \cdot X_{m/2}^{(m'/2)-1} + D_1^2$ ($R_{5c}$); $p + X_m^{m'} \rightarrow \gamma + 2 \cdot X_{m/2-1}^{(m'/2)-1} + 2p + n$ ($R_{5d}$); $p + X_m^{m'} \rightarrow \gamma + 2 \cdot X_{m/2-1}^{(m'/2)-1} + He_2^3$ ($R_{5e}$); $p + X_m^{m'} \rightarrow \gamma + 2 \cdot X_{m/2-1}^{(m'/2)-2} + 2p + 2n$ ($R_{5f}$); $p + X_m^{m'} \rightarrow \gamma + 2 \cdot X_{m/2-1}^{(m'/2)-2} + \alpha$ ($R_{5g}$).

**If m and m' odd ($Cu_{29}^{63}$, $Cs_{55}^{137}$):**

$p + X_m^{m'} \rightarrow \gamma + p + X_{(m-1)/2}^{(m'-1)/2} + X_{(m+1)/2}^{(m'+1)/2}$ ($R_{6a}$); $p + X_m^{m'} \rightarrow \gamma + p + n + X_{(m-1)/2}^{m'/2} + X_{(m+1)/2}^{m'/2}$ ($R_{6b}$);

$p + X_m^{m'} \rightarrow \gamma + D_1^2 + X_{(m-1)/2}^{m'/2} + X_{(m+1)/2}^{m'/2}$ ($R_{6c}$); $p + X_m^{m'} \rightarrow \gamma + 2p + n + X_{(m-1)/2}^{m'/2} + X_{(m-1)/2}^{m'/2}$ ($R_{6d}$);

$p + X_m^{m'} \rightarrow \gamma + He_2^3 + X_{(m-1)/2}^{m'/2} + X_{(m-1)/2}^{m'/2}$ ($R_{6e}$); $p + X_m^{m'} \rightarrow \gamma + 2p + 2n + X_{(m-1)/2}^{(m'-1)/2} + X_{(m-1)/2}^{(m'-1)/2}$ ($R_{6f}$);

$p + X_m^{m'} \rightarrow \gamma + \alpha + X_{(m-1)/2}^{(m'-1)/2} + X_{(m-1)/2}^{(m'-1)/2}$ ($R_{6g}$); $p + X_m^{m'} \rightarrow \gamma + 2p + 3n + X_{(m-1)/2}^{(m'-1)/2} + X_{(m-1)/2}^{(m'-3)/2}$ ($R_{6h}$);

$p + X_m^{m'} \rightarrow \gamma + \alpha + n + X_{(m-1)/2}^{(m'-1)/2} + X_{(m-1)/2}^{(m'-3)/2}$ ($R_{6i}$); $p + X_m^{m'} \rightarrow \gamma + 2p + 4n + X_{(m-1)/2}^{(m'-3)/2} + X_{(m-1)/2}^{(m'-3)/2}$ ($R_{6j}$);

$p + X_m^{m'} \rightarrow \gamma + \alpha + 2n + X_{(m-1)/2}^{(m'-3)/2} + X_{(m-1)/2}^{(m'-3)/2}$ ($R_{6k}$);

One has to be aware of that the sequences ($R_{5a}$) - ($R_{6d}$) may suitably be continued. However, we shall account for higher order terms by adding them to the nearest neighbor.

| Z1: $E_{proton}$/MeV | | |
|---|---|---|
| lower/upper boundaries | $w_1$ | channel type |
| $0 > E_p \leq 30.85$ | 0.10 | $R_0$ (0.08), $R_{1a}$ (0.02) |
| $30.85 > E_p \leq 68.13$ | 0.10 | $R_{1a}$ (0.02), $R_{1b}$ (0.02), $R_{1c}$ (0.02), $R_{1d}$ (0.02), $R_{1e}$ (0.02) |
| $68.13 > E_p \leq 155.39$ | 0.10 | $R_{1c}$ (0.01), $R_{1d}$ (0.01), $R_{1e}$ (0.02), $R_{2a}$ (0.02), $R_{2b}$ (0.02), $R_{2c}$ (0.02) |
| $155.39 > E_p \leq 199.64$ | 0.09 | $R_{2a}$ (0.01), $R_{2b}$ (0.01), $R_{2c}$ (0.01), $R_{2d}$ (0.01), $R_{2e}$ (0.01), $R_{2f}$ (0.02), $R_{2g}$ (0.01), $R_{2h}$ (0.01) |
| $199.64 > E_p \leq 242.03$ | 0.09 | $R_{2c}$ (0.01), $R_{2d}$ (0.01), $R_{2e}$ (0.01), $R_{2f}$ (0.01), $R_{2g}$ (0.01), $R_{2h}$ (0.01), $R_{3a}$ (0.02), $R_{3b}$ (0.01) |
| $242.03 > E_p \leq 300.00$ | 0.08 | $R_{2d}$ (0.01), $R_{2e}$ (0.01), $R_{2f}$ (0.01), $R_{2g}$ (0.01), $R_{2h}$ (0.01), $R_{3a}$ (0.01), $R_{3b}$ (0.01), $R_{3c}$ (0.01) |
| Z2: $E_{proton}$/MeV | | |
| lower/upper boundaries | $w_2$ | channel type |

| Z2 (cont.) $E_{proton}$/MeV lower/upper boundaries | $w_2$ | channel type |
|---|---|---|
| $0 > E_p / \leq 30.85$ | - | - |
| $30.85 > E_p \leq 68.13$ | 0.08 | $R_{1b}$ (0.01), $R_{1c}$ (0.01), $R_{1d}$ (0.01), $R_{1e}$ (0.01), $R_{2a}$ (0.01), $R_{2b}$ (0.01), $R_{2c}$ (0.02) |
| $68.13 > E_p \leq 155.39$ | 0.08 | $R_{2c}$ (0.01), $R_{2d}$ (0.01), $R_{2e}$ (0.01), $R_{2f}$ (0.01), $R_{2g}$ (0.01), $R_{2h}$ (0.01), $R_{3a}$ (0.01), $R_{3b}$ (0.01) |
| $155.39 > E_p \leq 199.64$ | 0.07 | $R_{3a}$ (0.01), $R_{3b}$ (0.01), $R_{3c}$ (0.01), $R_{3d}$ (0.01), $R_{4a}$ (0.01), $R_{4b}$ (0.01), $R_{4c}$ (0.01) |
| $199.64 > E_p \leq 242.03$ | 0.06 | $R_{4b}$ (0.01), $R_{4c}$ (0.01), $R_{4d}$ (0.01), $R_{4e}$ (0.01), $R_{4f}$ (0.005), $R_{4g}$ (0.005), $R_{4h}$ (0.005), $R_{4i}$ (0.005) |
| $242.03 > E_p \leq 300.00$ | 0.04 | $R_{5a}$ (0.01), $R_{5b}$ (0.005), $R_{5c}$ (0.005), $R_{5d}$ (0.005), $R_{5e}$ (0.005), $R_{5f}$ (0.005), $R_{5g}$ (0.005) |
| **Z3: $E_{proton}$/MeV** lower/upper boundaries | $w_3$ | channel type |
| $0 > E_p / \leq 30.85$ | - | - |
| $30.85 > E_p \leq 68.13$ | 0.020 | $R_{2a}$ (0.002), $R_{2b}$ (0.002), $R_{2c}$ (0.002), $R_{2d}$ (0.002), $R_{2e}$ (0.002), $R_{2f}$ (0.003), $R_{2g}$ (0.003), $R_{2h}$ (0.004) |
| $68.13 > E_p \leq 155.39$ | 0.015 | $R_{3a}$ (0.003), $R_{3b}$ (0.003), $R_{3c}$ (0.002), $R_{3d}$ (0.002), $R_{4a}$ (0.02), $R_{4b}$ (0.001), $R_{4c}$ (0.001), $R_{4d}$ (0.001) |
| $155.39 > E_p / \leq 199.64$ | 0.015 | $R_{4e}$ (0.001), $R_{4f}$ (0.001), $R_{4g}$ (0.001), $R_{4h}$ (0.001), $R_{4i}$ (0.001), $R_{5a}$ (0.004), $R_{5b}$ (0.003), $R_{5c}$ (0.003) |
| $199.64 > E_p \leq 242.03$ | 0.015 | $R_{5a}$ (0.001), $R_{5b}$ (0.001), $R_{5c}$ (0.001), $R_{5d}$ (0.003), $R_{5e}$ (0.003), $R_{5f}$ (0.002), $R_{5g}$ (0.002), $R_{5i}$ (0.002) |
| $242.03 > E_p \leq 300.00$ | 0.015 | $R_{5a}$ (0.001), $R_{5b}$ (0.001), $R_{5c}$ (0.001), $R_{5d}$ (0.001), $R_{5e}$ (0.001), $R_{5f}$ (0.001), $R_{5g}$ (0.001), $R_{5i}$ (0.002), $R_{5j}$ (0.003), $R_{5k}$ (0.003) |

**Table 13:** List of nuclear reactions of the proton - calcium interaction with $m = 20$ and $m' = 40$ yielding $p + Ca_{20}^{40} \to X_{m-q}^{m'-r}$ + *secondary particles*, weights and decay products. The overall weights of the zones Z1, Z2, Z3 amount to Z1: $w_1 = 0.56$; Z2: $w_2 = 0.33$; Z3: $w_3 = 0.11$.

| Z1: $E_{proton}$/MeV lower/upper boundaries | $w_1$ | channel type |
|---|---|---|
| $0 > E_p \leq 49.47$ | 0.15 | $R_0$ (0.1), $R_{1a-1e}$ (0.05) |
| $49.47 > E_p \leq 155.53$ | 0.14 | $R_{1a-1e}$ (0.03), $R_{2a-2h}$ (0.06), $R_{3a-3d}$ (0.06) |
| $155.53 > E_p \leq 219.62$ | 0.14 | $R_{3a-3d}$ (0.05), $R_{4a-4i}$ (0.09) |
| $219.62 > E_p \leq 300.00$ | 0.13 | $R_{3a-3d}$ (0.01), $R_{4a-4i}$ (0.12) |
| **Z2: $E_{proton}$/MeV** lower/upper boundaries | $w_2$ | channel type |
| $0 > E_p / \leq 49.47$ | 0.03 | $R_0$ (0.01), $R_{1a-1e}$ (0.02) |
| $49.47 > E_p \leq 155.53$ | 0.11 | $R_{1a-1e}$ (0.03), $R_{2a-2h}$ (0.04), $R_{3a-3d}$ (0.04) |
| $155.53 > / \leq 219.62$ | 0.1 | $R_{3a-3d}$ (0.02), $R_{4a-4i}$ (0.08) |
| $219.62 > E_p \leq 300.00$ | 0.09 | $R_{4a-4i}$ (0.09) |
| **Z3: $E_{proton}$/MeV** lower/upper boundaries | $w_3$ | channel type |
| $0 > E_p \leq 49.47$ | - | - |
| $49.47 > E_p \leq 155.53$ | 0.04 | $R_{3a-3d}$ (0.04) |
| $155.53 > E_p \leq 219.62$ | 0.04 | $R_{4a-4i}$ (0.04) |
| $219.62 > E_p \leq 300.00$ | 0.03 | $R_{6a-6k}$ (0.03) |

**Table 14:** List of nuclear reactions of the proton - copper interaction with $m = 29$ and $m' = 63$ yielding $p + Cu_{29}^{63} \to X_{m-q}^{m'-r}$ + *secondary particles*, weights and decay products. The overall weights of the zones Z1, Z2. Z3 amount to Z1: $w_1 = 0.56$; Z2: $w_2 = 0.33$; Z3: $w_3 = 0.11$.

| Z1: $E_{proton}$/MeV lower/upper boundaries | $w_1$ | channel type |
|---|---|---|
| $0 > E_p \leq 51.20$ | 0.20 | $R_0$ (0.12), $R_{1a-1e}$ (0.08) |
| $51.20 > E_p \leq 198.47$ | 0.18 | $R_{1a-1e}$ (0.02), $R_{2a-2h}$ (0.04), $R_{3a-3d}$ (0.06), $R_{4a-4i}$ (0.06) |
| $198.47 > E_p \leq 300.00$ | 0.18 | $R_{3a-3d}$ (0.02), $R_{4a-4i}$ (0.16) |
| Z2: $E_{proton}$/MeV lower/upper boundaries | $w_2$ | channel type |
| $0 > E_p \leq 51.20$ | - | - |
| $51.20 > E_p \leq 198.47$ | 0.17 | $R_{1a-1e}$ (0.02), $R_{2a-2h}$ (0.04), $R_{3a-3d}$ (0.04), $R_{4a-4i}$ (0.07) |
| $198.47 > E_p \leq 300.00$ | 0.16 | $R_{4a-4i}$ (0.10), $R_{6a-6k}$ (0.06) |
| Z3: $E_{proton}$/MeV lower/upper boundaries | $w_3$ | channel type |
| $0 > E_p \leq 51.20$ | - | |
| $51.20 > E_p \leq 198.47$ | 0.05 | $R_{1a-1e}$ (0.005), $R_{2a-2h}$ (0.015), $R_{3a-3d}$ (0.01), $R_{4a-4i}$ (0.02) |
| $198.47 > E_p \leq 300.00$ | 0.06 | $R_{6a-6k}$ (0.06) |

**Table 15:** List of nuclear reactions of the proton - cesium interaction with $m = 55$ and $m' = 137$ yielding $p + Cs_{55}^{137} \rightarrow X_{m-q}^{m'-r}$ + *secondary particles*, weights and decay products. The overall weights of the zones Z1, Z2. Z3 amount to Z1: $w_1 = 0.56$; Z2: $w_2 = 0.33$; Z3: $w_3 = 0.11$.

Figure 47 referring to $Cs_{55}^{137}$ certainly do not show any therapeutic relevance, since the purpose of this figure is to demonstrate that the methods presented in this study may have applications going beyond proton radiotherapy. Since it is rather known that $Cs_{55}^{137}$ belongs to the most critical reaction products of nuclear reactors due to its very long half-time of about 30 years. On the other hand, the nuclear reaction leading to the isotope $Cs_{55}^{136}$ represents a rather attractive reaction product due to its half-time of ≈ 20 days, and with regard to low impinging proton energies we should also like to mention the following reactions:

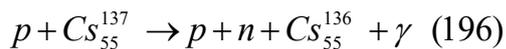
$$p + Cs_{55}^{137} \rightarrow p + n + Cs_{55}^{136} + \gamma \quad (196)$$
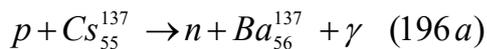
$$p + Cs_{55}^{137} \rightarrow n + Ba_{56}^{137} + \gamma \quad (196a)$$

This reaction is a rather nice one, since $Ba_{56}^{137}$ is a stable nucleus.

Nuclear reactions according to the channel (196) occur in the domain between the two maxima of 40.103 and 51.2004 MeV. If we have a glance along the complete cross-section function $Q^{red}(E)$, we are able to verify that there exist various reaction processes evoking many isotopes as reaction products with $Z < 55$ or $<< 55$ and $A_N < 137$ or $<< 137$. This possible technical application is referred to as *'transmutation'* of $Cs_{55}^{137}$ and has been taken into account as a further option besides the storage of this critical element.

With reference to the overall weight coefficients $w$ of the zones Z1, Z2, Z3 we have to form $w_1 \cdot \pi \cdot R_N^2$, $w_2 \cdot \pi \cdot R_N^2$ and $w_3 \cdot \pi \cdot R_N^2$ in order to obtain the real weights of the nuclear reactions. If the subscripts in Tables 11 and 12 start with the corresponding index and finish with the highest index, then all contributions incorporate approximately the identical weight factor, e.g. $R_{1a-1e}$, $R_{2a-2h}$, etc.

*Some final notes seem to be justified:*

1. It is evident and has already been mentioned that the release of neutrons and, by that, possible reaction channels evoked these neutrons need considerable improvement in therapy planning systems. This is particularly evident due to the high RBE of neutrons. The aspects of LET and RBE of proton beams has been discussed in [63].

2. Radiation effects occurring in connection with proton irradiation are not sufficiently accounted for. There three origins of radiation creation: Interaction of protons with nuclei by creating γ-quanta, emission of γ-quanta of the heavy recoil nuclei, and ß$^+$-decay of these nuclei yielding to production of γ-quanta via annihilation of positrons by collisions with electrons. These photons lead to a broader lateral scatter of dose profiles and reduce the LET and RBE in the environment of the Bragg peak.